\documentclass[english,aps,prl,twocolumn,amsmath,amssymb,showpacs,superscriptaddress,notitlepage,longbibliography]{revtex4-1}
\usepackage[T1]{fontenc}
\usepackage[latin9]{inputenc}
\usepackage{color}
\usepackage{amsmath}
\usepackage{amssymb}
\usepackage{graphicx}

\makeatletter

\usepackage{amssymb}
\usepackage{amsmath}
\usepackage{graphicx}
\usepackage[colorlinks=true,linkcolor=blue,anchorcolor=red,citecolor=blue,urlcolor=blue]{hyperref}
\usepackage[caption=false]{subfig}
\usepackage{lipsum}

\makeatother

\usepackage{babel}
\begin{document}
\renewcommand{\figurename}{Fig.}
\title{Symmetry breaking and spectral structure of the interacting Hatano-Nelson model}
\author{Song-Bo Zhang}
\address{Department of Physics, University of Z\"urich, Winterthurerstrasse 190, 8057, Z\"urich, Switzerland}
\author{M. Michael Denner}
\address{Department of Physics, University of Z\"urich, Winterthurerstrasse 190, 8057, Z\"urich, Switzerland}
\author{Tom\'a\v{s} Bzdu\v{s}ek}
\address{Condensed Matter Theory Group, Paul Scherrer Institute, 5232 Villigen PSI, Switzerland}
\address{Department of Physics, University of Z\"urich, Winterthurerstrasse 190, 8057, Z\"urich, Switzerland}
\author{Michael A. Sentef}
\address{Max Planck Institute for the Structure and Dynamics of Matter, Luruper Chaussee 149, 22761 Hamburg, Germany}
\author{Titus Neupert}
\address{Department of Physics, University of Z\"urich, Winterthurerstrasse 190, 8057, Z\"urich, Switzerland}
\date{\today}

\begin{abstract}
We study the Hatano-Nelson model, i.e., a one-dimensional non-Hermitian chain of spinless fermions with nearest-neighbor nonreciprocal hopping, in the presence of repulsive nearest-neighbor interactions. At half-filling, we find two $\mathcal{PT}$ transitions, as the interaction strength increases. The first transition is marked by an exceptional point between the first and the second excited state in a finite-size system and is a first-order symmetry-breaking transition into a charge-density wave regime. Persistent currents characteristic of the Hatano-Nelson model abruptly vanish at the transition. The second transition happens at a critical interaction strength that scales with the system size and can thus only be observed in finite-size systems. It is characterized by a collapse of all energy eigenvalues onto the real axis. We further show that in a strong interaction regime, but away from half-filling, the many-body spectrum shows point gaps with nontrivial winding numbers, akin to the topological properties of the single-particle spectrum of the Hatano-Nelson chain, which indicates the skin effect of extensive many-body eigenstates under open boundary conditions. Our results can be applied to other models such as the non-Hermitian Su-Schrieffer-Heeger-type model, and contribute to an understanding of fermionic many-body systems with non-Hermitian Hamiltonians.
\end{abstract}

\maketitle

\textit{Introduction}\textemdash Non-Hermitian topological phases constitute one of the most recent active research fields in condensed matter, cold atom, and photonic physics~\citep{ZPGong18PRX,HTShen18PRL,Kozzi17arXiv,Zyuzin18PRB,SYYao18prl,Kunst18prl,ELee16PRL,SYYao18prlb,Torres19JPM,Ashida20AP,Miri19Science, Bergholtz21RMP,Kawabata19PRX,KZhang20PRL,Okuma20PRL,Leykam17PRL,Martinez18PRB,Lieu18PRB,YXiong18JPC,Yoshida18PRB,WBRui19PRB,Longhi19PRL,CHLee19PRB, Kawabata19PRL,Yoshida19PRB,JYLee19PRL,Borgnia20PRL,LHLi20PRL,Wojcik20prb,Vecsei21PRB,Denner21NC,Schindler21PRB,Ryu20PRR}. They have been experimentally realized in different platforms of high controllability~\citep{WJChen17Nature,Hodaei17Nature,Weimann17Nmater,HYZhou18Science,HZhao19Science,Cerjan19Nphoto,Helbig20Nphys,Weidemann20science, Xiao20nphys,Ghatak20PNAS,WGZhang21PRL,SQXia21Science,RSu21SciAdv,KWang21Science,Guo20PRL,Liang22arXiv,Ren2022chiral}. So far, most previous efforts have been devoted to single-particle physics, with no or only perturbative many-body interactions. It is well known that in Hermitian systems strong interactions among particles give rise to many exotic phenomena, such as unconventional superconductivity, Mott insulators, and density-wave ordering~\citep{Haldane81JPC,Giamarchi03book,Fradkin13book}. Thus, it is of fundamental interest to explore non-Hermitian phenomena in many-body systems with strong interactions~\citep{Nakagawa18prl,Hamazaki19PRL,GQZhang20PRB,LJZhai20PRB,Fayard21PRR,Yamamoto19PRL,EWLee20PRB,SMu20PRB,DWZhang20PRB,TLiu20PRB, Panda20PRB,OKuma21PRL,Yoshida21PRB,Yoshida19SR,KCao21arXiv,Alsallom21arXiv,YGuo21arXiv,YGlIu21JPCM,Dora20PRL,WJXi21SB,PLei20PRA,ZHXu21PRB,Hyart21arXiv, Crippa21PRB,WZuo21arxiv,Banerjee21arxiv}. Most of the existing studies on this subject are focused on the issues of non-Hermitian many-body localization~\citep{Hamazaki19PRL,LJZhai20PRB,Fayard21PRR,GQZhang20PRB} and the non-Hermitian skin effect~\citep{EWLee20PRB,SMu20PRB,DWZhang20PRB,TLiu20PRB,Panda20PRB,Yoshida21PRB,Alsallom21arXiv,OKuma21PRL,KCao21arXiv}. However, many-body interaction effects, especially on bulk fermionic properties, remain largely unexplored even in simple models.

\begin{figure}[t]
\includegraphics[width=1\columnwidth]{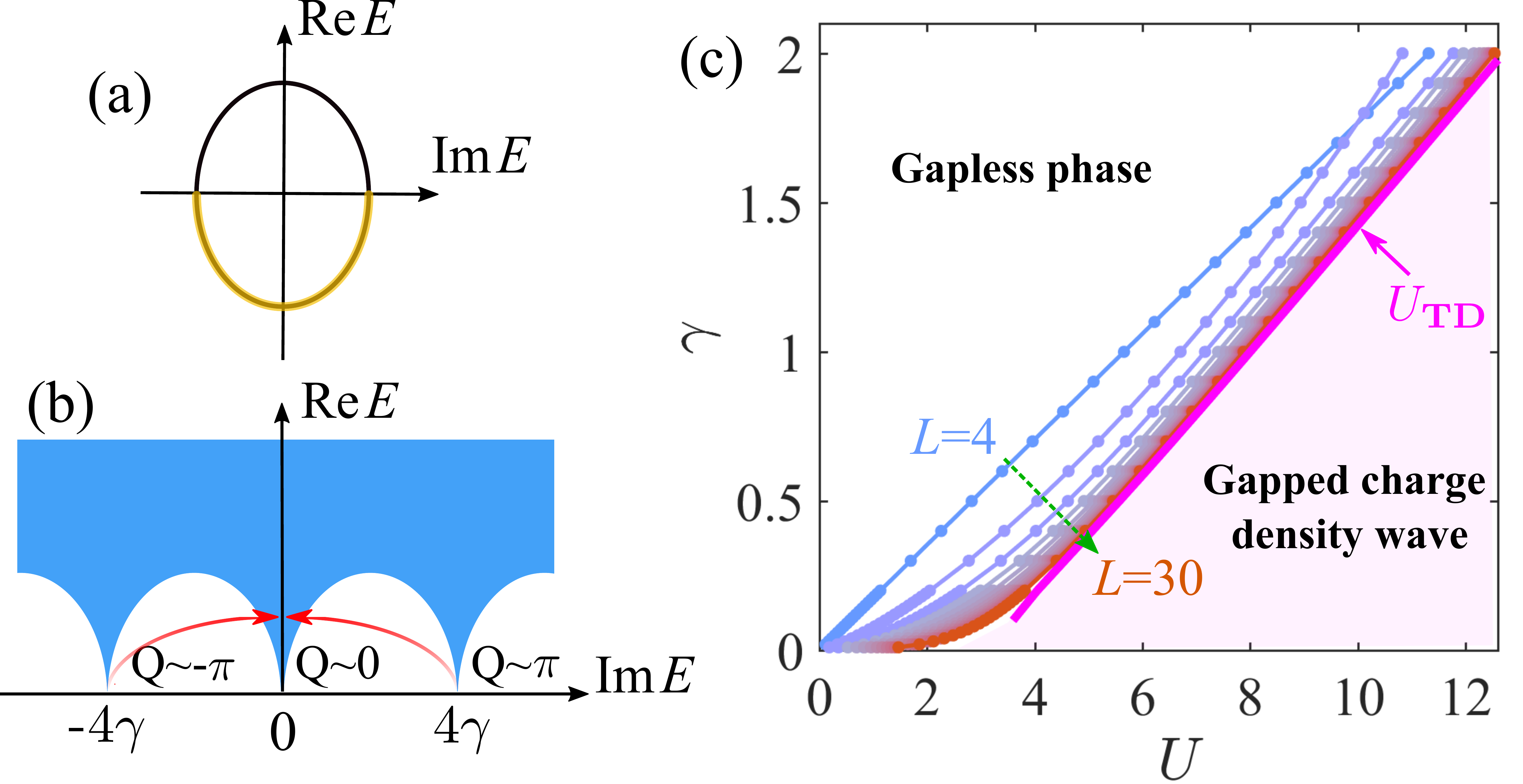}

\caption{(a) Single-particle spectrum and (b) low-excitation-energy many-body spectrum at half-filling in the absence of interactions. The yellow arc denotes the filled states. Two LES have momenta $Q=\pm \pi$ and identical $\mathrm{Re}(E)$ but opposite $\mathrm{Im}(E)\approx \pm 4i\gamma$. By increasing the interaction strength $U$ to a critical value $U_{c}$, the two LES collapse onto the real axis, as sketched by the arrows. (c) Phase diagram against $\gamma$ and $U$. The gapless phase is adiabatically connected to the Hatano-Nelson model at $U=0$, while the CDW phase is smoothly connected to the CDW phase with $\gamma=0$ and finite $U$. The magenta line is the extrapolated $U_{\text{TD}}$ in the TDL. It demarcates the gapless regime and the gapped CDW regime.}

\label{fig1:sketch}
\end{figure}

In this Letter, we study the Hatano-Nelson model of spinless fermions, a prototypical one-dimensional non-Hermitian system with nearest-neighbor nonreciprocal hopping~\citep{Hatano96PRL}, in a ring geometry and under the presence of the Pauli principle and strong Coulomb interactions. At half-filling [Fig.~\ref{fig1:sketch}(a)], we find that as the interaction strength increases, the imaginary energies of the many-body spectrum are substantially suppressed, giving rise to two $\mathcal{PT}$ transitions. The first transition is marked by an exceptional point between two lowest excited states (LES)~\citep{Note-defineGS} of a finite-size system [see a sketch in Fig.~\ref{fig1:sketch}(b)]. By employing exact diagonalization, we show that this transition corresponds to a first-order quantum phase transition of the ground state from a gapless phase to a gapped charge-density wave (CDW) that breaks translation symmetry spontaneously~[Fig.~\ref{fig1:sketch}(c)]. Moreover, it features a sudden disappearance of the characteristic persistent current of the Hatano-Nelson model in a low-temperature regime. The second transition corresponds to a full collapse of the many-body spectrum onto the real axis. Its critical interaction strength, however, increases as the system size grows. Thus, it can only be observed in finite-size systems.

For finite doping away from half-filling, the spectrum stratifies into clusters with states of a different number of simultaneously occupied nearest-neighbor sites. The energy clusters have nonzero extents along imaginary axis being largely unaffected by interactions. Furthermore, they exhibit point gaps with nontrivial winding numbers, thus indicating the skin effect of many-body states in the presence of open boundaries~\citep{Kawabata19PRX,KZhang20PRL,Okuma20PRL,KZhang21arXiv}, also for strong interactions. At half-filling, in contrast, the spectrum shrinks to open lines under strong interactions. Accordingly, the many-body states extend over the whole lattice chain with open boundaries.

\textit{Interacting Hatano-Nelson model}.\textemdash We consider the interacting Hatano-Nelson model of spinless fermions described by
\begin{equation}
\hat{H}=\sum_{\ell}[(t+\gamma)\hat{c}_{\ell}^{\dagger}\hat{c}_{\ell+1}^{}+(t-\gamma)\hat{c}_{\ell+1}^{\dagger}\hat{c}_{\ell}^{\vphantom{\dagger}}+U\hat{n}_{\ell}^{\vphantom{\dagger}}\hat{n}_{\ell+1}^{\vphantom{\dagger}}],\label{eq:model-Hamiltonian}
\end{equation}
where $\hat{c}_{\ell}^{\dagger}$ ($\hat{c}_{\ell}^{\vphantom{\dagger}})$ is the creation (annihilation) operator of a fermion at lattice site $\ell$, and $\hat{n}_{\ell}^{\vphantom{\dagger}}=\hat{c}_{\ell}^{\dagger}\hat{c}_{\ell}^{\vphantom{\dagger}}$ is the fermion number operator with eigenvalues $\{0,1\}$. The fermionic operators $\hat{c}_{\ell}^{\dagger}$ and $\hat{c}_{\ell}^{\vphantom{\dagger}}$ obey the anticommutation relations, thus imposing the Pauli principle to the system~\cite{SuppInf}. The real parameters $t$ and $\gamma$ denote the reciprocal and nonreciprocal components of the hopping between neighboring sites, respectively~\cite{Note-nonreciprocal}. The last term describes the Coulomb interaction with strength $U\geqslant0$ between two fermions at adjacent sites. Without loss of generality, we set $t>0$ to be our unit of energy.

To investigate bulk many-body properties, we consider the system in a ring geometry with $L$ sites and $N$ particles. For periodic boundary conditions (PBC) or anti-PBC, the system respects a combined space-time-reversal $(\mathcal{PT}$) symmetry~\citep{SuppInf}; thus, the eigenenergies of the system are either real or come in complex-conjugate pairs. For single particles without interactions, the spectrum reduces to a closed orbit with a point gap [Fig~\ref{fig1:sketch}(a)], resulting in the non-Hermitian skin effect of single-particle states at open boundaries~\citep{ELee16PRL,SYYao18prl,Kunst18prl}. Moreover, the system has a particle-hole symmetry~\citep{SuppInf}. Thus, the spectrum for $N$ particles is essentially the same (up to an overall shift in energy) as that for $L-N$ particles. Both of these spectral relations are reproduced by our exact-diagonalization calculations presented below.

\textit{Low-energy $\mathcal{PT}$ transition and phase diagram}.\textemdash It is instructive to first analyze the case without interactions ($U=0$). In this case, the many-body spectrum displays a scatter distribution pattern centered at the origin of the complex-energy plane. When $n_{\text{cl}}=\text{min}(N,L-N)\gg1$, its extent along real and imaginary axes can be estimated as $\Xi_{\text{R}}\approx t\alpha_{\{N,L\}}$ and $\Xi_{\text{I}}\approx\gamma\alpha_{\{N,L\}}$, respectively, where $\alpha_{\{N,L\}}=2L\sin(\pi n_{\text{cl}}/L)/\pi$~\citep{SuppInf}. Clearly, the spectrum is larger when the system is larger and filled closer to half-filling. For fixed finite $n_{\text{cl}}(\ll L)$, however, its extent is approximately independent of $L$ and determined by $\alpha_{\{N,L\}}\approx 2n_{\text{cl}}$.

\begin{figure}[t]
\includegraphics[width=1\columnwidth]{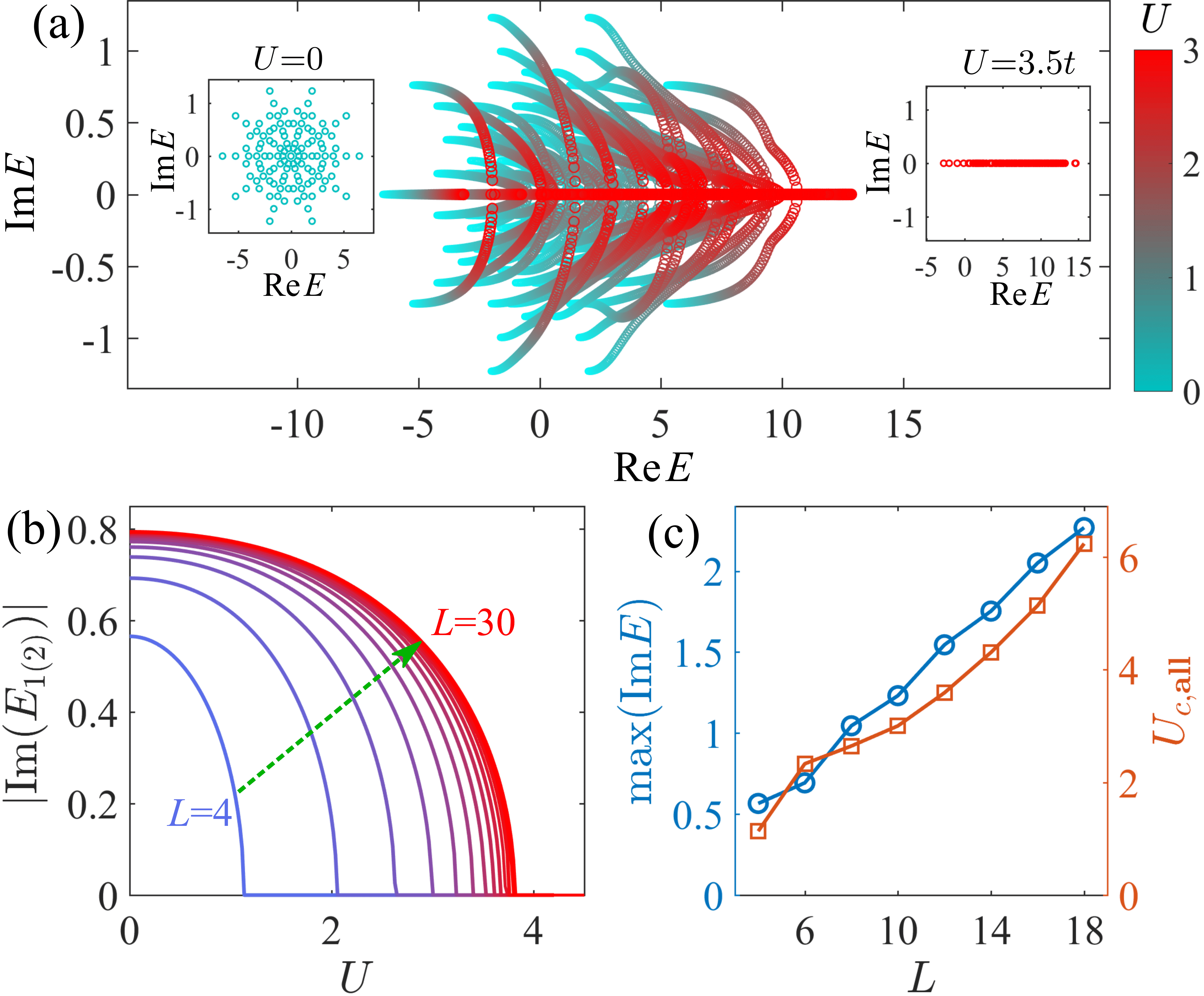}

\caption{(a) Flow of the spectrum at half-filling as $U$ increases from $0$ (cyan) to $3.5t$ (red). Insets: spectra at $U=0$ and $U$=$3t$, respectively. All eigenenergies become real when $U\geqslant U_{c,\text{all}}$. (b) $|\text{Im}(E_{1(2)})|$ as a function of $U$ for increasing $L$. The dependence of $|\text{Im}(E_{1(2)})|$ on $U$ converges to the red curve as $L$ grows.
(c) Maximum imaginary energy $\text{max(Im}\thinspace E)$ (blue) and $U_{c,\text{all}}$ (orange) as functions of $L$. Both quantities diverge as $L\rightarrow \infty$. We consider $L=10$ in (a), $\gamma=0.2t$ in all panels, adopt (anti-)PBC for odd (even) $N$.}

\label{fig3:fulltransition}
\end{figure}

More intriguing features arise when the interaction is present. We first consider the half-filled ($N=L/2$) scenario and show that a $\mathcal{PT}$ transition between $\mathcal{PT}$-symmetry-broken and -unbroken phases occurs at low excitation energies, as $U$ increases. As illustrated in Fig.~\ref{fig1:sketch}(b) and the inset of Fig.~\ref{fig3:fulltransition}(a), at $U=0$, there is one ground state with real energy $E_0$ and two LES with complex-conjugate energies $E_2=E_1^*$ in systems with odd (even) $N$ and (anti-)PBC. When $L\gg1$, we have $E_0\approx 2tL/\pi$ and $E_1=E_2^*\approx E_0+4t\sin(\pi/L)+4i\gamma\cos(\pi/L)$. As $U$ increases, we find that the energies of the LES approach each other and merge at an exceptional point on the real axis at a critical strength $U=U_c$, and split along the real axis for $U>U_c$.

To better understand the \emph{low-energy} $\mathcal{PT}$ transition, we determine $U_{c}$ for varying $\gamma$ and increasing $L$ [Figs.~\ref{fig1:sketch}(c) and \ref{fig3:fulltransition}(b)]. Evidently, $|\text{Im}(E_{1(2)})|$ decays monotonically with increasing $U(<U_{c})$ and completely vanishes when $U>U_{c}$. As $L$ grows, the dependence of $|\text{Im}(E_{1(2)})|$ on $U$ converges to a curve [red line in Fig.~\ref{fig3:fulltransition}(b)]. In the thermodynamic limit (TDL) $L\to \infty$, $U_{c}$ converges to a finite value $U_{\text{TD}}$. Specifically, for fixed $\gamma,$ $U_{c}$ shows a power-law scaling as $L$ grows, i.e.,
\begin{equation}
U_{c}=U_{\text{TD}}-\beta L^{-\alpha},\label{eq:fitting}
\end{equation}
where $\alpha$ and $\beta$ are positive numbers depending on $\gamma$. This feature enables us to extrapolate $U_{\text{TD}}$. The phase diagram parametrized by $U$ and $\gamma$ is given in Fig.~\ref{fig1:sketch}(c), where the red line is $U_{\text{TD}}$ which marks the phase boundary in the TDL~\citep{Note-failatsmallgamma}. We observe that
$U_{\text{TD}}$ grows monotonically with $\gamma$, indicating that the $\mathcal{PT}$ transition occurs even in the TDL and for ultrastrong nonreciprocity ($|\gamma|\geqslant t$). More details of the calculation are given in the Supplemental Material~\citep{SuppInf}.

\begin{figure}[t]
\includegraphics[width=1\columnwidth]{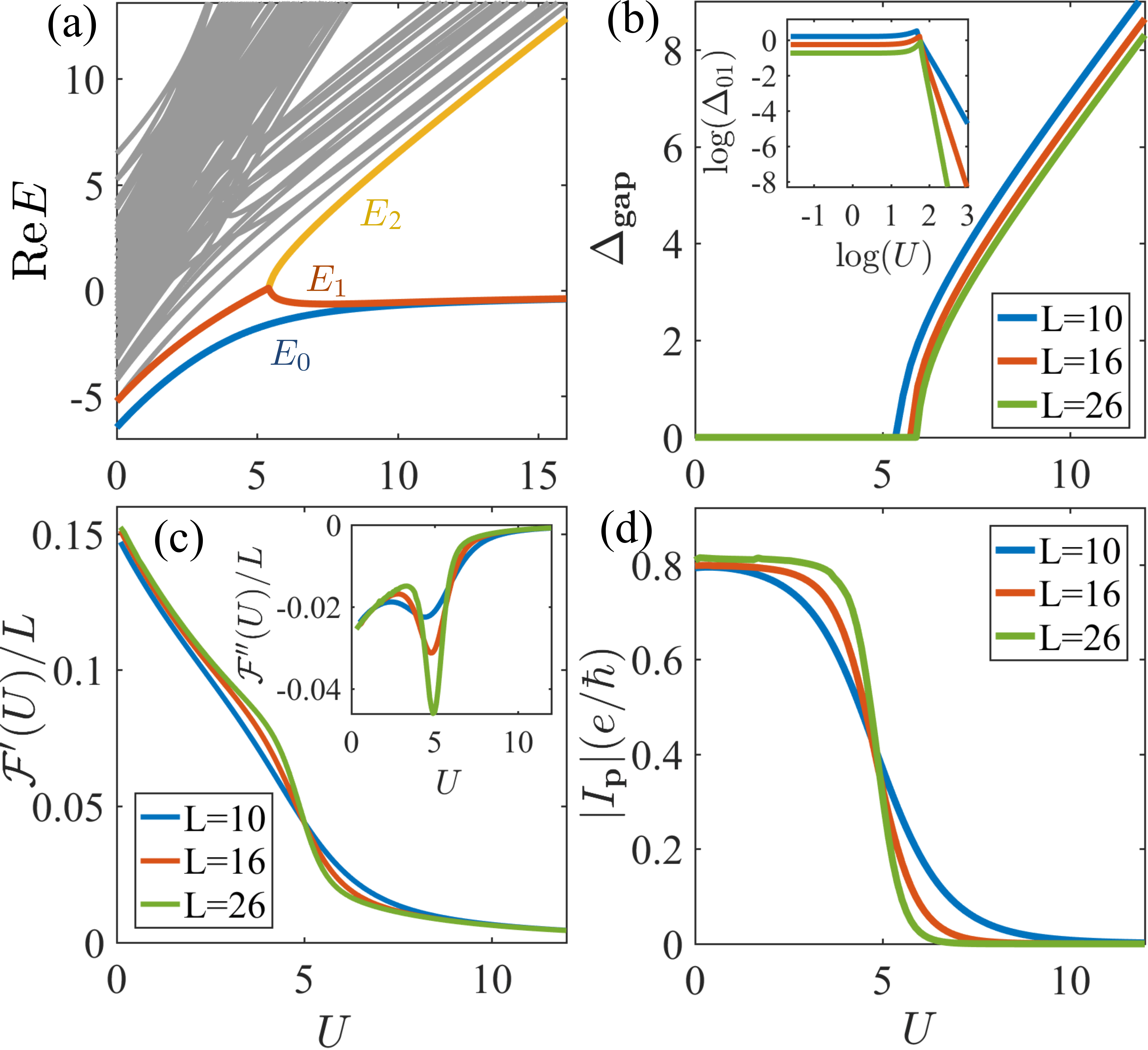}

\caption{(a) Real part of the spectrum at half-filling as a function of $U$. Three lowest energy levels are respectively denoted as $E_{0}$, $E_{1}$ and $E_{2}$. (b) $\Delta_{\text{gap}}$ as a function of $U$. Inset: $\text{log}(\Delta_{01})$ as a function of $\text{log}(U)$. (c) The first and second (inset) derivatives of the free energy as functions of $U$. (d) $|I_{\text{P}}|$ (in units of $e/\hbar$) as a function of $U$. We consider $\gamma=0.6t$ in all panels, $L=10$ in (a), $k_{\mathrm{B}}T=0.4t$ and $50$ lowest-$\mathrm{Re}(E)$ eigenstates in (c) and (d).}

\label{fig2:1storder}
\end{figure}

To further understand the physics behind the phase diagram, we analyze the real part of the low-excitation-energy spectrum [Fig.~\ref{fig2:1storder}(a)]. For $U<U_{c}$, the two LES are degenerate in real energy. The finite-size level spacing $\Delta_{01}\equiv\text{Re}(E_{1}-E_{0})$ is approximately constant [$\approx4t\sin(2\pi/L)$] for a wide range of $U$ and increases subtly when approaching $U_{c}$ [inset of Fig.~\ref{fig2:1storder}(b)]. However, $\Delta_{01}$ decreases as $L$ grows and it vanishes in the TDL, indicating a gapless phase when $U<U_c$. For $U>U_{c}$, the LES have vanishing imaginary energy while being split in real energy. Upon further increasing $U$, the energy splitting $\Delta_{\text{gap}}\equiv\text{Re}(E_{2}-E_{1})$ increases whereas $\Delta_{01}$ decreases sharply [Fig.~\ref{fig2:1storder}(b)]. More explicitly, $\Delta_{01}$ follows a power-law dependence $\Delta_{01}\propto U^{1-L/2}$ on $L$. Thus, in large systems, one of the LES rapidly becomes degenerate with the ground state and separated from the excited states by a large gap $\Delta_{\text{gap}}$, implying a transition of the system into a gapped regime. The degenerate ground states break translation symmetry spontaneously, forming a CDW with long-range density-density correlation~\citep{SuppInf}. The $\mathcal{PT}$ transition may also be related to the breakdown of the Mott insulator which instead considers two spin species and on-site interactions~\cite{Fukui98PRB}.


\textit{Free energy and persistent current}.\textemdash Next, we consider the free energy of the system, which reads $\mathcal{F} = -k_{\textrm{B}}T\text{log}(\sum_{j}e^{-E_{j}/k_{B}T})$, where $T$ is the temperature, $k_{\textrm{B}}$ is the Boltzmann constant and $\sum_j$ sums over all eigenenergies. Since the eigenenergies are real or come in complex-conjugate pairs due to the $\mathcal{PT}$ symmetry, $\mathcal{F}$ is always purely real. At low temperatures, $\mathcal{F}$ is determined mainly by the low-$\mathrm{Re}(E)$ eigenstates in an energy window of magnitude $k_{B}T$. As discussed above, in the TDL, the system is gapless for $U<U_{c}$, whereas it quickly develops a large energy gap after $U>U_{c}$. As a result, $\mathcal{F}$ and its derivatives (with respect to $U$) change significantly at $U_{c}$, provided that $\Delta_{01}<k_{B}T\ll U_{c}$.  In Fig.~\ref{fig2:1storder}(c), we calculate the first and second derivatives of $\mathcal{F}$ at low temperatures as functions of $U$. We observe that for $L/2\gg 1$, the first derivative $\mathcal{F}'\equiv d\mathcal{F}/dU$ shows a sudden drop while the second derivative $\mathcal{F}''\equiv d^2\mathcal{F}/dU^2$ diverges at $U_{c}$. These features are more pronounced in larger systems, suggesting that the low-energy $\mathcal{PT}$ transition is of first order. This is in sharp contrast to the Hermitian limit ($\gamma=0$), where the transition is of Berezinskii-Kosterlitz-Thouless type~\citep{Affleck86CMP,Dalmonte15PRB,Note-BKT-transition} (see also~\citep{SuppInf}).

In a metallic ring, a persistent current $I_{\text{p}}$ can be induced as the response of $\mathcal{F}$ to a small change of magnetic flux $\phi$ through the ring, i.e., $I_{\text{p}}=-(e/\hbar)\partial\mathcal{F}/\partial\phi$~\citep{Byers61PRL}. Notably, by virtue of its non-Hermitian hopping, the Hatano-Nelson model supports an imaginary current $I_{\text{p}}$ at zero flux for $U<U_{c}$~\citep{Pi-current}. In the TDL and for $U=0$ and $T=0$, $I_{\text{p}}$ can be derived as $I_{\text{p}}=4ie\gamma/h$~\citep{SuppInf}. Moreover, when $\Delta_{01}<k_{B}T\ll U_{c}$, $I_{\text{p}}$ is approximately constant for $U<U_{c}$, whereas it suddenly drops to zero for $U>U_{c}$, as shown numerically in Fig.~\ref{fig2:1storder}(d). For small $U$, $I_p$ saturates for large $L$, and it exhibits a sudden drop at the transition that sharpens with increasing $L$. This is in contrast to the persistent current in Hermitian systems that requires a finite flux and vanishes in the TDL~\citep{Cheung88PRB}. Note that the imaginary current characterizes the delocalization of eigenstates~\citep{Hatano96PRL}. The sudden disappearance of $I_{\text{p}}$ thus constitutes another indicator of the metal-insulator transition in the low-$\text{Re}(E)$ regime.

\textit{$\mathcal{PT}$ transition in the full spectrum}.\textemdash The full many-body spectrum can also exhibit a $\mathcal{PT}$ transition at half-filling. As shown in Fig.~\ref{fig3:fulltransition}(a) for $|\gamma|<t$, the imaginary part of the spectrum is dramatically suppressed by increasing $U$ and, more remarkably, all eigenenergies collapse onto the real axis after a critical strength $U_{c,\text{all}}$ in a system with odd (even) $N$ and (anti-\nolinebreak{)}PBC. This \emph{full} $\mathcal{PT}$ transition can be understood as follows. In the presence of Coulomb interactions, the many-body Fock states of the system acquire different Coulomb potentials determined by their occupation configurations~\citep{Note-configuration}, forming different groups with different Coulomb potentials (we term them Fock components for convenience). At half-filling, we find that the imaginary energies of the spectrum mainly stem from the nonreciprocal hopping between different Fock components. By increasing $U$, the energy separation between the Fock components grows and the coupling between them becomes weaker. Thus, the imaginary energies are suppressed. We stress that the complex-real transition discovered here emerges in the many-body spectrum and is driven by two-particle interactions, distinctively different from the complex-real transition in the single-particle spectrum driven by disorders~\citep{Hatano96PRL}.

The value of $U_{c,\text{all}}$ depends on the system size $L$ and nonreciprocity $\gamma$. In larger systems, there are more excited states with larger imaginary energies at $U=0$. When $L\gg1$, the maximum imaginary energy $\mathrm{max[Im}(E)]$ is approximately $2\gamma L/\pi$, which grows linearly with $L$ [in contrast to the imaginary energies of LES, which is bounded by $\text{Im}(E_1)<4\gamma$]. Thus, in order to completely suppress the imaginary energies, a stronger $U_{c,\text{all}}$ is required. Explicitly, $U_{c,\text{all}}$ scales with the system size and can thus only be observed in finite-size systems [Fig.~\ref{fig3:fulltransition}(c)]. Similarly, for larger $\gamma$, we have larger imaginary energies at $U=0$ and thus larger $U_{c,\text{all}}$.

For $|\gamma|\geqslant t$, the imaginary energies stem not only from the nonreciprocal hopping between different Fock components but also from those between the states within the same component. Since these hoppings are not suppressed by the $U$-driven separation of the Fock components, the imaginary energies are a robust property, and the full $\mathcal{PT}$ transition is not realized at any finite $U$.

We also note that the above-discussed $\mathcal{PT}$ transitions occur only in the case with odd (even) $N$ and (anti-)PBC while they are absent in the case with even (add) $N$ and (anti-)PBC. However, we expect the phase diagram in the TDL to be identical for all the cases~\cite{SuppInf}.

\begin{figure}[t]
\includegraphics[width=1\columnwidth]{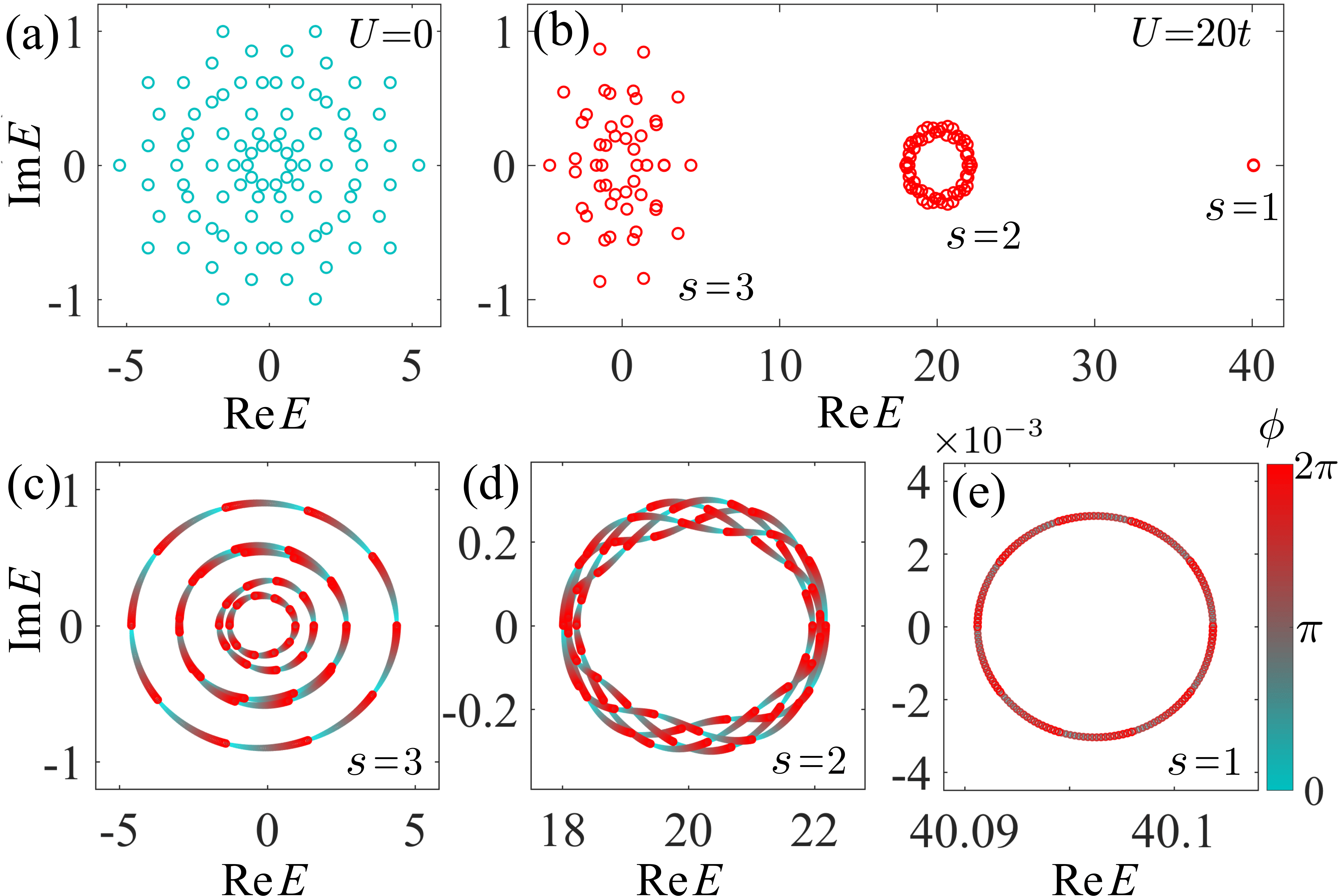}

\caption{Many-body spectrum at (a) $U=0$ and (b) $20t$, respectively. (c--e) Movement of the spectral clusters as $\phi$ varies from $0$ (cyan) to $2\pi$ (red). Parameters: $L=10$, $N=3$ and $\gamma=0.2t$.}

\label{fig4:cluster}
\end{figure}

\textit{Spectral clusters with nontrivial windings away from half-filling}.\textemdash  Finally, we turn to the non-half-filled case where additional interesting properties emerge in the presence of strong interactions. First, the many-body spectrum is substantially redistributed and dispersed from a connected area in the complex-energy plane into $n_{\text{cl}}$ clusters when $U>\Xi_{\text{R}}$ [Figs.~\ref{fig4:cluster}(a-b)]. Each cluster corresponds to a Fock component with a given Coulomb potential. Accordingly, the clusters are centered respectively around the energies $\varepsilon_{s}=(N-s)U$ with $1\leq s\leq n_{\text{cl}}$ labeling the clusters. Second, each cluster by itself also exhibits a symmetric pattern in the complex-energy plane. Due to particle-hole symmetry, the clusters for $L-N$ particles are the same as those for $N$ particles up to an overall energy shift $|2N-L|U$. Finally, in the non-half-filled case, the imaginary energies always comprise nonreciprocal hopping between states within the same Fock component. Thus, the clusters are insensitive to $U$ in the strong $U$ regime, except for the cluster centered at $\varepsilon_{1}$, whose extent shrinks with increasing $U$.

The spectral clusters can be characterized by nontrivial topological invariants. To see this, for finite-size systems, we introduce a twist angle $\phi$ to the PBC~\citep{ZPGong18PRX}. When $\phi$ increases from $0$ to $2\pi$, all the eigenenergies belonging to cluster $s$ wind around the center $\varepsilon_{s}$ in one direction determined by $\text{sgn}(\gamma)$ [Figs.~\ref{fig4:cluster}(c--e)]. Thus, a nontrivial winding number can be found as
\begin{align}
\nu_{s} & =\int_{0}^{2\pi}\dfrac{d\phi}{2\pi i}\sum_{j}\partial_{\phi}\text{log}\{E_{j}(\phi)-\varepsilon_{s}\},
\end{align}
where $\sum_{j}$ sums over all eigenenergies. Specifically, we find $\nu_1=\text{sgn}(\gamma)N$ for $s=1$. For other clusters $s\geq2$, $\nu_s$ depend also on $L$ and diverge as $L\rightarrow \infty$. Furthermore, for fixed $N$ in the TDL, the clusters form at least one continuous orbit of eigenenergies surrounding each energy center~\citep{SuppInf}. Hence, the winding numbers of these orbits can also define nontrivial topological invariants. Note that the winding numbers defined for the spectral clusters are general for strong interactions $U>\Xi_{\text{R}}$~\cite{Note-winding}. The spectrum under open boundary conditions collapses to open lines without windings. The nonzero winding numbers under PBC indicate the localization of the many-body eigenstates at the Fock basis states with particles accumulated at an open boundary (if present) and hence the localization of the many-body eigen wavefunctions towards the boundary~\citep{SMu20PRB,Alsallom21arXiv,Note-Lee20PRB}, regardless of the Pauli principle and strong interactions in the system. We show this explicitly in the supplemental materials~\citep{SuppInf}. Our work thus constitutes a many-body interacting generalization of the spectral winding number~\citep{Kawabata19PRX,KZhang20PRL,Okuma20PRL,KZhang21arXiv} used to characterize the non-Hermitian skin effect.

By contrast, at half-filling, the spectrum of a finite-size system under PBC shrinks to open lines in the strong interaction regime, as discussed before. In this case, the spectral winding numbers become ill-defined. The many-body wavefunctions extend over the whole lattice even with boundaries~\citep{SuppInf}.

\textit{Summary and discussion}.\textemdash We have revealed two $\mathcal{PT}$ transitions in the interacting Hatano-Nelson model at half-filling upon increasing interaction strength: one is marked by an exceptional point between two LES, and the other one is characterized by a full collapse of the many-body spectrum onto the real axis. The former transition corresponds to a symmetry-breaking transition into a gapped CDW regime and features a sudden disappearance of the persistent current at low temperatures. We have further shown that with strong interactions but away from half-filling, the many-body spectrum stratifies into multiple clusters characterized by nontrivial winding numbers.

It is important to note that our main results from the Hatano-Nelson model are general.  They can also be applied to other models, such as Su-Schrieffer-Heeger (SSH) type model that has nonreciprocal hopping for every two nearest-neighbor bonds, as we have verified in the supplemental materials~\citep{SuppInf}. Our theory may be implemented, for instance, in open quantum dot~\cite{Ferry12FP,LLZhang18SM,Zajac16PRApplied,Hensgens17Nature}, cold-atom~\cite{LHLi20PRL,Guo20PRL,Liang22arXiv,Ren2022chiral,Chin10RMP,Zhou21arXiv,Guo21arXiv}, and monitored quantum circuit systems~\cite{Kells21arXiv,Fleckenstein21arXiv}.

\begin{acknowledgments}We thank F. Alsallom, N. Astrakhantsev, M. Brzezi\'nska, J. Budich, X.-D. Cao, S.-J. Choi, C.-A. Li, L. Herviou, W. Rui, G. Tang, A. Tiwari, and O. V. Yazyev for valuable discussions. This work was supported by the European Research Council (ERC) under the European Union's Horizon 2020 research and innovation programm (ERC-StG-Neupert-757867-PARATOP) and from NCCR MARVEL funded by the SNSF. M.~S. was supported by DFG through the Emmy Noether program (SE 2558/2). T.~B. was supported by the Ambizione grant No.~185806 by the Swiss National Science Foundation.
\end{acknowledgments}

{\textit{Note added in proof.}} Recently, we noticed the related work \cite{Kawabata22PRR}, which focuses on the spectral winding and non-Hermitian skin effect.

\onecolumngrid

\clearpage
\appendix

\setcounter{figure}{0}
\setcounter{page}{1}
\renewcommand\thefigure{S\arabic{figure}}
\renewcommand{\thepage}{S\arabic{page}}
\global\long\def\thesection{\Roman{section}}
\global\long\def\thesubsection{\Alph{subsection}}

\onecolumngrid

In this Supplemental Material, we present the analysis of the system symmetries (Sec.~\ref{sec:symmetry}), Pauli exclusion principle (Sec.~\ref{sec:Pauli}),  dimensions of the many-body spectrum (Sec.~\ref{Sec:estimation-of-cluster}), more discussions of critical interaction strength (Sec.~\ref{sec:Uc}), density-density correlation (Sec.~\ref{Sec:correlation-function}), free energy and persistent current (Sec.~\ref{Sec:Free-energy}), spectral winding numbers and charge-density profiles (Sec.~\ref{Sec:windingnumber}), other complementary half-filled cases (Sec.~\ref{sec:other-half-filling}), and the non-Hermitian SSH-type model (Sec.~\ref{Sec:other-model}).

\section{$\mathcal{PT}$ and particle-hole symmetries} \label{sec:symmetry}

For general twisted periodic boundary conditions (PBC) characterized by a twist boundary angle $\phi$, the Hamiltonian reads
\begin{align}
\hat{H} & =\sum_{\ell=1}^{L-1}[(t+\gamma)\hat{c}_{\ell}^{\dagger}\hat{c}_{\ell+1}^{\vphantom{\dagger}}+(t-\gamma)\hat{c}_{\ell+1}^{\dagger}\hat{c}_{\ell}^{\vphantom{\dagger}}+U\hat{n}_{\ell}^{\vphantom{\dagger}}\hat{n}_{\ell+1}^{\vphantom{\dagger}}]+e^{i\phi}(t+\gamma)\hat{c}_{L}^{\dagger}\hat{c}_{1}^{\vphantom{\dagger}}+e^{-i\phi}(t-\gamma)\hat{c}_{1}^{\dagger}\hat{c}_{L}^{\vphantom{\dagger}}+U\hat{n}_{L}^{\vphantom{\dagger}}\hat{n}_{1}^{\vphantom{\dagger}},
\end{align}
Particularly, $\phi=0$ and $\pi$ correspond to PBC and anti-PBC, respectively. We define the $\mathcal{PT}$ symmetry operation as
\begin{equation}
\mathcal{PT}\hat{c}_{\ell}^{\vphantom{\dagger}}(\mathcal{PT})^{-1}=\hat{c}_{L+1-\ell}^{\vphantom{\dagger}},\ \ \ \mathcal{PT}i(\mathcal{PT})^{-1}=-i.
\end{equation}
Acting with $\mathcal{PT}$ on $\hat{H}$, we find
\begin{align}
 & \mathcal{PT}\hat{H}(\mathcal{PT})^{-1}\nonumber \\
= & \mathcal{PT}\Big\{\sum_{\ell=1}^{L-1}[(t+\gamma)\hat{c}_{\ell}^{\dagger}\hat{c}_{\ell+1}^{\vphantom{\dagger}}+(t-\gamma)\hat{c}_{\ell+1}^{\dagger}\hat{c}_{\ell}^{\vphantom{\dagger}}+U\hat{n}_{\ell}^{\vphantom{\dagger}}\hat{n}_{\ell+1}^{\vphantom{\dagger}}]+e^{i\phi}(t+\gamma)\hat{c}_{L}^{\dagger}\hat{c}_{1}^{\vphantom{\dagger}}+e^{-i\phi}(t-\gamma)\hat{c}_{1}^{\dagger}\hat{c}_{L}^{\vphantom{\dagger}}+U\hat{n}_{L}^{\vphantom{\dagger}}\hat{n}_{1}^{\vphantom{\dagger}}\Big\}(\mathcal{PT})^{-1}\nonumber \\
= & \sum_{\ell=1}^{L-1}[(t+\gamma)\hat{c}_{L+1-\ell}^{\dagger}\hat{c}_{L-\ell}^{\vphantom{\dagger}}+(t-\gamma)\hat{c}_{L-\ell}^{\dagger}\hat{c}_{L+1-\ell}^{\vphantom{\dagger}}+U\hat{n}_{L+1-\ell}^{\vphantom{\dagger}}\hat{n}_{L-\ell}^{\vphantom{\dagger}}]+e^{-i\phi}(t+\gamma)\hat{c}_{1}^{\dagger}\hat{c}_{L}^{\vphantom{\dagger}}+e^{i\phi}(t-\gamma)\hat{c}_{L}^{\dagger}\hat{c}_{1}^{\vphantom{\dagger}}+U\hat{n}_{1}^{\vphantom{\dagger}}\hat{n}_{L}^{\vphantom{\dagger}}\nonumber \\
= & \sum_{\jmath=1}^{L-1}[(t+\gamma)\hat{c}_{\jmath+1}^{\dagger}\hat{c}_{\jmath}^{\vphantom{\dagger}}+(t-\gamma)\hat{c}_{\jmath}^{\dagger}\hat{c}_{\jmath+1}^{\vphantom{\dagger}}+U\hat{n}_{\jmath+1}^{\vphantom{\dagger}}\hat{n}_{\jmath}^{\vphantom{\dagger}}]+[e^{i\phi}(t+\gamma)\hat{c}_{L}^{\dagger}\hat{c}_{1}^{\vphantom{\dagger}}+e^{-i\phi}(t-\gamma)\hat{c}_{1}^{\dagger}\hat{c}_L^{\vphantom{\dagger}}]^{\dagger}+U\hat{n}_{1}^{\vphantom{\dagger}}\hat{n}_{L}^{\vphantom{\dagger}}.
\end{align}
In the last line, we have replaced $\jmath=L-\ell$. For PBC and anti-PBC, it follows that
\begin{align}
\mathcal{PT}\hat{H}(\mathcal{PT})^{-1} & =\hat{H}^{\dagger}.
\end{align}
This relation indicates that the eigenenergies of $\hat{H}$ must either be real or come in complex-conjugate pairs.

To see the particle-hole symmetry explicitly, we divide the lattices into two sublattices $A$ and $B$ of even and odd sites. We perform the transformation $\hat{c}_{\ell}^{\dagger}\rightarrow\hat{c}_{\ell}^{\vphantom{\dagger}}$ and $\hat{c}_{\ell}^{\vphantom{\dagger}}\rightarrow\hat{c}_{\ell}^{\dagger}$ on sublattice $A$ while $\hat{c}_{\ell}^{\dagger}\rightarrow-\hat{c}_{\ell}^{\vphantom{\dagger}}$ and $\hat{c}_{\ell}^{\vphantom{\dagger}}\rightarrow-\hat{c}_{\ell}^{\dagger}$ on sublattice $B$. Accordingly, the transformation for the occupation number operators reads $\hat{n}_{\ell}\rightarrow1-\hat{n}_{\ell}$. Then, it is clear to see that the transformed Hamiltonian takes the same form up to a constant energy shift.

\section{Pauli exclusion principle} \label{sec:Pauli}

In this work, we consider the interacting Hatano-Nelson model
which can be written in the basis of conventional fermionic
operators $\hat{c}_{\ell}^{\dagger}$ and $\hat{c}_{\ell}^{{\color{white}\dagger}}$
as
\begin{equation}
\hat{H}=\sum_{\ell=1}^{L}[(t+\gamma)\hat{c}_{\ell}^{\dagger}\hat{c}_{\ell+1}^{{\color{white}\dagger}}+(t-\gamma)\hat{c}_{\ell+1}^{\dagger}\hat{c}_{\ell}^{{\color{white}\dagger}}+U\hat{n}_{\ell}\hat{n}_{\ell+1}].
\end{equation}
The fermionic operators $\hat{c}_{\ell}^{\dagger}$ and $\hat{c}_{\ell}^{{\color{white}\dagger}}$ follow the anticommutation relations
\begin{equation}
\{\hat{c}_{\ell}^{\dagger},\hat{c}_{\jmath}^{{\color{white}\dagger}}\}=\delta_{\ell,\jmath},\ \ \{\hat{c}_{\ell},\hat{c}_{\jmath}\}=0.
\end{equation}
With these relations, we find that the number operator
$\hat{n}_{\jmath}\equiv\hat{c}_{\jmath}^{\dagger}\hat{c}_{\jmath}^{{\color{white}\dagger}}$
at site $\jmath$ follows the relation
\begin{equation}
\hat{n}_{\jmath}^{2}=\hat{c}_{\jmath}^{\dagger}\hat{c}_{\jmath}^{{\color{white}\dagger}}\hat{c}_{\jmath}^{\dagger} \hat{c}_{\jmath}^{{\color{white}\dagger}}=\hat{c}_{\jmath}^{\dagger}\left(1-\hat{c}_{\jmath}^{\dagger} \hat{c}_{\jmath}^{{\color{white}\dagger}}\right)\hat{c}_{\jmath}^{{\color{white}\dagger}}=\hat{c}_{\jmath}^{\dagger} \hat{c}_{\jmath}^{{\color{white}\dagger}}=\hat{n}_{\jmath},
\end{equation}
which gives $\hat{n}_{\jmath}=0$ or $1$. This means that an arbitrary site $\jmath$ is at most occupied by one electron, which
is the Pauli exclusion principle considered in this work.

The singe-particle eigenstates also follow the Pauli exclusion principle.
To illustrate this, we rewrite the single-particle Hamiltonian (with $U=0$) as
\begin{equation}
\hat{H}=\sum_{n}E_{n}|R_{n}\rangle\langle L_{n}|,
\end{equation}
where $|L_{n}\rangle$ and $|R_{n}\rangle$ are the left and right
eigenstates of $\hat{H}$, satisfying $\hat{H}|R_{n}\rangle=E_{n}|R_{n}\rangle$
and $\hat{H}^{\dagger}|L_{n}\rangle=E_{n}^{*}|R_{n}\rangle$, respectively.
The left and right eigenstates have the biorthogonal relations
$\langle L_{m}|R_{n}\rangle=\delta_{m,n}$. If we define $d_{n,R}^{\dagger}$
($d_{n,L}^{\dagger}$) as the creation operator related to the eigenstate
$|R_{n}\rangle$ ($|L_{n}\rangle$),
\begin{equation}
d_{n,R}^{\dagger}=\sum_{\ell}\langle\ell|R_{n}\rangle\hat{c}_{\ell}^{\dagger},
\end{equation}
then we find that these operators have the following modified anticommutation
relations
\begin{align}
\{d_{m,R}^{\dagger},d_{n,L}^{{\color{white}\dagger}}\} & =\delta_{m,n},\nonumber \\
\{d_{m,R}^{\dagger},d_{n,R}^{\dagger}\} & =0,\nonumber \\
\{d_{m,R}^{\dagger},d_{n,R}^{{\color{white}\dagger}}\} & =\langle R_{n}|R_{m}\rangle.\label{eq:modified-relation}
\end{align}
Note that for non-Hermitian system, not all the eigenstates $|R_{m}\rangle$
are orthogonal, i.e., $\langle R_{n}|R_{m}\rangle$ could be finite
even for $m\neq n$. Using the modified anticommutation relations in Eq.\ (\ref{eq:modified-relation}),
we find that the number operator $\hat{\aleph}_{m}\equiv d_{m,R}^{\dagger}d_{m,R}$
for the eigenstate $|R_{m}\rangle$ follows
\begin{equation}
\hat{\aleph}_{m}^{2}=d_{m,R}^{\dagger}\left(\langle R_{m}|R_{m}\rangle-d_{m,R}^{\dagger}d_{m,R}\right)d_{m,R}=\langle R_{m}|R_{m}\rangle\hat{\aleph}_{m}.
\end{equation}
Thus, $\hat{\aleph}_{m}$ can only have the eigenvalues 0 and $\langle R_{m}|R_{m}\rangle$, where the latter can be different from 1 (in particular larger). Similar results can be obtained for the left eigenstates. These results show that the eigenstates also obey the Pauli exclusion principle, i.e., a fermionic state can be occupied at most by one particle simultaneously.

\section{Dimension of the many-body spectrum \label{Sec:estimation-of-cluster}}

The extent of the spectrum at $U=0$ on the real and imaginary axes can be written as
\begin{equation}
\Xi_{\text{R}}\approx t\alpha_{\{N,L\}}\ \text{and }\ \Xi_{\text{I}}\approx\gamma\alpha_{\{N,L\}},\label{eq:ground-state}
\end{equation}
where $\kappa$ are integers, $n_{\text{cl}}\equiv \text{min}(N,L-N)$ and
\begin{equation}
\alpha_{\{N,L\}}=2\sum_{|\kappa|\leqslant n_{\text{cl}}/2}\cos(2\pi\kappa/L).\label{eq:alpha}
\end{equation}
For $n_{\text{cl}}\gg1$, we transform the summation in Eq.~(\ref{eq:alpha}) to an integral and obtain
\begin{equation}
\alpha_{\{N,L\}}=\dfrac{2L}{\pi}\sin\Big(\dfrac{n_{\text{cl}}\pi}{L}\Big).\label{eq:alpha-1}
\end{equation}
At half-filling, $n_{\text{cl}}=L/2$. Thus, $\alpha_{\{N,L\}}$ simplifies to
\begin{equation}
\alpha_{\{N,L\}}=\dfrac{2L}{\pi}.\label{eq:alpha-1-1}
\end{equation}

\section{Critical interaction strength in the thermodynamic limit \label{sec:Uc}}
\begin{figure}[ht]
\includegraphics[width=0.8\columnwidth]{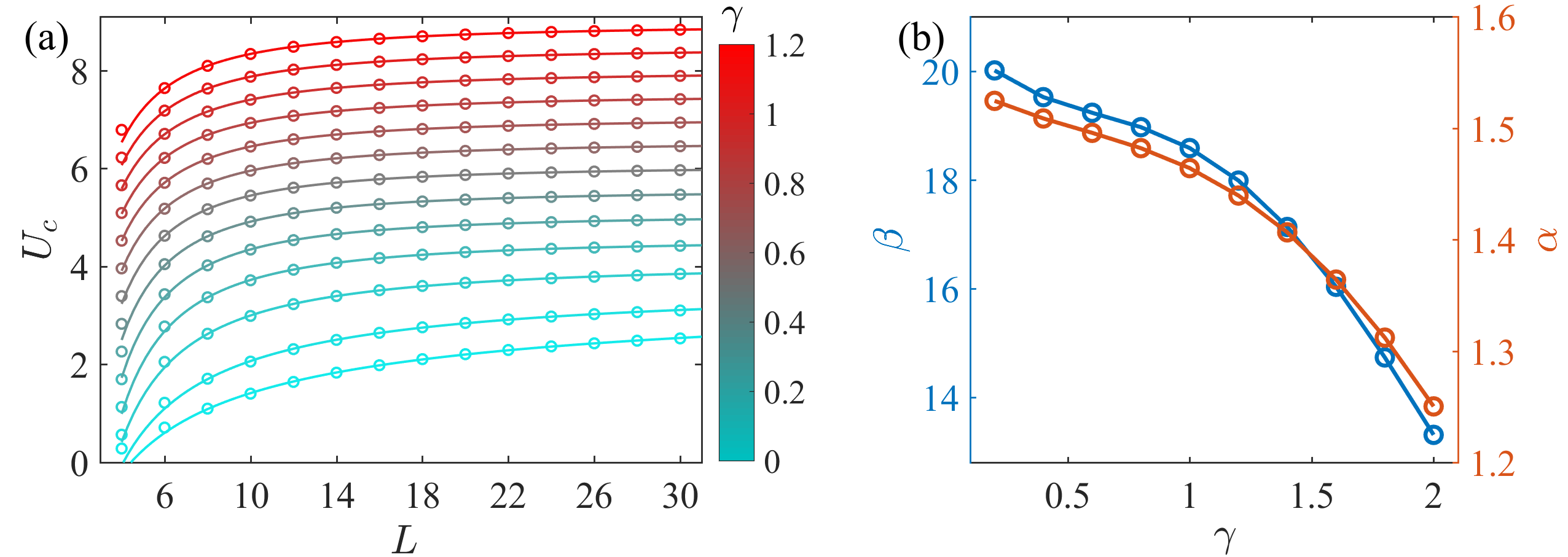}

\caption{(a) Critical interaction strength $U_{c}$ as a function of $L$ for different nonreciprocity $\gamma$. The color changes from cyan to red when $\gamma$ increases from 0 to $1.2t$. The circle dots are exact diagonalization results and the curves are the fitting with the power-law relation, Eq.\ (2), in the main text. (b) Extrapolated index $\alpha$ (right orange) and decaying strength $\beta$ (left blue) as functions of $\gamma$.}

\label{figSM-fitting}
\end{figure}

The dependence of $U_c$ on $L$ follows a power-law scaling, as shown in Fig.~\ref{figSM-fitting}(a). The circle dots are numerical results obtained by exact diagonalization while the curves are the fitting with the power-law relation, Eq.~(2) in the main text. The extrapolated $U_c=U_{\text{TD}}$ in the TDL is shown by the red line in Fig.~1(c) of the main text. The fitting parameters $\alpha$ and $\beta$ depend explicitly on $\gamma$, as shown in Fig.~\ref{figSM-fitting}(b).

\section{Density-density correlation function \label{Sec:correlation-function}}
\begin{figure}[b]
\includegraphics[width=0.66\columnwidth]{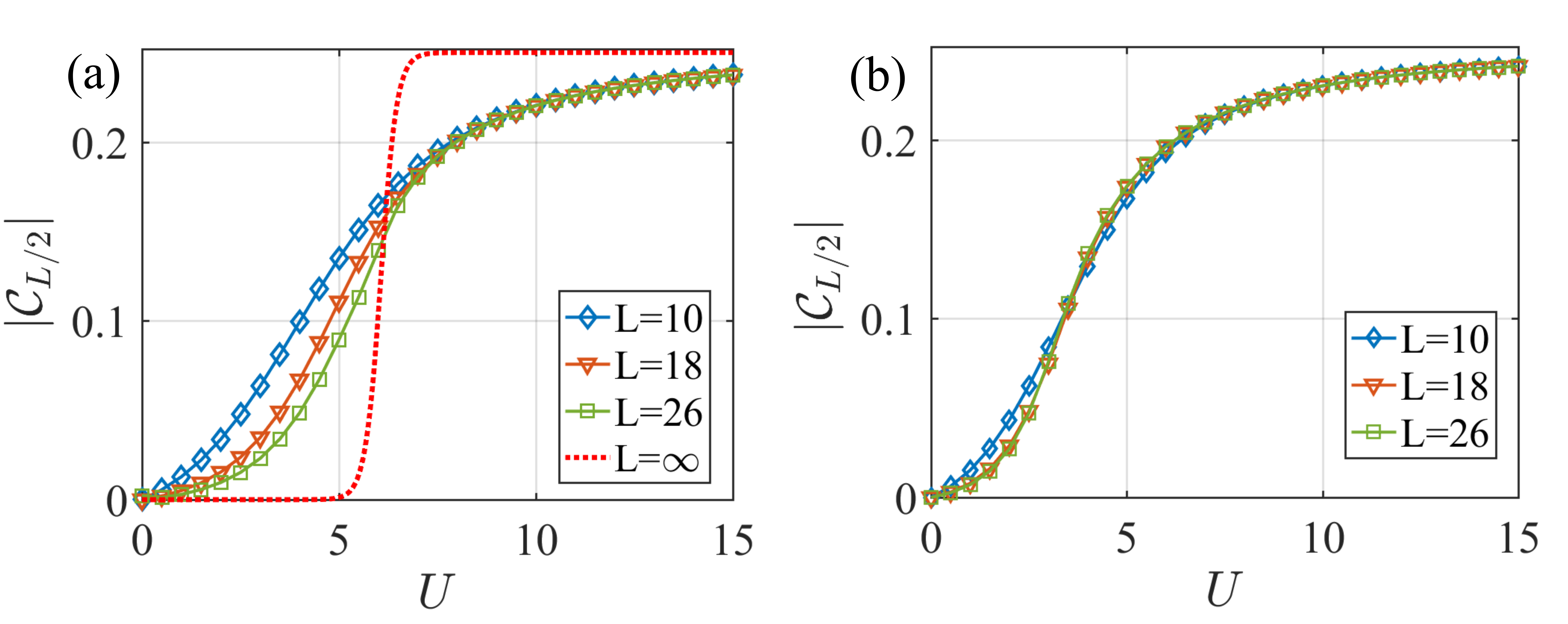}

\caption{(a) Density-density correlation function as a function of $U$ for different $L$. The red dotted curve sketches the result in the TDL. Other parameters are $\gamma=0.6t$ and $k_{B}T=0.1t$. (b) is the same as (a) but for the Hermitian limit $\gamma=0$.}

\label{fig:correlations}
\end{figure}
To characterize the phase diagram, we compute the density-density correlation function by~\citep{Kalthoff19PRB}
\begin{align}
\mathcal{C}_{\ell} & =\langle(\hat{n}_{0}-1/2)(\hat{n}_{\ell}-1/2)\rangle\nonumber \\
 & =\dfrac{\sum_{j}\langle j|(\hat{n}_{0}-1/2)(\hat{n}_{\ell}-1/2)e^{-\beta E_{j}}|j\rangle}{\mathcal{Z}},\label{eq:correlation}
\end{align}
where $|j\rangle$ are the right eigenstates of the many-body Hamiltonian with eigenvalues $E_{j}$, $\mathcal{Z}$ is the partition function given by Eq.~(\ref{eq:partition-function}). For concreteness, we consider the long-range correlation function with $\ell=L/2$ and at low temperatures $1/\beta=k_{B}T\ll t$. In the low-temperature limit $\beta\rightarrow\infty$, Eq.~(\ref{eq:correlation}) describes the density-density correlation of the ground state. The correlation function as a function of interaction strength $U$ for increasing $L$ is shown in Fig.\ \ref{fig:correlations}. At $U=0$, we find $\mathcal{C}_{L/2}=0$. When increasing $U$, $|\mathcal{C}_{L/2}|$ increases slowly in the small $U$ regime but rapidly around $U=U_{c}$. In the large $U$ regime, $|\mathcal{C}_{L/2}|$ saturates slowly to a universal value $0.25$. For larger systems$,$ these features are more pronounced. In the TDL, we can expect $|\mathcal{C}_{L/2}|$ to suddenly jump from zero to the universal value 0.25 at the transition, as sketched by the red dotted curve in Fig.~\ref{fig:correlations}(a). A similar behavior of $|\mathcal{C}_{L/2}|$ happens for $\gamma=0$ but with the rapid increment around a smaller $U$, as shown in Fig.~\ref{fig:correlations}(b). Finally, we note that for odd $L/2$, we have always $\mathcal{C}_{L/2}\geq0$, while for even $L/2$, we have always $\mathcal{C}_{L/2}\leq0$. This reflects the fact that the ground state of the system is likely to have staggered charge-density distribution, namely, different charge densities at odd and even lattice sites.

\section{Free energy and persistent current \label{Sec:Free-energy}}

\begin{figure}[t]
\includegraphics[width=0.56\columnwidth]{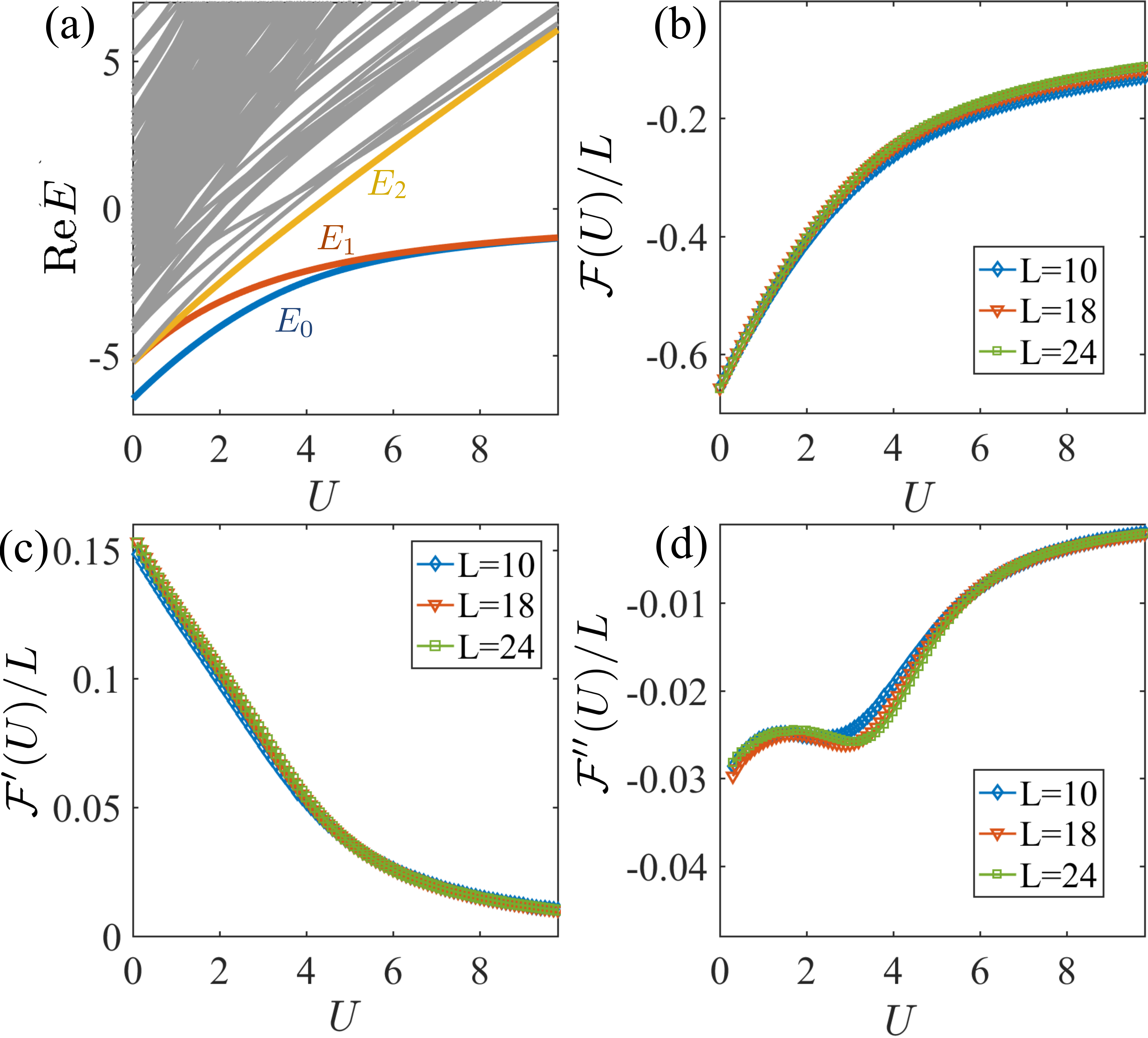}

\caption{(a) Real part of the low-Re$(E)$ spectrum as a function of $U$. We take $L=10$ for illustration. (b) Free energy $\mathcal{F}(U)$, its first (b) $\mathcal{F}'(U)$ and second (c) derivative $\mathcal{F}''(U)$ as functions of $U$. In contrast to the case with finite $\gamma\protect\neq0$, $\mathcal{F}$ and its derivatives change smoothly when increasing $U$. $\gamma=0$ and other parameters are the same as Fig.~3 in the main text.}

\label{fig:trivial-FreeEnergy}
\end{figure}

The free energy $\mathcal{F}$ of the interacting system at temperature $T$ is given by
\begin{equation}
\mathcal{F}=-\dfrac{1}{\beta}\text{\ensuremath{\log}}\mathcal{Z},
\end{equation}
where $\beta=1/k_{\textrm{B}}T$ with $k_{\textrm{B}}$ being the Boltzmann constant, and $\mathcal{Z}$ is the partition function
\begin{equation}
\mathcal{Z}=\sum_{j}e^{-\beta E_{j}}.\label{eq:partition-function}
\end{equation}
The sum in Eq.~\eqref{eq:partition-function} runs over all eigenenergies. In the Hermitian limit $\gamma=0$, all eigenenergies change smoothly with increasing $U$ [Fig.~\ref{fig:trivial-FreeEnergy}(a)]. Consequently, $\mathcal{F}$ and its derivative also change smoothly, as shown in Figs.~\ref{fig:trivial-FreeEnergy}(b--d).

The application of a magnetic flux $\phi$ thread through the ring changes the eigenenergies and hence $\mathcal{F}$ periodically. The persistent current along the ring can be found as the derivative of $\mathcal{F}$ with respect to $\phi$~\citep{Byers61PRL}, i.e.,
\begin{equation}
I_{\text{p}}=-\dfrac{e}{\hbar}\dfrac{\partial\mathcal{F}}{\partial\phi},\label{eq:persistent-current}
\end{equation}
where $e$ is the elementary charge and $\hbar$ the reduced Planck constant.

At zero temperature, the free energy is equal to the ground-state energy, i.e., $\mathcal{F}=E_{0}$. In the absence of interactions, the ground-state energy can be found as
\begin{equation}
E_{0}(\phi)=-2\sum_{|\kappa|\leqslant L/4}\Big[t\cos\Big(\dfrac{2\pi\kappa}{L}+\dfrac{\phi}{L}\Big)+i\gamma\sin\Big(\dfrac{2\pi\kappa}{L}+\dfrac{\phi}{L}\Big)\Big],\label{eq:Ground-state-Energy}
\end{equation}
where $\kappa$ are integers. For $L\gg1$, we approximate the summation in Eq.~(\ref{eq:Ground-state-Energy}) as an integral and obtain
\begin{align}
E_{0}(\phi) & =-\dfrac{L}{\pi}\int_{_{-\pi/2}}^{\pi/2}dx\Big[t\cos(x)\cos\Big(\dfrac{\phi}{L}\Big)-t\sin(x)\sin\Big(\dfrac{\phi}{L}\Big)+i\gamma\sin(x)\cos\Big(\dfrac{\phi}{L}\Big)+i\gamma\cos(x)\sin\Big(\dfrac{\phi}{L}\Big)\Big]\nonumber \\
 & =-\dfrac{2L}{\pi}\Big[t\cos\Big(\dfrac{\phi}{L}\Big)+i\gamma\sin\Big(\dfrac{\phi}{L}\Big)\Big].
\end{align}
Plugging $\mathcal{F}=E_{0}$ into Eq.~(\ref{eq:persistent-current}), we find
\begin{equation}
I_{\text{p}}=\dfrac{2e}{\hbar\pi}\Big[-t\sin\Big(\dfrac{\phi}{L}\Big)+i\gamma\cos\Big(\dfrac{\phi}{L}\Big)\Big].\label{eq:persistent-current-1}
\end{equation}
At zero flux $\phi=0$, the persistent current is given by
\begin{equation}
I_{\text{p}}=i\dfrac{2e}{\hbar}\dfrac{\gamma}{\pi}.\label{eq:persistent-current-1-1}
\end{equation}

\section{Nontrivial winding numbers and charge density profiles of the many-body eigenstates \label{Sec:windingnumber}}

\begin{figure}[h]
\includegraphics[width=1\columnwidth]{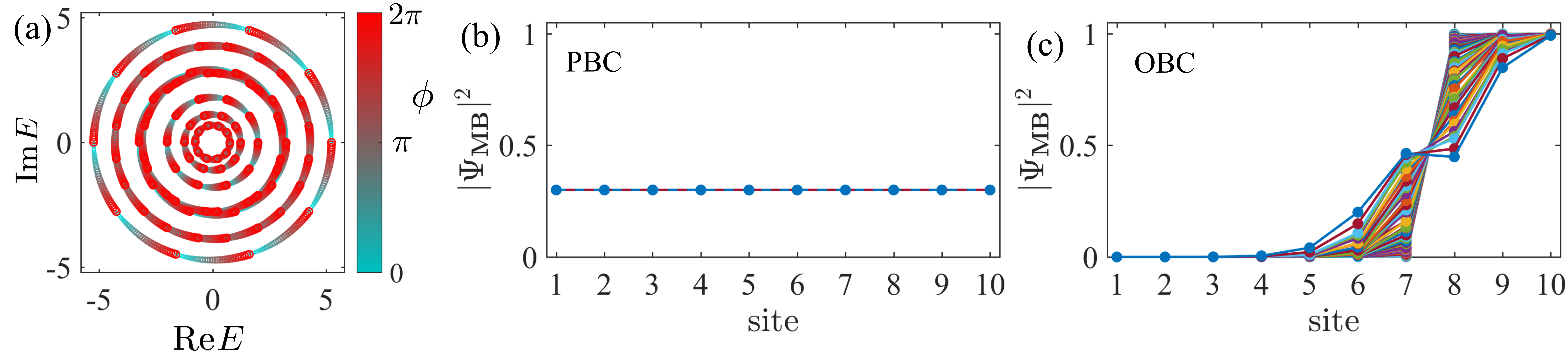}
\caption{(a) Movement of the eigenenergies at $U=0$ as the twist boundary angle $\phi$ increases from $0$ (cyan) to $2\pi$ (red). All eigenenergies wind around $E=0$ in one direction. The spectral winding number for the spectrum is found as $\nu=12$. (b) Charge density profiles of all many-body states under PBC. (c) the same as (b) but for OBC. We consider $L=10$, $N=3$, $U=20t$ and a relatively large $\gamma=0.9t$ for better illustration.}
\label{figSM-windingnumber-1}
\end{figure}

\begin{figure}[t]
\includegraphics[width=1\textwidth]{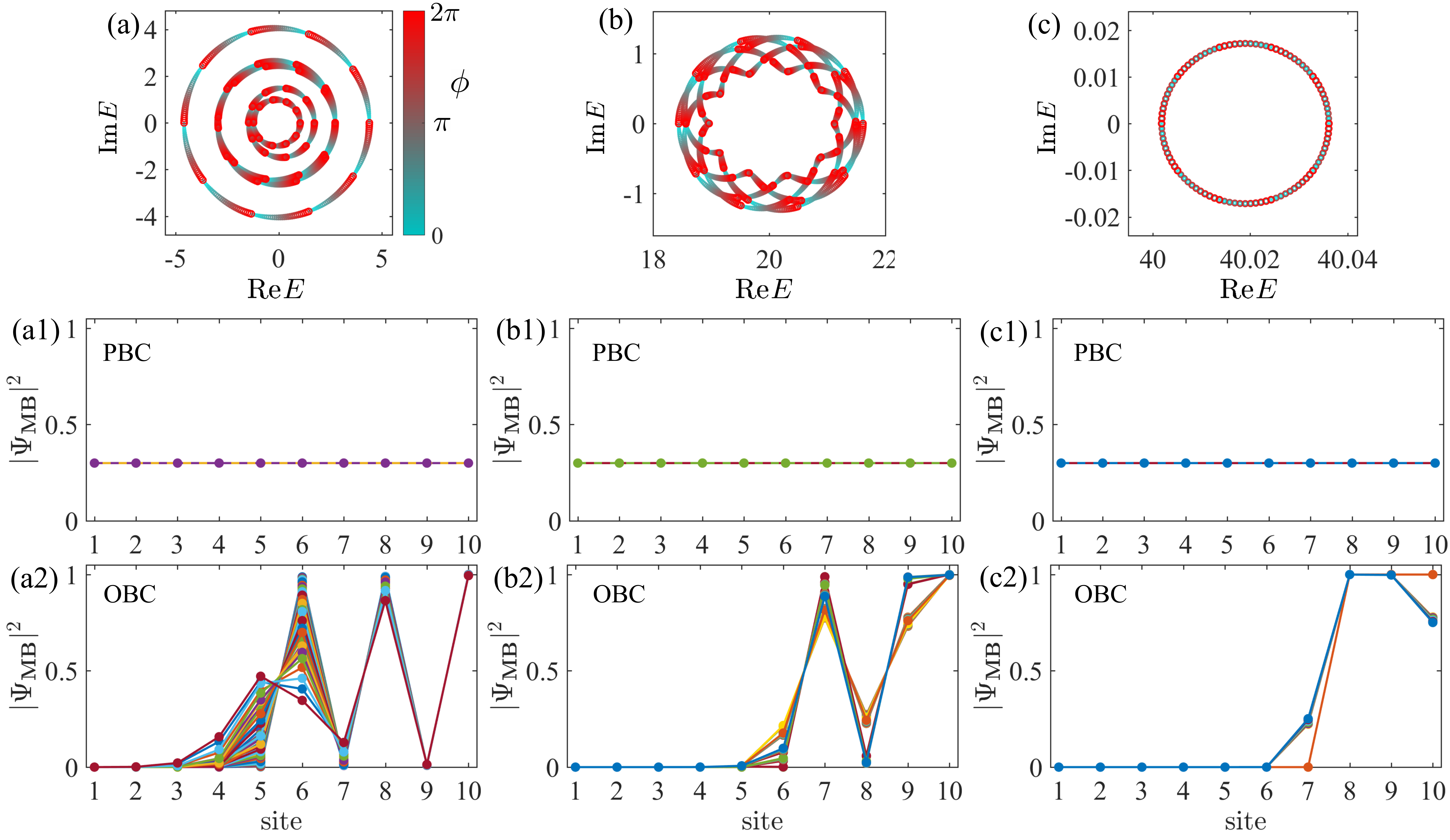}

\caption{(a-c) Movement of the eigenenergies of the clusters at $\varepsilon_{s}=0$, $U$ and $2U$, respectively, when the twist boundary angle $\phi$ increases from $0$ (cyan) to $2\pi$ (red). All eigenenergies belonging to the same cluster wind around the center in one direction. The spectral winding numbers for the three clusters are found as $\nu_{s}=5$, $4$ and $3,$ respectively. (a1-c1) Charge density profiles of the eigenstates with energies around $\varepsilon_{s}=0$, $U$ and $2U$, respectively, when PBC are imposed. (a2-c2) the same as (a1-c1) but for OBC. The charge-density profiles are markedly different for PBC and OBC. Moreover, the charge densities tend to localize at one open boundary when OBC are imposed. Other parameters are the same as Fig.~\ref{figSM-windingnumber-1}.}

\label{figSM-windingnumber}
\end{figure}

At $U=0$, the flow of the eigenenergies as varying the twist boundary angle $\phi$  in a period $[0,2\pi)$, and charge-density profiles of all many-body sates are shown in Fig.~\ref{figSM-windingnumber-1}. In the strong interaction regime, the flow of the eigenenergies and charge-density profiles of the many-body eigenstates associated with the spectral clusters (i.e., with the energies close to the respective cluster energy centers) are shown in Fig.~\ref{figSM-windingnumber}. Clearly, all the eigenenergies wind around the corresponding energy centers when $\phi$ increases from $0$ to $2\pi$. Correspondingly, the charge densities of the many-body eigenstates, which are evenly distributed in the system under PBC, tend to localize to an open boundary. We note that the winding numbers defined by Eq.~(3) in the main text are applicable to finite-size systems. Due to the finite-size effect, all eigenenergies are away from the energy centers of clusters, provided that $U$ is finite. Thus, the clusters have well-defined point gaps in finite-size systems. In the $L\rightarrow \infty$ limit but with fixed $N$, each clusters has at least one continuous orbit surrounding their energy centers. Nontrivial topological invariants can be defined as the winding numbers along these orbits.

\begin{figure}[h]
\includegraphics[width=1\columnwidth]{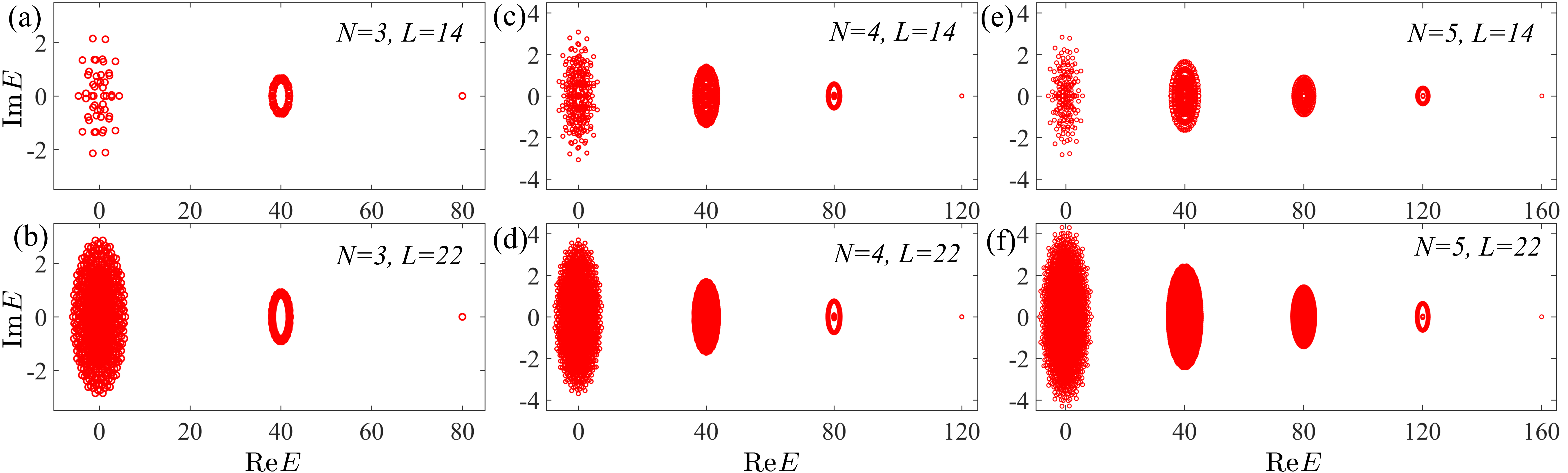}

\caption{Spectral clusters for fixed $N$ and increasing $L$. The right, middle and left panels are for $N=3$, 4 and 5, respectively. The upper and lower panels are $L=14$ and $22$, respectively. As $L$ grows, the eigenenergies fill the clusters more densely. In the TDL, the clusters occupy continuous areas and orbits around their energy centers in the complex-energy plane. }

\label{fig:SM-TDlimit}
\end{figure}

Figure~\ref{fig:SM-TDlimit} plots the clusters for fixed $N$ and increasing $L$. We see that as $L$ grows, the eigenenergies fill the clusters more densely. In the TDL, the clusters form continuous areas and orbits in the complex-energy plane. Thus, we can always find a continuous orbit of eigenenergies that surround the energy center of any cluster. A nonzero topological invariant can be obtained as the winding numbers of these orbits. We note that the extents of the clusters (including the one at $\varepsilon_1=(N-1)U$) on real and imaginary axes are independent of large $L$. The extents of the clusters are finite, provided that $U$ is finite.

For single particles, it has been demonstrated that a nonzero
spectral winding number inevitably lead to the non-Hermitian skin
effect \citep{KZhang20PRL,Okuma20PRL}. In the following, we show that
this relationship can be generalized to many-body strongly interacting
systems. The point-gap topology with nonzero winding numbers in the PBC spectrum
ensures that its difference as compared to the OBC one, thus leading to
the non-Hermitian skin effect even in the many-body interacting systems.

We focus on the strong interaction ($U\gg t$) regime where the PBC spectrum develops multiple clusters which are centered at $\varepsilon_{s}\in\{0,U,...,(N-1)U\}$
and have nonzero winding numbers, as we discussed before. In this strong interaction regime,
we consider the kinetic hopping term
\begin{equation}
\hat{V}_{h}=\sum_{\ell=1}^{L}[(t+\gamma)\hat{c}_{\ell}^{\dagger}\hat{c}_{\ell+1}^{{\color{white}\dagger}}+(t-\gamma)\hat{c}_{\ell+1}^{\dagger}\hat{c}_{\ell}^{{\color{white}\dagger}}]
\end{equation}
as a perturbation to the unperturbed Hamiltonian which consists of only the interaction
term
\begin{equation}
\hat{H}_{\text{int}}=U\sum_{\ell=1}^{L}\hat{n}_{\ell}\hat{n}_{\ell+1}.
\end{equation}
Then, up to the $N$-order perturbations in $\hat{V}_{h}$, the effective
Hamiltonian projected within cluster $s\in\{1,2,...,N\}$ may be written
as
\begin{equation}
\hat{H}_{\text{eff}}^{(s)}=(N-s)U+\hat{P}_{s}\sum_{j=1}^{N}\hat{V}_{h}[(E-\hat{H}_{\text{int}})^{-1}\hat{V}_{h}]^{j-1}\hat{P}_{s}+\mathcal{O}(\hat{V}_{h}^{N+1}),
\end{equation}
where $\hat{P}_{s}$ is the orthogonal projection operator onto the
many-body Fock subspace associated with cluster $s$. Using the perturbation theory, we can derive effective Hamiltonians under PBC and OBC, respectively, for each cluster. From these Hamiltonians, we can find again that the PBC spectrum  has a nonzero nonzero winding number whereas the OBC spectrum  is always real, which is consistent with our numerical calculations. Indeed, the effective Hamiltonian under OBC can be converted into a Hermitian one via a similarity transformation. Such a similarity transformation does not change the spectrum, and clearly shows the concentration of the many-body eigenstates to the Fock basis state with its particles accumulated close to an open boundary. Therefore, the many-body eigen wavefunctions are localized to the boundary, thus exhibiting the many-body non-Hermitian skin effect.

As an illustration, we consider the cluster $s=1$. For this cluster,
we denote and order the orthogonal many-body basis with the position
as
\begin{equation}
|\overline{\ell}\rangle=\hat{c}_{\ell}^{\dagger}\hat{c}_{\ell+1}^{\dagger}...\hat{c}_{\ell+N-1}^{\dagger}|\text{vac}\rangle,\ \ \ \ell\in\{1,...,L\},
\end{equation}
where $|\text{vac}\rangle$ is the vaccum state. The effective Hamiltonian
under PBC can be derived as
\begin{equation}
\langle\overline{\jmath}|\hat{H}_{\text{eff}}^{(1)}|\overline{\ell}\rangle=\text{\ensuremath{\tilde{E}}}_{1}\delta_{\jmath,\ell}+\dfrac{(t-\gamma)^{N}}{U^{N-1}}\delta_{\jmath,\ell+1}+\dfrac{(t+\gamma)^{N}}{U^{N-1}}\delta_{\jmath,\ell-1},\label{eq:effective-Hamiltonian}
\end{equation}
where $\text{\ensuremath{\tilde{\varepsilon}}}_{1}=(N-1)U+E_{1,c}$.
Thus, the effective Hamiltonian is given by
\begin{equation}
\hat{H}_{\text{PBC}}^{(1)}=\text{\ensuremath{\tilde{\varepsilon}}}_{1}+\sum_{\ell=1}^{L}\Big[\dfrac{(t-\gamma)^{N}}{U^{N-1}}|\overline{\ell+1}\rangle\langle\overline{\ell}|+\dfrac{(t+\gamma)^{N}}{U^{N-1}}|\overline{\ell}\rangle\langle\overline{\ell+1}|\Big].\label{eq:effective-Hamiltonian-2}
\end{equation}
This Hamiltonian takes a similar form as the single-particle Hatano-Nelson
model but is defined on the basis of Fock states. In the $N=1$ limit, it recovers the single-particle
Hamiltonian. The $E_{\text{shift}}$ in Eq.\ (\ref{eq:effective-Hamiltonian-2})
stems from the even-order corrections. It shifts the energy of the
cluster globally in real axis and thus does not affect the topology
of the system of interest. To the second-order corrections, the energy
shift can be estimated as $E_{1,c}\approx2(t^{2}-\gamma^{2})/U$.
The hopping terms are dominated by the $N$-order correction. They
are non-reciprocal in the orthogonal Fock subspace: the hopping amplitude
from the many-body state $|\overline{\ell}\rangle$ to the next one
$|\overline{\ell+1}\rangle$ is $(t-\gamma)^{N}/U^{N-1}$, while the
hopping amplitude for the inverse process is instead $(t+\gamma)^{N}/U^{N-1}$.
By Fourier transformation, the energy spectrum can be found as
\begin{equation}
\widetilde{E}_{\text{PBC}}^{(1)}=\text{\ensuremath{\tilde{\varepsilon}}}_{1}+ \dfrac{(t-\gamma)^{N}}{U^{N-1}}e^{-iq}+\dfrac{(t+\gamma)^{N}}{U^{N-1}}e^{iq},\label{eq:spectrum}
\end{equation}
where $q\in\{0,2\pi/L,4\pi/L,...,2\pi\}$ can be viewed
as the momentum of the many-body eigenstates. This spectrum is consistent
with the numerical calculations, as shown in Fig.~S7.
From Eq.\ (\ref{eq:spectrum}), we also see that in the TDL ($L\rightarrow\infty$),
the spectrum forms a closed loop in the complex-energy plane, similar
to that of the single-particle Hatano-Nelson model {[}cf. Fig.\ S7(b){]}.
Taking into account the magnetic flux, the spectrum becomes
\begin{equation}
\widetilde{E}_{\text{PBC}}^{(1)}(\phi)=\text{\ensuremath{\tilde{\varepsilon}}}_{1}
+e^{-iN\phi/L}e^{-iq}+e^{iN\phi/L}e^{iq}\dfrac{(t+\gamma)^{N}}{U^{N-1}}.\label{eq:effective-spectrum}
\end{equation}
Using Eq.\ (3) in the main text, we obtain the winding number as
$\nu_{1}=\text{sgn}(\gamma)N$, which again is consistent with our
numerical calculations {[}c.f. Fig.\ S7(c){]}.

\begin{figure}[th]
\includegraphics[width=0.68\textwidth]{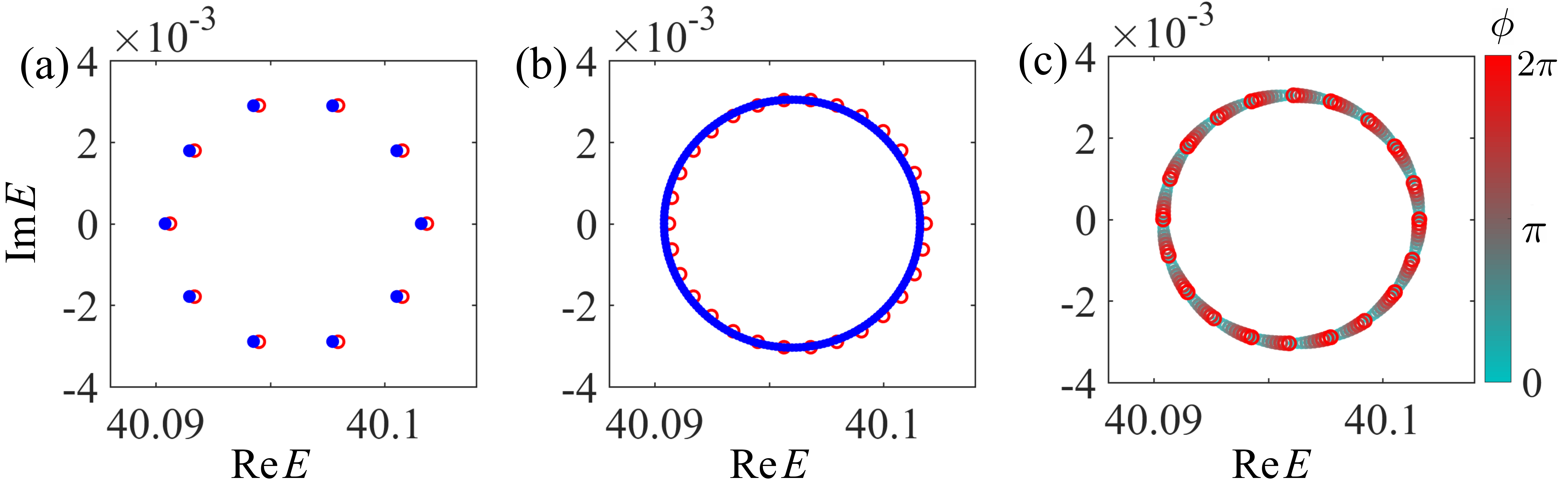}

\caption{(a) Many-body spectrum (blue) obtained by the effective model in Eq.\ (\ref{eq:effective-Hamiltonian-2}).
The red circles are obtained by exact diagonalizing the full many-body
Hamiltonian and are plotted for comparison. (c) Many-body spectrum
(blue) obtained by the effective model in Eq.\ (\ref{eq:spectrum})
in the TDL. The red circles are obtained by exact diagonalizing the
full many-body Hamiltonian with $L=30$ sites and are plotted for
comparison. (c) Evolution of many-body spectrum (blue) obtained by
the effective model in Eq.\ (\ref{eq:effective-spectrum}) by
increasing the twist angle $\phi$ from 0 to $2\pi$. We consider $N=3$, $\gamma=0.2t$, $U=20t$, $t=1$, $s=1$ and PBC in all panels, and $\phi=0$ in (a) and (b).}

\label{fig:comparison}

\end{figure}

Next, we impose OBC (i.e., termination of the coupling between the $\ell=1$
and $L$ sites) and derive the effective Hamiltonian as
\begin{equation}
\hat{H}_{\text{OBC}}^{(1)}=\text{\ensuremath{\tilde{\varepsilon}}}_{1}-\sum_{\ell=1}^{N-1}E_{\ell}(|\overline{\ell}\rangle\langle\overline{\ell}|+|\overline{L-\ell}\rangle\langle\overline{L-\ell}|)+\sum_{\ell=1}^{L-N}\dfrac{(t-\gamma)^{N}}{U^{N-1}}|\overline{\ell+1}\rangle\langle\overline{\ell}|+\dfrac{(t+\gamma)^{N}}{U^{N-1}}|\overline{\ell}\rangle\langle\overline{\ell+1}|.\label{eq:effective-Hamiltonian-2-2}
\end{equation}
Note that under OBC, we have $L-N+1$ many-body states in cluster
$s=N$. Due to the termination, the Fock basis states close to the
two ends have different potential corrections compared to those deep
in the bulk. Following the spirit of Refs.\ \citep{KZhang20PRL,Okuma20PRL},
one can show that the Hamiltonian with open boundaries is always topological
trivial in terms of point gaps. To be explicit, we can apply a similarity
transformation
\begin{equation}
|\overline{\ell}\rangle\rightarrow\eta^{\ell}|\overline{\ell}\rangle,\ \ \langle\overline{\ell}|\rightarrow\eta^{-\ell}\langle\overline{\ell}|,\ \ \ \ \ \ \ \ (\ell\in\{1,...,L-N+1\}),
\end{equation}
where $\eta=|(t+\gamma)/(t-\gamma)|{}^{N/2}$, and transfer the Hamiltonian\ (\ref{eq:effective-Hamiltonian-2-2})
to a Hermitian one
\begin{equation}
\hat{H}_{\text{OBC}}^{(1)\prime}=\text{\ensuremath{\tilde{\varepsilon}}}_{1}-\sum_{\ell=1}^{N-1}E_{\ell}(|\overline{\ell}\rangle\langle\overline{\ell}|+|\overline{L-\ell}\rangle\langle\overline{L-\ell}|)+\dfrac{(t^{2}-\gamma^{2})^{N/2}}{U^{N-1}}\sum_{\ell=1}^{L-N}(|\overline{\ell+1}\rangle\langle\overline{\ell}|+|\overline{\ell}\rangle\langle\overline{\ell+1}|).\label{eq:effective-Hamiltonian-2-2-1}
\end{equation}
The similarity transformation does not change the spectrum. Thus,
the effective Hamiltonian\ (\ref{eq:effective-Hamiltonian-2-2})
under OBC has a purely real spectrum, topologically different from
that under PBC. In the TDL, we approximate the spectrum as
\begin{equation}
\widetilde{E}_{\text{OBC}}^{(1)}\approx\text{\ensuremath{\tilde{\varepsilon}}}_{1}+2\dfrac{(t^{2}-\gamma^{2})^{N/2}}{U^{N-1}}\cos q,\label{eq:effectivespectrum-1}
\end{equation}
where $q\in\{0,2\pi/(L-N+1),4\pi/(L-N+1),...,2\pi\}$.
The OBC spectrum forms a line in real axis inside the PBC spectrum
(i.e., closed loop) (see Fig.\ S8). This result indicates
that most and extensive many-body eigenstates under OBC are localized to
one open boundary. These results are also consistent with our numerical
calculations.

\begin{figure}[h]
\includegraphics[width=0.28\textwidth]{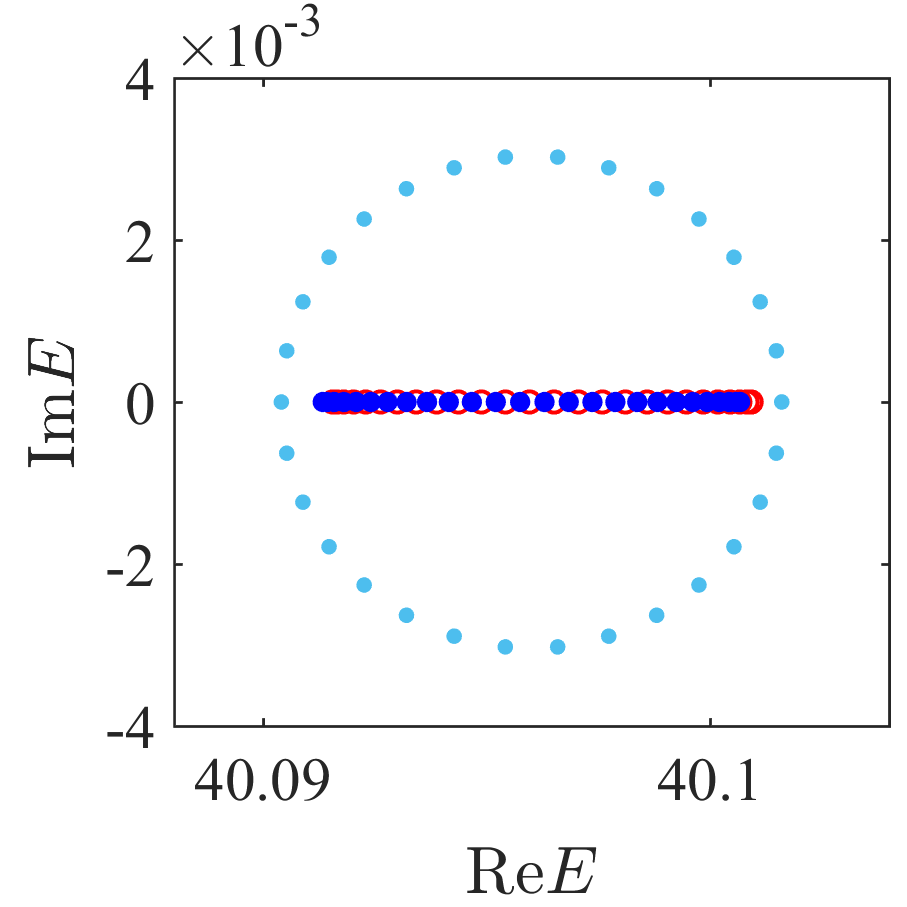}

\caption{(a) Many-body spectrum (blue dots) obtained by the effective model in Eq.\ (\ref{eq:effective-Hamiltonian-2-2}).
The red circles are obtained by exact diagonalizing the full many-body
Hamiltonian. The cyan dots are the spectrum under OBC. Other parameters
are the same as those in Fig.\ S7.}

\label{fig2}
\end{figure}

In fact, from the similarity transformation, we can further see that
for $N\ll L$, the many-body eigenstates tend to concentrate to the
Fock basis state $|\overline{L-N+1}\rangle$ if $|\eta|>1$ while
to the state $|\overline{1}\rangle$ if $\eta<1$. Note that the wavefunction
of the basis state $|\overline{\ell}\rangle$ are accumulated from
sites $\ell$ to $\ell+N-1$. Thus, the many-body eigen wavefunctions
are localized at the right end if $\eta>1$ while at the left end
if $\eta<1$. When $\gamma=0$, the spectral winding numbers become
ill-defined. Accordingly, the similarity transformation become trivial
with $\eta=1$, and thus the many-body skin effect disappears.
From this aspect, we see clearly that the nontrivial spectral
winding number and the many-body skin effect are intimately related.

The above derivation can be generalized to other clusters, which is relatively more complicate. However, the essential results are the same: the many-body system can be understood as a "single-particle" model defined on the basis of Fock states, and the OBC effectively terminate the direct coupling between the "boundary" Fock basis states whose particles accumulated towards the boundaries. The nontrivial spectral topology of the PBC spectrum leads to the localization of the many-body eigenstates to the boundary Fock states, in the same manner as that in single-particle system, which hence exhibit the localization of the many-body eigen wavefunctions towards the boundaries. Finally, it may be also worth noting that for other clusters, the non-reciprocal hopping between the Fock basis states can be obtained by lower-order corrections. Thus, we find stronger effective nonreciprocity and hence larger spectral areas for these clusters. This is also confirmed by our numerical observations [Fig. 4(b) in the main text].

\begin{figure}[h]
\includegraphics[width=0.8\columnwidth]{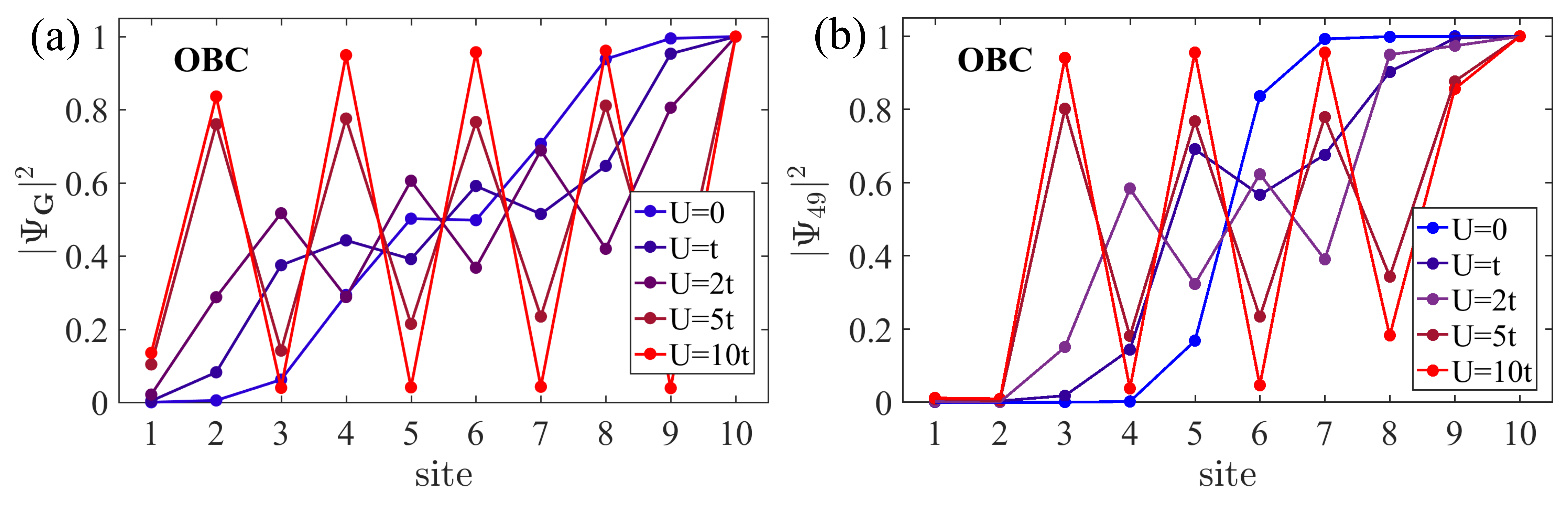}

\caption{Charge-density profiles of many-body states for $U=0$, $t$, $2t$, $5t$ and $10t$, respectively. Two states, namely, (a) the ground state and (b) the 50th lowest-$\text{Re}(E)$ states, are taken for illustrations. With increasing $U$, the charge-density profile of the many-body states become extended over the whole lattice chain with open boundaries. $L=10$, $N=5$ and other parameters are the same as Fig.~\ref{figSM-windingnumber}}

\label{figSM-windingnumber-1-1}
\end{figure}

In contrast, at half-filling and for $|\gamma|<t$ ($|\gamma|>t$), the many-body spectrum of a finite system under PBC shrinks onto the real axis (open lines parallel to the imaginary axis) as $U$ increase. Accordingly, we observe that upon increasing $U$, the charge-density profile of the many-body states become extended over the whole lattice chain with open boundaries, as shown in Fig.~\ref{figSM-windingnumber-1-1}. For concreteness and brevity, we show the results for two states in Fig.~\ref{figSM-windingnumber-1-1}. We note that all other many-body states exhibit similar behaviors.

\section{Other half-filled cases\label{sec:other-half-filling}}
\begin{figure}[h]
\includegraphics[width=0.68\columnwidth]{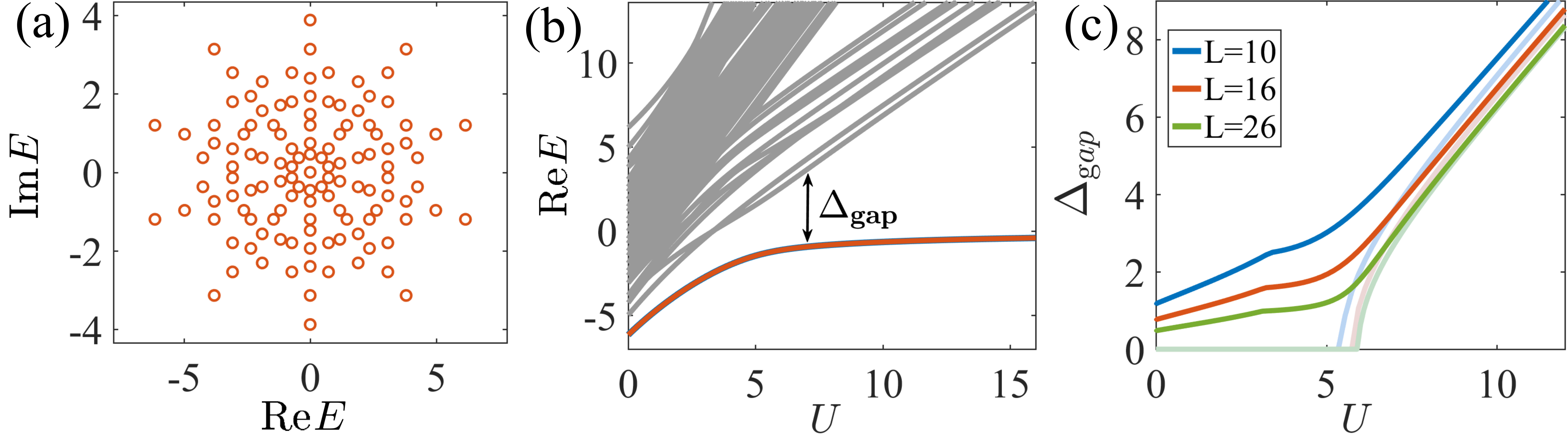}

\caption{Results for the cases with even (odd) $N=L/2$ and (anti-)PBC. (a) Many-body spectrum at half-filling in the absence of interactions ($U=0$). (b) Real part of the low-energy many-body spectrum as a function of $U$. We take $L=10$ for illustration. (c) Excitation gap $\Delta_{\text{gap}}$ as a function of $U$. The light colored lines are the results for cases (I) and presented here for comparison. We see that as $L\rightarrow\infty$, the metal-insulator transition in cases (II) resemble that in cases (I). Other parameters are the same as that in Fig.~3 in the main text, namely $L=10$ in panels (a) and (b), and $\gamma=0.6t$ in all panels.}

\label{fig8:other-cases}
\end{figure}

In the main text, we have discussed (I) the half-filled cases with odd (even) $N=L/2$ and (anti-)PBC. In this section, we discuss the results for the complementary half-filled cases (II), namely, with even (odd) $N$ and (anti-)PBC. In the cases (II), the system has always two ground states with complex-conjugate energies; and there is no exceptional point between two lowest excited states, as shown in Figs.~\ref{fig8:other-cases}(a) and (b). Thus, we do not observe the two $\mathcal{PT}$ transitions. However, we observe that as the size of the system $L$ grows, the excitation gap $\Delta_{\text{gap}}$ becomes vanishingly small for $U<U_c'$ and increases rapidly for $U>U_c'$, and the critical interaction strength $U_c'$ approaches with that one ($U_c$) characterizing the low-energy $\mathcal{TP}$ transition in the cases (I) [cf. Fig.~\ref{fig8:other-cases}(c)]. This result indicates that in the TDL, the phase transition from the gapless regime to the gapped CDW regime happens also in the cases (II) and the corresponding critical interaction strength coincides with that in the cases (I).

{
\section{Calculations for the non-Hermitian SSH-type model} \label{Sec:other-model}
Our main results, such as the symmetry-breaking phase transitions and nontrivial spectral topology, revealed from the interacting Hatano-Nelson model are general. They can also be applied to other models. As an example, we consider the Su-Schrieffer-Heeger (SSH) type model in which non-reciprocal hopping occurs every two nearest-neighbour bonds, as sketched in Fig.~\ref{fig9:SSH-model-transition}(a).
The non-Hermitian SSH-type Hamiltonian reads
\begin{align}
H
 & =\sum_{\ell}\Big\{(\hat{C}_{\ell}^{\dagger}(t\sigma_{x}- i\gamma\sigma_{y})\hat{C}_{\ell}^{{\color{white}\dagger}}+ \dfrac{t}{2}[\hat{C}_{\ell+1}^{\dagger}(\sigma_{x}+i\sigma_{y})\hat{C}_{\ell}^{{\color{white}\dagger}}+h.c.]\Big\}, \label{SSH-model}
\end{align}
where $\hat{C}_{\ell}\equiv(\hat{c}_{A,\ell},\hat{c}_{B,\ell})^{T}=(\hat{c}_{2\ell-1},\hat{c}_{2\ell})^{T}$
with $A$ and $B$ indicating odd and even lattice sites, respectively.
The Pauli matrices act on sublattice space consisting of $A$ and
$B$.  $h.c.$ indicates Hermitian conjugation of the previous term.
Figure~\ref{fig9:SSH-model-transition}(c) illustrates the flow of the full spectrum of the model with size $L=12$ and half-filling under anti-PBC. We see clearly that both the low-energy and full $\mathcal{PT}$ phase transitions occurs in the non-Hermitian SSH-type model as we increase the strength of nearest neighbor interaction. We have also checked that as $L$ increases, while the critical strength $U_{c,all}$ for the full $\mathcal{PT}$ transition increases monotonically, the critical strength $U_c$ for the low-energy $\mathcal{PT}$ transition saturates to a finite value. The non-Hermitian SSH-type model also exhibits the nontrivial winding in the spectrum, as shown in Fig.~\ref{fig10:SSH-model-winding}. All these features are the same as those in the interacting Hatano-Nelson model (up to that the imaginary energies are half of those of the Hatano-Nelson model).

It may be worthy noting that through a local unitary transformation $\hat{C}_{\ell}\rightarrow1/\sqrt{2}(\sigma_{0}+i\sigma_{x})\hat{C}_{\ell}$, the model \eqref{SSH-model} can be converted to a Creutz-ladder-like model~\citep{Bergholtz21RMP}
\begin{align}
H & =\sum_{\ell}\Big\{(\hat{C}_{\ell}^{\dagger}(t\sigma_{x}+i\gamma\sigma_{z})\hat{C}_{\ell}+\dfrac{t}{2}[\hat{C}_{\ell+1}^{\dagger} (\sigma_{x}-i\sigma_{z})\hat{C}_{\ell}+h.c.]\Big\}.\label{eq:Creutz-model}
\end{align}
Notably, the non-Hermiticity in the Creutz-ladder-like model contains pure
onsite gain ($i\gamma$) and loss ($-i\gamma$) [see Fig.~\ref{fig9:SSH-model-transition}(a)]. The interaction term can be transformed accordingly and remains Hermitian after the transformation.

\begin{figure}[h]
\includegraphics[width=0.56\columnwidth]{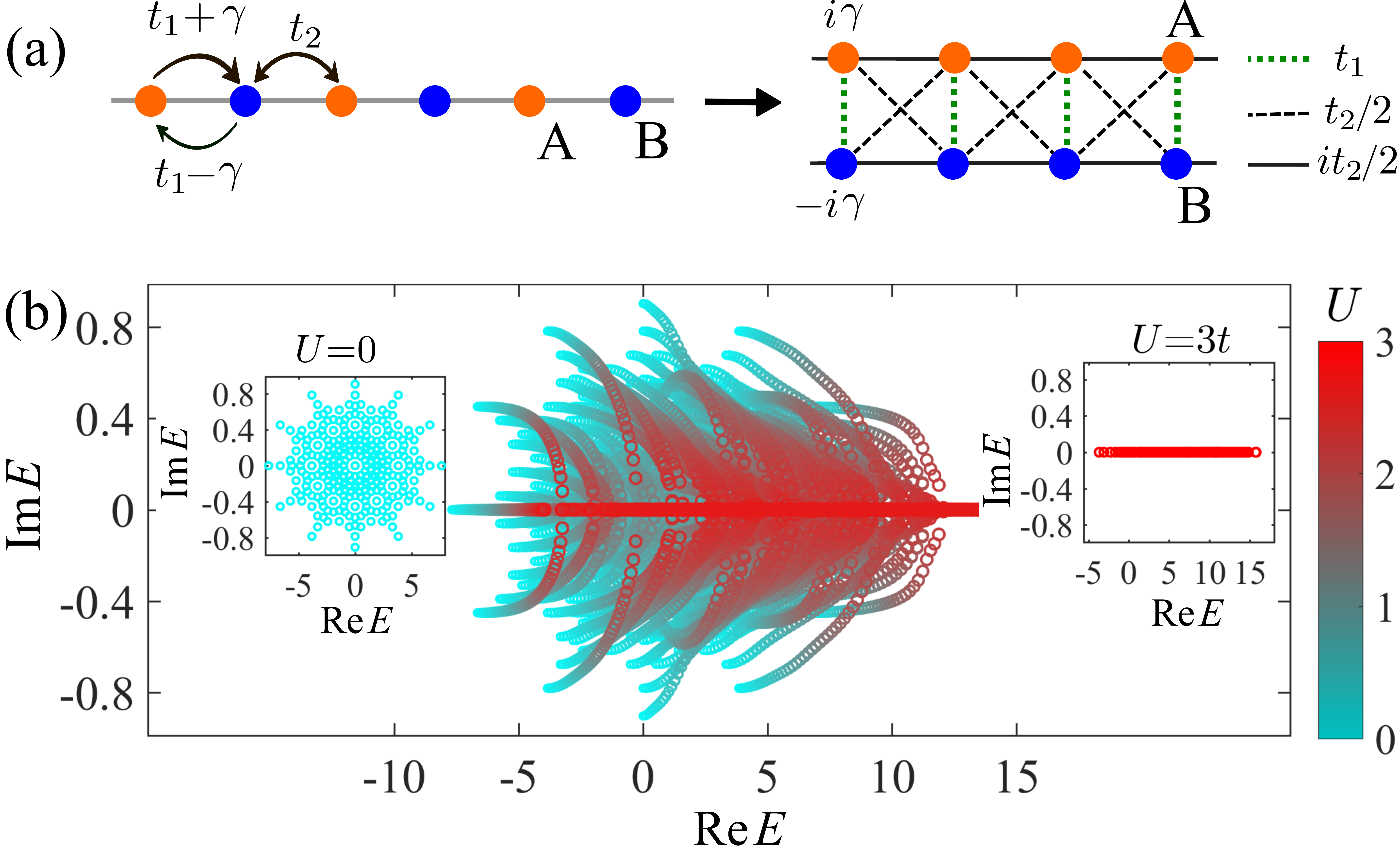}

\caption{{(a) Schematic for the non-Hermitian SSH-type model (left panel). The orange and blue dots indicate two sublattices ($A$ and $B$), respectively. The nonreciprocal hopping occurs only within unit cells. The model can be converted to a Creutz-ladder-like model with imaginary hopping and onsite gain and loss (right panel). (b) $\mathcal{PT}$ transitions in the interacting non-Hermitian SSH-type lattice chain with $L=12$, $N=6$ and $\gamma=0.2t$. The changing colour (from cyan to red) indicates the change of interaction strength from $U=0$ to $3t$. The insets show the spectra at $U=0$ and $3t$, respectively. We consider anti-PBC for illustration.}}

\label{fig9:SSH-model-transition}
\end{figure}

\begin{figure}[h]
\includegraphics[width=0.6\columnwidth]{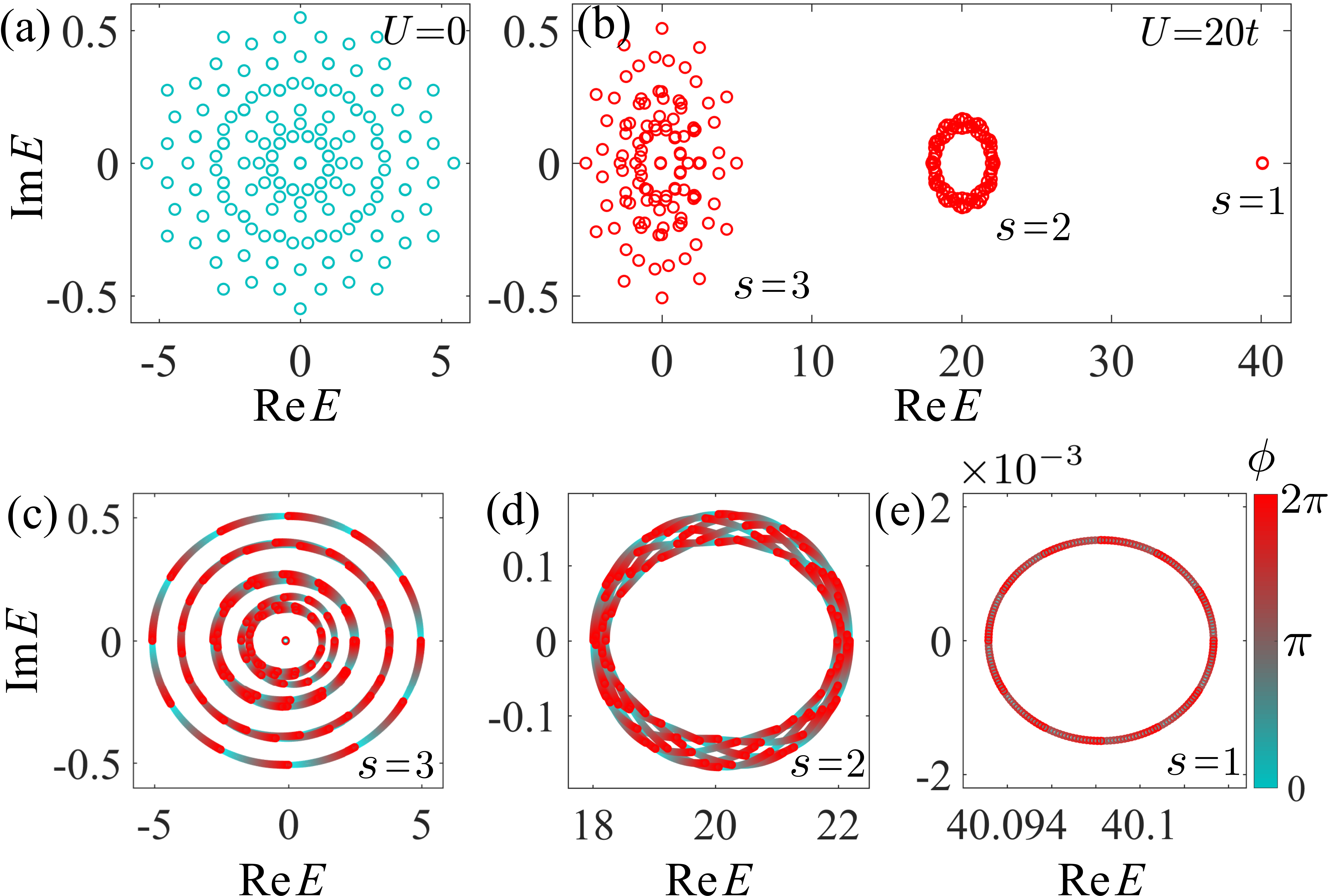}

\caption{{ Energy spectrum of the interacting SSH-type lattice chain with $L=12$, $N=3$ and $\gamma=0.2t$. (a) and (b) are the spectra at $U=0$ and $20t$ respectively. (c), (d) and (e) display respectively the movement of the three spectral clusters (labeled by $s=1,2,3$) as varies $\phi$ from $0$ (cyan) to $2\pi$ (red). }}

\label{fig10:SSH-model-winding}
\end{figure}

The dissipative Hamiltonian (\ref{eq:Creutz-model}) can be directly derived
from the Lindblad equation as we show in the following. The Lindblad equation under the
Markov approximation reads \citep{Lindblad76CMP}
\begin{equation}
\dfrac{d\rho}{dt}=-i[H_{0},\rho]+\sum_{\ell}\Big(L_{\ell}\rho L_{\ell}^{\dagger}-\dfrac{1}{2}\{L_{\ell}^{\dagger}L_{\ell},\rho\}\Big),
\end{equation}
where $\rho(t)$ is the density matrix of the system at time $t$.
$H_{0}$ is the system Hamiltonian in the absence of coupling to environment.
It is Hermitian and prepared as
\begin{equation}
H_{0}=\sum_{\ell}t\Big\{\hat{C}_{\ell}^{\dagger}\sigma_{x}\hat{C}_{\ell}+\dfrac{1}{2}[\hat{C}_{\ell+1}^{\dagger}
(\sigma_{x}-i\sigma_{z})\hat{C}_{\ell}+h.c.]\Big\}.
\end{equation}
$L_{\ell}$ are the Lindblad jump operators in terms of fermion operators.
We consider one-body loss events described by $L_{\ell}=g_{\ell}\hat{c}_{\ell}$
with $g_{\ell}=i^{\ell}\sqrt{2\gamma}$. The coefficients $g_{\ell}$
are determined from the loss rates of atoms. Following the quantum
trajectory method \citep{Daley2014AP}, the dynamics of the system
can be decomposed into a Schr\"odinger evolution under an effective
non-Hermitian Hamiltonian and quantum-jump processes which induce
particle losses with the jump operators $L_{\ell}$,
\begin{equation}
\dfrac{d\rho}{dt}=-i(H_{\text{eff}}\rho-\rho H_{\text{eff}}^{\dagger})+\sum_{\ell}L_{\ell}\rho L_{\ell}.
\end{equation}
We find that the resulting  effective Hamiltonian
\begin{equation}
H_{\text{eff}}=H_{0}-\dfrac{i}{2}\sum_{\ell}L_{\ell}^{\dagger}L_{\ell}=H_{0}+i\gamma\sum_{\ell}\hat{C}_{\ell}^{\dagger}
\sigma_{z}\hat{C}_{\ell}
\end{equation}
gives exactly the dissipative Creutz model in Eq.~\eqref{eq:Creutz-model}. It describes the
evolution of the system during a time interval between quantum jump
events. If we consider a situation where the equilibration time in
the measured many-body system is shorter than a typical time between
quantum jumps, it is justified to consider only $H_{\text{eff}}$
for the short-time evolution.
}


\begin{thebibliography}{90}%
\makeatletter
\providecommand \@ifxundefined [1]{%
 \@ifx{#1\undefined}
}%
\providecommand \@ifnum [1]{%
 \ifnum #1\expandafter \@firstoftwo
 \else \expandafter \@secondoftwo
 \fi
}%
\providecommand \@ifx [1]{%
 \ifx #1\expandafter \@firstoftwo
 \else \expandafter \@secondoftwo
 \fi
}%
\providecommand \natexlab [1]{#1}%
\providecommand \enquote  [1]{``#1''}%
\providecommand \bibnamefont  [1]{#1}%
\providecommand \bibfnamefont [1]{#1}%
\providecommand \citenamefont [1]{#1}%
\providecommand \href@noop [0]{\@secondoftwo}%
\providecommand \href [0]{\begingroup \@sanitize@url \@href}%
\providecommand \@href[1]{\@@startlink{#1}\@@href}%
\providecommand \@@href[1]{\endgroup#1\@@endlink}%
\providecommand \@sanitize@url [0]{\catcode `\\12\catcode `\$12\catcode
  `\&12\catcode `\#12\catcode `\^12\catcode `\_12\catcode `\%12\relax}%
\providecommand \@@startlink[1]{}%
\providecommand \@@endlink[0]{}%
\providecommand \url  [0]{\begingroup\@sanitize@url \@url }%
\providecommand \@url [1]{\endgroup\@href {#1}{\urlprefix }}%
\providecommand \urlprefix  [0]{URL }%
\providecommand \Eprint [0]{\href }%
\providecommand \doibase [0]{http://dx.doi.org/}%
\providecommand \selectlanguage [0]{\@gobble}%
\providecommand \bibinfo  [0]{\@secondoftwo}%
\providecommand \bibfield  [0]{\@secondoftwo}%
\providecommand \translation [1]{[#1]}%
\providecommand \BibitemOpen [0]{}%
\providecommand \bibitemStop [0]{}%
\providecommand \bibitemNoStop [0]{.\EOS\space}%
\providecommand \EOS [0]{\spacefactor3000\relax}%
\providecommand \BibitemShut  [1]{\csname bibitem#1\endcsname}%
\let\auto@bib@innerbib\@empty
\bibitem [{\citenamefont {Shen}\ \emph {et~al.}(2018)\citenamefont {Shen},
  \citenamefont {Zhen},\ and\ \citenamefont {Fu}}]{HTShen18PRL}%
  \BibitemOpen
  \bibfield  {author} {\bibinfo {author} {\bibfnamefont {H.}~\bibnamefont
  {Shen}}, \bibinfo {author} {\bibfnamefont {B.}~\bibnamefont {Zhen}}, \ and\
  \bibinfo {author} {\bibfnamefont {L.}~\bibnamefont {Fu}},\ }\bibfield
  {title} {\enquote {\bibinfo {title} {{Topological Band Theory for
  Non-Hermitian Hamiltonians}}}, }\href {\doibase
  10.1103/PhysRevLett.120.146402} {\bibfield  {journal} {\bibinfo  {journal}
  {Phys. Rev. Lett.}\ }\textbf {\bibinfo {volume} {120}},\ \bibinfo {pages}
  {146402} (\bibinfo {year} {2018})}\BibitemShut {NoStop}%
\bibitem [{\citenamefont {{Kozii}}\ and\ \citenamefont
  {{Fu}}()}]{Kozzi17arXiv}%
  \BibitemOpen
  \bibfield  {author} {\bibinfo {author} {\bibfnamefont {V.}~\bibnamefont
  {{Kozii}}}\ and\ \bibinfo {author} {\bibfnamefont {L.}~\bibnamefont {{Fu}}},\
  }\bibfield  {title} {\enquote {\bibinfo {title} {{Non-Hermitian Topological
  Theory of Finite-Lifetime Quasiparticles: Prediction of Bulk Fermi Arc Due to
  Exceptional Point}}}, }\href@noop {} {\ }\Eprint
  {http://arxiv.org/abs/1708.05841} {arXiv:1708.05841} \BibitemShut {NoStop}%
\bibitem [{\citenamefont {Zyuzin}\ and\ \citenamefont
  {Zyuzin}(2018)}]{Zyuzin18PRB}%
  \BibitemOpen
  \bibfield  {author} {\bibinfo {author} {\bibfnamefont {A.~A.}\ \bibnamefont
  {Zyuzin}}\ and\ \bibinfo {author} {\bibfnamefont {A.~Y.}\ \bibnamefont
  {Zyuzin}},\ }\bibfield  {title} {\enquote {\bibinfo {title} {{Flat band in
  disorder-driven non-Hermitian Weyl semimetals}}}, }\href {\doibase
  10.1103/PhysRevB.97.041203} {\bibfield  {journal} {\bibinfo  {journal} {Phys.
  Rev. B}\ }\textbf {\bibinfo {volume} {97}},\ \bibinfo {pages} {041203}
  (\bibinfo {year} {2018})}\BibitemShut {NoStop}%
\bibitem [{\citenamefont {Yao}\ and\ \citenamefont {Wang}(2018)}]{SYYao18prl}%
  \BibitemOpen
  \bibfield  {author} {\bibinfo {author} {\bibfnamefont {S.}~\bibnamefont
  {Yao}}\ and\ \bibinfo {author} {\bibfnamefont {Z.}~\bibnamefont {Wang}},\
  }\bibfield  {title} {\enquote {\bibinfo {title} {{Edge States and Topological
  Invariants of Non-Hermitian Systems}}}, }\href {\doibase
  10.1103/PhysRevLett.121.086803} {\bibfield  {journal} {\bibinfo  {journal}
  {Phys. Rev. Lett.}\ }\textbf {\bibinfo {volume} {121}},\ \bibinfo {pages}
  {086803} (\bibinfo {year} {2018})}\BibitemShut {NoStop}%
\bibitem [{\citenamefont {Kunst}\ \emph {et~al.}(2018)\citenamefont {Kunst},
  \citenamefont {Edvardsson}, \citenamefont {Budich},\ and\ \citenamefont
  {Bergholtz}}]{Kunst18prl}%
  \BibitemOpen
  \bibfield  {author} {\bibinfo {author} {\bibfnamefont {F.~K.}\ \bibnamefont
  {Kunst}}, \bibinfo {author} {\bibfnamefont {E.}~\bibnamefont {Edvardsson}},
  \bibinfo {author} {\bibfnamefont {J.~C.}\ \bibnamefont {Budich}}, \ and\
  \bibinfo {author} {\bibfnamefont {E.~J.}\ \bibnamefont {Bergholtz}},\
  }\bibfield  {title} {\enquote {\bibinfo {title} {{Biorthogonal Bulk-Boundary
  Correspondence in Non-Hermitian Systems}}}, }\href {\doibase
  10.1103/PhysRevLett.121.026808} {\bibfield  {journal} {\bibinfo  {journal}
  {Phys. Rev. Lett.}\ }\textbf {\bibinfo {volume} {121}},\ \bibinfo {pages}
  {026808} (\bibinfo {year} {2018})}\BibitemShut {NoStop}%
\bibitem [{\citenamefont {Lee}(2016)}]{ELee16PRL}%
  \BibitemOpen
  \bibfield  {author} {\bibinfo {author} {\bibfnamefont {T.~E.}\ \bibnamefont
  {Lee}},\ }\bibfield  {title} {\enquote {\bibinfo {title} {Anomalous edge
  state in a non-hermitian lattice}}, }\href {\doibase
  10.1103/PhysRevLett.116.133903} {\bibfield  {journal} {\bibinfo  {journal}
  {Phys. Rev. Lett.}\ }\textbf {\bibinfo {volume} {116}},\ \bibinfo {pages}
  {133903} (\bibinfo {year} {2016})}\BibitemShut {NoStop}%
\bibitem [{\citenamefont {Yao}\ \emph {et~al.}(2018)\citenamefont {Yao},
  \citenamefont {Song},\ and\ \citenamefont {Wang}}]{SYYao18prlb}%
  \BibitemOpen
  \bibfield  {author} {\bibinfo {author} {\bibfnamefont {S.}~\bibnamefont
  {Yao}}, \bibinfo {author} {\bibfnamefont {F.}~\bibnamefont {Song}}, \ and\
  \bibinfo {author} {\bibfnamefont {Z.}~\bibnamefont {Wang}},\ }\bibfield
  {title} {\enquote {\bibinfo {title} {Non-hermitian chern bands}}, }\href
  {\doibase 10.1103/PhysRevLett.121.136802} {\bibfield  {journal} {\bibinfo
  {journal} {Phys. Rev. Lett.}\ }\textbf {\bibinfo {volume} {121}},\ \bibinfo
  {pages} {136802} (\bibinfo {year} {2018})}\BibitemShut {NoStop}%
\bibitem [{\citenamefont {Gong}\ \emph {et~al.}(2018)\citenamefont {Gong},
  \citenamefont {Ashida}, \citenamefont {Kawabata}, \citenamefont {Takasan},
  \citenamefont {Higashikawa},\ and\ \citenamefont {Ueda}}]{ZPGong18PRX}%
  \BibitemOpen
  \bibfield  {author} {\bibinfo {author} {\bibfnamefont {Z.}~\bibnamefont
  {Gong}}, \bibinfo {author} {\bibfnamefont {Y.}~\bibnamefont {Ashida}},
  \bibinfo {author} {\bibfnamefont {K.}~\bibnamefont {Kawabata}}, \bibinfo
  {author} {\bibfnamefont {K.}~\bibnamefont {Takasan}}, \bibinfo {author}
  {\bibfnamefont {S.}~\bibnamefont {Higashikawa}}, \ and\ \bibinfo {author}
  {\bibfnamefont {M.}~\bibnamefont {Ueda}},\ }\bibfield  {title} {\enquote
  {\bibinfo {title} {{Topological Phases of Non-Hermitian Systems}}}, }\href
  {\doibase 10.1103/PhysRevX.8.031079} {\bibfield  {journal} {\bibinfo
  {journal} {Phys. Rev. X}\ }\textbf {\bibinfo {volume} {8}},\ \bibinfo {pages}
  {031079} (\bibinfo {year} {2018})}\BibitemShut {NoStop}%
\bibitem [{\citenamefont {Torres}(2019)}]{Torres19JPM}%
  \BibitemOpen
  \bibfield  {author} {\bibinfo {author} {\bibfnamefont {L.~E.~F.}\
  \bibnamefont {Foa Torres}},\ }\bibfield  {title} {\enquote {\bibinfo {title}
  {{Perspective on topological states of non-Hermitian lattices}}}, }\href
  {\doibase 10.1088/2515-7639/ab4092} {\bibfield  {journal} {\bibinfo
  {journal} {J. Phys. Mater.}\ }\textbf {\bibinfo {volume} {3}},\ \bibinfo
  {pages} {014002} (\bibinfo {year} {2020})}\BibitemShut {NoStop}%
\bibitem [{\citenamefont {Ashida}\ \emph {et~al.}(2020)\citenamefont {Ashida},
  \citenamefont {Gong},\ and\ \citenamefont {Ueda}}]{Ashida20AP}%
  \BibitemOpen
  \bibfield  {author} {\bibinfo {author} {\bibfnamefont {Y.}~\bibnamefont
  {Ashida}}, \bibinfo {author} {\bibfnamefont {Z.}~\bibnamefont {Gong}}, \ and\
  \bibinfo {author} {\bibfnamefont {M.}~\bibnamefont {Ueda}},\ }\bibfield
  {title} {\enquote {\bibinfo {title} {Non-hermitian physics}}, }\href
  {\doibase 10.1080/00018732.2021.1876991} {\bibfield  {journal} {\bibinfo
  {journal} {Adv. Phys.}\ }\textbf {\bibinfo {volume} {69}},\ \bibinfo {pages}
  {249} (\bibinfo {year} {2020})}\BibitemShut {NoStop}%
\bibitem [{\citenamefont {Miri}\ and\ \citenamefont
  {Alu}(2019)}]{Miri19Science}%
  \BibitemOpen
  \bibfield  {author} {\bibinfo {author} {\bibfnamefont {M.-A.}\ \bibnamefont
  {Miri}}\ and\ \bibinfo {author} {\bibfnamefont {A.}~\bibnamefont {Alu}},\
  }\bibfield  {title} {\enquote {\bibinfo {title} {Exceptional points in optics
  and photonics}}, }\href {\doibase 10.1126/science.aar7709} {\bibfield
  {journal} {\bibinfo  {journal} {Science}\ }\textbf {\bibinfo {volume}
  {363}},\ \bibinfo {pages} {eaar7709} (\bibinfo {year} {2019})}\BibitemShut
  {NoStop}%
\bibitem [{\citenamefont {Bergholtz}\ \emph {et~al.}(2021)\citenamefont
  {Bergholtz}, \citenamefont {Budich},\ and\ \citenamefont
  {Kunst}}]{Bergholtz21RMP}%
  \BibitemOpen
  \bibfield  {author} {\bibinfo {author} {\bibfnamefont {E.~J.}\ \bibnamefont
  {Bergholtz}}, \bibinfo {author} {\bibfnamefont {J.~C.}\ \bibnamefont
  {Budich}}, \ and\ \bibinfo {author} {\bibfnamefont {F.~K.}\ \bibnamefont
  {Kunst}},\ }\bibfield  {title} {\enquote {\bibinfo {title} {{Exceptional
  topology of non-Hermitian systems}}}, }\href {\doibase
  10.1103/RevModPhys.93.015005} {\bibfield  {journal} {\bibinfo  {journal}
  {Rev. Mod. Phys.}\ }\textbf {\bibinfo {volume} {93}},\ \bibinfo {pages}
  {015005} (\bibinfo {year} {2021})}\BibitemShut {NoStop}%
\bibitem [{\citenamefont {Kawabata}\ \emph
  {et~al.}(2019{\natexlab{a}})\citenamefont {Kawabata}, \citenamefont
  {Shiozaki}, \citenamefont {Ueda},\ and\ \citenamefont
  {Sato}}]{Kawabata19PRX}%
  \BibitemOpen
  \bibfield  {author} {\bibinfo {author} {\bibfnamefont {K.}~\bibnamefont
  {Kawabata}}, \bibinfo {author} {\bibfnamefont {K.}~\bibnamefont {Shiozaki}},
  \bibinfo {author} {\bibfnamefont {M.}~\bibnamefont {Ueda}}, \ and\ \bibinfo
  {author} {\bibfnamefont {M.}~\bibnamefont {Sato}},\ }\bibfield  {title}
  {\enquote {\bibinfo {title} {{Symmetry and Topology in Non-Hermitian
  Physics}}}, }\href {\doibase 10.1103/PhysRevX.9.041015} {\bibfield  {journal}
  {\bibinfo  {journal} {Phys. Rev. X}\ }\textbf {\bibinfo {volume} {9}},\
  \bibinfo {pages} {041015} (\bibinfo {year} {2019}{\natexlab{a}})}\BibitemShut
  {NoStop}%
\bibitem [{\citenamefont {Zhang}\ \emph
  {et~al.}(2020{\natexlab{a}})\citenamefont {Zhang}, \citenamefont {Yang},\
  and\ \citenamefont {Fang}}]{KZhang20PRL}%
  \BibitemOpen
  \bibfield  {author} {\bibinfo {author} {\bibfnamefont {K.}~\bibnamefont
  {Zhang}}, \bibinfo {author} {\bibfnamefont {Z.}~\bibnamefont {Yang}}, \ and\
  \bibinfo {author} {\bibfnamefont {C.}~\bibnamefont {Fang}},\ }\bibfield
  {title} {\enquote {\bibinfo {title} {{Correspondence between Winding Numbers
  and Skin Modes in Non-Hermitian Systems}}}, }\href {\doibase
  10.1103/PhysRevLett.125.126402} {\bibfield  {journal} {\bibinfo  {journal}
  {Phys. Rev. Lett.}\ }\textbf {\bibinfo {volume} {125}},\ \bibinfo {pages}
  {126402} (\bibinfo {year} {2020}{\natexlab{a}})}\BibitemShut {NoStop}%
\bibitem [{\citenamefont {Okuma}\ \emph {et~al.}(2020)\citenamefont {Okuma},
  \citenamefont {Kawabata}, \citenamefont {Shiozaki},\ and\ \citenamefont
  {Sato}}]{Okuma20PRL}%
  \BibitemOpen
  \bibfield  {author} {\bibinfo {author} {\bibfnamefont {N.}~\bibnamefont
  {Okuma}}, \bibinfo {author} {\bibfnamefont {K.}~\bibnamefont {Kawabata}},
  \bibinfo {author} {\bibfnamefont {K.}~\bibnamefont {Shiozaki}}, \ and\
  \bibinfo {author} {\bibfnamefont {M.}~\bibnamefont {Sato}},\ }\bibfield
  {title} {\enquote {\bibinfo {title} {Topological origin of non-hermitian skin
  effects}}, }\href {\doibase 10.1103/PhysRevLett.124.086801} {\bibfield
  {journal} {\bibinfo  {journal} {Phys. Rev. Lett.}\ }\textbf {\bibinfo
  {volume} {124}},\ \bibinfo {pages} {086801} (\bibinfo {year}
  {2020})}\BibitemShut {NoStop}%
\bibitem [{\citenamefont {Leykam}\ \emph {et~al.}(2017)\citenamefont {Leykam},
  \citenamefont {Bliokh}, \citenamefont {Huang}, \citenamefont {Chong},\ and\
  \citenamefont {Nori}}]{Leykam17PRL}%
  \BibitemOpen
  \bibfield  {author} {\bibinfo {author} {\bibfnamefont {D.}~\bibnamefont
  {Leykam}}, \bibinfo {author} {\bibfnamefont {K.~Y.}\ \bibnamefont {Bliokh}},
  \bibinfo {author} {\bibfnamefont {C.}~\bibnamefont {Huang}}, \bibinfo
  {author} {\bibfnamefont {Y.~D.}\ \bibnamefont {Chong}}, \ and\ \bibinfo
  {author} {\bibfnamefont {F.}~\bibnamefont {Nori}},\ }\bibfield  {title}
  {\enquote {\bibinfo {title} {{Edge Modes, Degeneracies, and Topological
  Numbers in Non-Hermitian Systems}}}, }\href {\doibase
  10.1103/PhysRevLett.118.040401} {\bibfield  {journal} {\bibinfo  {journal}
  {Phys. Rev. Lett.}\ }\textbf {\bibinfo {volume} {118}},\ \bibinfo {pages}
  {040401} (\bibinfo {year} {2017})}\BibitemShut {NoStop}%
\bibitem [{\citenamefont {Martinez~Alvarez}\ \emph {et~al.}(2018)\citenamefont
  {Martinez~Alvarez}, \citenamefont {Barrios~Vargas},\ and\ \citenamefont
  {Foa~Torres}}]{Martinez18PRB}%
  \BibitemOpen
  \bibfield  {author} {\bibinfo {author} {\bibfnamefont {V.~M.}\ \bibnamefont
  {Martinez~Alvarez}}, \bibinfo {author} {\bibfnamefont {J.~E.}\ \bibnamefont
  {Barrios~Vargas}}, \ and\ \bibinfo {author} {\bibfnamefont {L.~E.~F.}\
  \bibnamefont {Foa~Torres}},\ }\bibfield  {title} {\enquote {\bibinfo {title}
  {{Non-Hermitian robust edge states in one dimension: Anomalous localization
  and eigenspace condensation at exceptional points}}}, }\href {\doibase
  10.1103/PhysRevB.97.121401} {\bibfield  {journal} {\bibinfo  {journal} {Phys.
  Rev. B}\ }\textbf {\bibinfo {volume} {97}},\ \bibinfo {pages} {121401}
  (\bibinfo {year} {2018})}\BibitemShut {NoStop}%
\bibitem [{\citenamefont {Lieu}(2018)}]{Lieu18PRB}%
  \BibitemOpen
  \bibfield  {author} {\bibinfo {author} {\bibfnamefont {S.}~\bibnamefont
  {Lieu}},\ }\bibfield  {title} {\enquote {\bibinfo {title} {{Topological
  phases in the non-Hermitian Su-Schrieffer-Heeger model}}}, }\href {\doibase
  10.1103/PhysRevB.97.045106} {\bibfield  {journal} {\bibinfo  {journal} {Phys.
  Rev. B}\ }\textbf {\bibinfo {volume} {97}},\ \bibinfo {pages} {045106}
  (\bibinfo {year} {2018})}\BibitemShut {NoStop}%
\bibitem [{\citenamefont {Xiong}(2018)}]{YXiong18JPC}%
  \BibitemOpen
  \bibfield  {author} {\bibinfo {author} {\bibfnamefont {Y.}~\bibnamefont
  {Xiong}},\ }\bibfield  {title} {\enquote {\bibinfo {title} {{Why does bulk
  boundary correspondence fail in some non-Hermitian topological models}}},
  }\href {\doibase 10.1088/2399-6528/aab64a} {\bibfield  {journal} {\bibinfo
  {journal} {J. Phys. Commun.}\ }\textbf {\bibinfo {volume} {2}},\ \bibinfo
  {pages} {035043} (\bibinfo {year} {2018})}\BibitemShut {NoStop}%
\bibitem [{\citenamefont {Yoshida}\ \emph {et~al.}(2018)\citenamefont
  {Yoshida}, \citenamefont {Peters},\ and\ \citenamefont
  {Kawakami}}]{Yoshida18PRB}%
  \BibitemOpen
  \bibfield  {author} {\bibinfo {author} {\bibfnamefont {T.}~\bibnamefont
  {Yoshida}}, \bibinfo {author} {\bibfnamefont {R.}~\bibnamefont {Peters}}, \
  and\ \bibinfo {author} {\bibfnamefont {N.}~\bibnamefont {Kawakami}},\
  }\bibfield  {title} {\enquote {\bibinfo {title} {{Non-Hermitian perspective
  of the band structure in heavy-fermion systems}}}, }\href {\doibase
  10.1103/PhysRevB.98.035141} {\bibfield  {journal} {\bibinfo  {journal} {Phys.
  Rev. B}\ }\textbf {\bibinfo {volume} {98}},\ \bibinfo {pages} {035141}
  (\bibinfo {year} {2018})}\BibitemShut {NoStop}%
\bibitem [{\citenamefont {Rui}\ \emph {et~al.}(2019)\citenamefont {Rui},
  \citenamefont {Hirschmann},\ and\ \citenamefont {Schnyder}}]{WBRui19PRB}%
  \BibitemOpen
  \bibfield  {author} {\bibinfo {author} {\bibfnamefont {W.~B.}\ \bibnamefont
  {Rui}}, \bibinfo {author} {\bibfnamefont {M.~M.}\ \bibnamefont {Hirschmann}},
  \ and\ \bibinfo {author} {\bibfnamefont {A.~P.}\ \bibnamefont {Schnyder}},\
  }\bibfield  {title} {\enquote {\bibinfo {title} {{$\mathcal{PT}$-symmetric
  non-Hermitian Dirac semimetals}}}, }\href {\doibase
  10.1103/PhysRevB.100.245116} {\bibfield  {journal} {\bibinfo  {journal}
  {Phys. Rev. B}\ }\textbf {\bibinfo {volume} {100}},\ \bibinfo {pages}
  {245116} (\bibinfo {year} {2019})}\BibitemShut {NoStop}%
\bibitem [{\citenamefont {Longhi}(2019)}]{Longhi19PRL}%
  \BibitemOpen
  \bibfield  {author} {\bibinfo {author} {\bibfnamefont {S.}~\bibnamefont
  {Longhi}},\ }\bibfield  {title} {\enquote {\bibinfo {title} {{Topological
  Phase Transition in non-Hermitian Quasicrystals}}}, }\href {\doibase
  10.1103/PhysRevLett.122.237601} {\bibfield  {journal} {\bibinfo  {journal}
  {Phys. Rev. Lett.}\ }\textbf {\bibinfo {volume} {122}},\ \bibinfo {pages}
  {237601} (\bibinfo {year} {2019})}\BibitemShut {NoStop}%
\bibitem [{\citenamefont {Lee}\ and\ \citenamefont
  {Thomale}(2019)}]{CHLee19PRB}%
  \BibitemOpen
  \bibfield  {author} {\bibinfo {author} {\bibfnamefont {C.~H.}\ \bibnamefont
  {Lee}}\ and\ \bibinfo {author} {\bibfnamefont {R.}~\bibnamefont {Thomale}},\
  }\bibfield  {title} {\enquote {\bibinfo {title} {{Anatomy of skin modes and
  topology in non-Hermitian systems}}}, }\href {\doibase
  10.1103/PhysRevB.99.201103} {\bibfield  {journal} {\bibinfo  {journal} {Phys.
  Rev. B}\ }\textbf {\bibinfo {volume} {99}},\ \bibinfo {pages} {201103}
  (\bibinfo {year} {2019})}\BibitemShut {NoStop}%
\bibitem [{\citenamefont {Kawabata}\ \emph
  {et~al.}(2019{\natexlab{b}})\citenamefont {Kawabata}, \citenamefont
  {Bessho},\ and\ \citenamefont {Sato}}]{Kawabata19PRL}%
  \BibitemOpen
  \bibfield  {author} {\bibinfo {author} {\bibfnamefont {K.}~\bibnamefont
  {Kawabata}}, \bibinfo {author} {\bibfnamefont {T.}~\bibnamefont {Bessho}}, \
  and\ \bibinfo {author} {\bibfnamefont {M.}~\bibnamefont {Sato}},\ }\bibfield
  {title} {\enquote {\bibinfo {title} {{Classification of Exceptional Points
  and Non-Hermitian Topological Semimetals}}}, }\href {\doibase
  10.1103/PhysRevLett.123.066405} {\bibfield  {journal} {\bibinfo  {journal}
  {Phys. Rev. Lett.}\ }\textbf {\bibinfo {volume} {123}},\ \bibinfo {pages}
  {066405} (\bibinfo {year} {2019}{\natexlab{b}})}\BibitemShut {NoStop}%
\bibitem [{\citenamefont {Yoshida}\ \emph
  {et~al.}(2019{\natexlab{b}})\citenamefont {Yoshida}, \citenamefont
  {Bessho},\ and\ \citenamefont {Sato}}]{Yoshida19PRB}%
  \BibitemOpen
  \bibfield  {author} {\bibinfo {author} {\bibfnamefont {T.}~\bibnamefont
  {Yoshida}}, \bibinfo {author} {\bibfnamefont {R.}~\bibnamefont {Peters}},
  \bibinfo {author} {\bibfnamefont {N.}~\bibnamefont {Kawakami}}, \
  and\ \bibinfo {author} {\bibfnamefont {Y.}~\bibnamefont {Hatsugai}},\ }\bibfield
  {title} {\enquote {\bibinfo {title} {{Symmetry-protected exceptional rings in
  two-dimensional correlated systems with chiral symmetry}}}, }\href {\doibase
  10.1103/PhysRevB.99.121101} {\bibfield  {journal} {\bibinfo  {journal}
  {Phys. Rev. B}\ }\textbf {\bibinfo {volume} {99}},\ \bibinfo {pages}
  {121101} (\bibinfo {year} {2019}{\natexlab{b}})}\BibitemShut {NoStop}%
\bibitem [{\citenamefont {Lee}\ \emph {et~al.}(2019)\citenamefont {Lee},
  \citenamefont {Ahn}, \citenamefont {Zhou},\ and\ \citenamefont
  {Vishwanath}}]{JYLee19PRL}%
  \BibitemOpen
  \bibfield  {author} {\bibinfo {author} {\bibfnamefont {J.~Y.}\ \bibnamefont
  {Lee}}, \bibinfo {author} {\bibfnamefont {J.}~\bibnamefont {Ahn}}, \bibinfo
  {author} {\bibfnamefont {H.}~\bibnamefont {Zhou}}, \ and\ \bibinfo {author}
  {\bibfnamefont {A.}~\bibnamefont {Vishwanath}},\ }\bibfield  {title}
  {\enquote {\bibinfo {title} {{Topological Correspondence between Hermitian
  and Non-Hermitian Systems: Anomalous Dynamics}}}, }\href {\doibase
  10.1103/PhysRevLett.123.206404} {\bibfield  {journal} {\bibinfo  {journal}
  {Phys. Rev. Lett.}\ }\textbf {\bibinfo {volume} {123}},\ \bibinfo {pages}
  {206404} (\bibinfo {year} {2019})}\BibitemShut {NoStop}%
\bibitem [{\citenamefont {Borgnia}\ \emph {et~al.}(2020)\citenamefont
  {Borgnia}, \citenamefont {Kruchkov},\ and\ \citenamefont
  {Slager}}]{Borgnia20PRL}%
  \BibitemOpen
  \bibfield  {author} {\bibinfo {author} {\bibfnamefont {D.~S.}\ \bibnamefont
  {Borgnia}}, \bibinfo {author} {\bibfnamefont {A.~J.}\ \bibnamefont
  {Kruchkov}}, \ and\ \bibinfo {author} {\bibfnamefont {R.-J.}\ \bibnamefont
  {Slager}},\ }\bibfield  {title} {\enquote {\bibinfo {title} {{Non-Hermitian
  Boundary Modes and Topology}}}, }\href {\doibase
  10.1103/PhysRevLett.124.056802} {\bibfield  {journal} {\bibinfo  {journal}
  {Phys. Rev. Lett.}\ }\textbf {\bibinfo {volume} {124}},\ \bibinfo {pages}
  {056802} (\bibinfo {year} {2020})}\BibitemShut {NoStop}%
\bibitem [{\citenamefont {Li}\ \emph {et~al.}(2020)\citenamefont {Li},
  \citenamefont {Lee},\ and\ \citenamefont {Gong}}]{LHLi20PRL}%
  \BibitemOpen
  \bibfield  {author} {\bibinfo {author} {\bibfnamefont {L.}~\bibnamefont
  {Li}}, \bibinfo {author} {\bibfnamefont {C.~H.}\ \bibnamefont {Lee}}, \ and\
  \bibinfo {author} {\bibfnamefont {J.}~\bibnamefont {Gong}},\ }\bibfield
  {title} {\enquote {\bibinfo {title} {{Topological Switch for Non-Hermitian
  Skin Effect in Cold-Atom Systems with Loss}}}, }\href {\doibase
  10.1103/PhysRevLett.124.250402} {\bibfield  {journal} {\bibinfo  {journal}
  {Phys. Rev. Lett.}\ }\textbf {\bibinfo {volume} {124}},\ \bibinfo {pages}
  {250402} (\bibinfo {year} {2020})}\BibitemShut {NoStop}%
\bibitem [{\citenamefont {Wojcik}\ \emph {et~al.}(2020)\citenamefont {Wojcik},
  \citenamefont {Sun}, \citenamefont {Bzdu\ifmmode~\check{s}\else
  \v{s}\fi{}ek},\ and\ \citenamefont {Fan}}]{Wojcik20prb}%
  \BibitemOpen
  \bibfield  {author} {\bibinfo {author} {\bibfnamefont {C.~C.}\ \bibnamefont
  {Wojcik}}, \bibinfo {author} {\bibfnamefont {X.-Q.}\ \bibnamefont {Sun}},
  \bibinfo {author} {\bibfnamefont {T.}\ \bibnamefont
  {Bzdu\ifmmode~\check{s}\else \v{s}\fi{}ek}}, \ and\ \bibinfo {author}
  {\bibfnamefont {S.}~\bibnamefont {Fan}},\ }\bibfield  {title} {\enquote
  {\bibinfo {title} {{Homotopy characterization of non-Hermitian
  Hamiltonians}}}, }\href {\doibase 10.1103/PhysRevB.101.205417} {\bibfield
  {journal} {\bibinfo  {journal} {Phys. Rev. B}\ }\textbf {\bibinfo {volume}
  {101}},\ \bibinfo {pages} {205417} (\bibinfo {year} {2020})}\BibitemShut
  {NoStop}%
\bibitem [{\citenamefont {Denner}\ \emph {et~al.}(2021)\citenamefont {Denner},
  \citenamefont {Skurativska}, \citenamefont {Schindler}, \citenamefont
  {Fischer}, \citenamefont {Thomale}, \citenamefont {Bzdu{\v{s}}ek},\ and\
  \citenamefont {Neupert}}]{Denner21NC}%
  \BibitemOpen
  \bibfield  {author} {\bibinfo {author} {\bibfnamefont {M.~M.}\ \bibnamefont
  {Denner}}, \bibinfo {author} {\bibfnamefont {A.}~\bibnamefont {Skurativska}},
  \bibinfo {author} {\bibfnamefont {F.}~\bibnamefont {Schindler}}, \bibinfo
  {author} {\bibfnamefont {M.~H.}\ \bibnamefont {Fischer}}, \bibinfo {author}
  {\bibfnamefont {R.}~\bibnamefont {Thomale}}, \bibinfo {author} {\bibfnamefont
  {T.}~\bibnamefont {Bzdu{\v{s}}ek}}, \ and\ \bibinfo {author} {\bibfnamefont
  {T.}~\bibnamefont {Neupert}},\ }\bibfield  {title} {\enquote {\bibinfo
  {title} {Exceptional topological insulators}}, }\href {\doibase
  10.1038/s41467-021-25947-z} {\bibfield  {journal} {\bibinfo  {journal}
  {Nature Commun.}\ }\textbf {\bibinfo {volume} {12}},\ \bibinfo {pages} {5681}
  (\bibinfo {year} {2021})}\BibitemShut {NoStop}%
\bibitem [{\citenamefont {Vecsei}\ \emph {et~al.}(2021)\citenamefont {Vecsei},
  \citenamefont {Denner}, \citenamefont {Neupert},\ and\
  \citenamefont {Schindler}}]{Vecsei21PRB}%
  \BibitemOpen
  \bibfield  {author} {\bibinfo {author} {\bibfnamefont {P.~M.}\ \bibnamefont
  {Vecsei}}, \bibinfo {author} {\bibfnamefont {M.~M.}~\bibnamefont {Denner}},
  \bibinfo {author} {\bibfnamefont {T.}~\bibnamefont {Neupert}}, \ and\ \bibinfo {author} {\bibfnamefont
  {F.}~\bibnamefont {Schindler}},\ }\bibfield  {title} {\enquote {\bibinfo
  {title} {Symmetry indicators for inversion-symmetric non-Hermitian topological band structures}}, }\href {\doibase
   10.1103/PhysRevB.103.L201114} {\bibfield  {journal} {\bibinfo  {journal}
  {Phys. Rev. B}\ }\textbf {\bibinfo {volume} {103}},\ \bibinfo {pages} {L201114}
  (\bibinfo {year} {2021})}\BibitemShut {NoStop}%
\bibitem [{\citenamefont {Schindler}\ \emph {et~al.}(2021)\citenamefont {Schindler} \ and\
  \citenamefont {Prem}}]{Schindler21PRB}%
  \BibitemOpen
  \bibfield  {author} {\bibinfo {author} {\bibfnamefont {F.}\ \bibnamefont
  {Schindler}} \ and\ \bibinfo {author} {\bibfnamefont
  {A.}~\bibnamefont {Prem}},\ }\bibfield  {title} {\enquote {\bibinfo
  {title} {Dislocation non-Hermitian skin effect}}, }\href {\doibase
   10.1103/PhysRevB.104.L161106} {\bibfield  {journal} {\bibinfo  {journal}
  {Phys. Rev. B}\ }\textbf {\bibinfo {volume} {104}},\ \bibinfo {pages} {L161106}
  (\bibinfo {year} {2021})}\BibitemShut {NoStop}%
\bibitem [{\citenamefont {Ryu}\ \emph {et~al.}(2017)\citenamefont {Ryu},
  \citenamefont {Myoung}, \citenamefont {Go}, \citenamefont {Woo},\
  and\ \citenamefont {Park}}]{Ryu20PRR}%
  \BibitemOpen
  \bibfield  {author} {\bibinfo {author} {\bibfnamefont {J.-W.}~\bibnamefont
  {Ryu}}, \bibinfo {author} {\bibfnamefont {N}\ \bibnamefont
  {Myoung}}, \bibinfo {author} {\bibfnamefont {A.}~\bibnamefont {Go}},
  \bibinfo {author} {\bibfnamefont {S.}~\bibnamefont {Woo}},   \bibinfo {author} {\bibfnamefont {S.-J.}~\bibnamefont {Choi}}, \ and\
  \bibinfo {author} {\bibfnamefont {H. C.}~\bibnamefont {Park}},\ }\bibfield
  {title} {\enquote {\bibinfo {title} {Emergent localized states at the interface of a twofold $\mathcal{PT}$-symmetric lattice}}, }\href {\doibase 10.1103/PhysRevResearch.2.033149} {\bibfield
  {journal} {\bibinfo  {journal} {Phys. Rev. Research}\ }\textbf {\bibinfo {volume} {2}},\
  \bibinfo {pages} {033149} (\bibinfo {year} {2020})}\BibitemShut {NoStop}%
\bibitem [{\citenamefont {Chen}\ \emph {et~al.}(2017)\citenamefont {Chen},
  \citenamefont {{\"O}zdemir}, \citenamefont {Zhao}, \citenamefont {Wiersig},\
  and\ \citenamefont {Yang}}]{WJChen17Nature}%
  \BibitemOpen
  \bibfield  {author} {\bibinfo {author} {\bibfnamefont {W.}~\bibnamefont
  {Chen}}, \bibinfo {author} {\bibfnamefont {{\c{S}}.~K.}\ \bibnamefont
  {{\"O}zdemir}}, \bibinfo {author} {\bibfnamefont {G.}~\bibnamefont {Zhao}},
  \bibinfo {author} {\bibfnamefont {J.}~\bibnamefont {Wiersig}}, \ and\
  \bibinfo {author} {\bibfnamefont {L.}~\bibnamefont {Yang}},\ }\bibfield
  {title} {\enquote {\bibinfo {title} {Exceptional points enhance sensing in an
  optical microcavity}}, }\href {\doibase 10.1038/nature23281} {\bibfield
  {journal} {\bibinfo  {journal} {Nature}\ }\textbf {\bibinfo {volume} {548}},\
  \bibinfo {pages} {192} (\bibinfo {year} {2017})}\BibitemShut {NoStop}%
\bibitem [{\citenamefont {Hodaei}\ \emph {et~al.}(2017)\citenamefont {Hodaei},
  \citenamefont {Hassan}, \citenamefont {Wittek}, \citenamefont
  {Garcia-Gracia}, \citenamefont {El-Ganainy}, \citenamefont
  {Christodoulides},\ and\ \citenamefont {Khajavikhan}}]{Hodaei17Nature}%
  \BibitemOpen
  \bibfield  {author} {\bibinfo {author} {\bibfnamefont {H.}~\bibnamefont
  {Hodaei}}, \bibinfo {author} {\bibfnamefont {A.~U.}\ \bibnamefont {Hassan}},
  \bibinfo {author} {\bibfnamefont {S.}~\bibnamefont {Wittek}}, \bibinfo
  {author} {\bibfnamefont {H.}~\bibnamefont {Garcia-Gracia}}, \bibinfo {author}
  {\bibfnamefont {R.}~\bibnamefont {El-Ganainy}}, \bibinfo {author}
  {\bibfnamefont {D.~N.}\ \bibnamefont {Christodoulides}}, \ and\ \bibinfo
  {author} {\bibfnamefont {M.}~\bibnamefont {Khajavikhan}},\ }\bibfield
  {title} {\enquote {\bibinfo {title} {Enhanced sensitivity at higher-order
  exceptional points}}, }\href {\doibase 10.1038/nature23280} {\bibfield
  {journal} {\bibinfo  {journal} {Nature}\ }\textbf {\bibinfo {volume} {548}},\
  \bibinfo {pages} {187} (\bibinfo {year} {2017})}\BibitemShut {NoStop}%
\bibitem [{\citenamefont {Weimann}\ \emph {et~al.}(2017)\citenamefont
  {Weimann}, \citenamefont {Kremer}, \citenamefont {Plotnik}, \citenamefont
  {Lumer}, \citenamefont {Nolte}, \citenamefont {Makris}, \citenamefont
  {Segev}, \citenamefont {Rechtsman},\ and\ \citenamefont
  {Szameit}}]{Weimann17Nmater}%
  \BibitemOpen
  \bibfield  {author} {\bibinfo {author} {\bibfnamefont {S.}~\bibnamefont
  {Weimann}}, \bibinfo {author} {\bibfnamefont {M.}~\bibnamefont {Kremer}},
  \bibinfo {author} {\bibfnamefont {Y.}~\bibnamefont {Plotnik}}, \bibinfo
  {author} {\bibfnamefont {Y.}~\bibnamefont {Lumer}}, \bibinfo {author}
  {\bibfnamefont {S.}~\bibnamefont {Nolte}}, \bibinfo {author} {\bibfnamefont
  {K.~G.}\ \bibnamefont {Makris}}, \bibinfo {author} {\bibfnamefont
  {M.}~\bibnamefont {Segev}}, \bibinfo {author} {\bibfnamefont {M.~C.}\
  \bibnamefont {Rechtsman}}, \ and\ \bibinfo {author} {\bibfnamefont
  {A.}~\bibnamefont {Szameit}},\ }\bibfield  {title} {\enquote {\bibinfo
  {title} {Topologically protected bound states in photonic
  parity--time-symmetric crystals}}, }\href {\doibase 10.1038/nmat4811}
  {\bibfield  {journal} {\bibinfo  {journal} {Nature Mater.}\ }\textbf
  {\bibinfo {volume} {16}},\ \bibinfo {pages} {433} (\bibinfo {year}
  {2017})}\BibitemShut {NoStop}%
\bibitem [{\citenamefont {Zhou}\ \emph {et~al.}(2018)\citenamefont {Zhou},
  \citenamefont {Peng}, \citenamefont {Yoon}, \citenamefont {Hsu},
  \citenamefont {Nelson}, \citenamefont {Fu}, \citenamefont {Joannopoulos},
  \citenamefont {Solja{\v{c}}i{\'c}},\ and\ \citenamefont
  {Zhen}}]{HYZhou18Science}%
  \BibitemOpen
  \bibfield  {author} {\bibinfo {author} {\bibfnamefont {H.}~\bibnamefont
  {Zhou}}, \bibinfo {author} {\bibfnamefont {C.}~\bibnamefont {Peng}}, \bibinfo
  {author} {\bibfnamefont {Y.}~\bibnamefont {Yoon}}, \bibinfo {author}
  {\bibfnamefont {C.~W.}\ \bibnamefont {Hsu}}, \bibinfo {author} {\bibfnamefont
  {K.~A.}\ \bibnamefont {Nelson}}, \bibinfo {author} {\bibfnamefont
  {L.}~\bibnamefont {Fu}}, \bibinfo {author} {\bibfnamefont {J.~D.}\
  \bibnamefont {Joannopoulos}}, \bibinfo {author} {\bibfnamefont
  {M.}~\bibnamefont {Solja{\v{c}}i{\'c}}}, \ and\ \bibinfo {author}
  {\bibfnamefont {B.}~\bibnamefont {Zhen}},\ }\bibfield  {title} {\enquote
  {\bibinfo {title} {{Observation of bulk Fermi arc and polarization half
  charge from paired exceptional points}}}, }\href {\doibase
  10.1126/science.aap9859} {\bibfield  {journal} {\bibinfo  {journal}
  {Science}\ }\textbf {\bibinfo {volume} {359}},\ \bibinfo {pages} {1009}
  (\bibinfo {year} {2018})}\BibitemShut {NoStop}%
\bibitem [{\citenamefont {Zhao}\ \emph {et~al.}(2019)\citenamefont {Zhao},
  \citenamefont {Qiao}, \citenamefont {Wu}, \citenamefont {Midya},
  \citenamefont {Longhi},\ and\ \citenamefont {Feng}}]{HZhao19Science}%
  \BibitemOpen
  \bibfield  {author} {\bibinfo {author} {\bibfnamefont {H.}~\bibnamefont
  {Zhao}}, \bibinfo {author} {\bibfnamefont {X.}~\bibnamefont {Qiao}}, \bibinfo
  {author} {\bibfnamefont {T.}~\bibnamefont {Wu}}, \bibinfo {author}
  {\bibfnamefont {B.}~\bibnamefont {Midya}}, \bibinfo {author} {\bibfnamefont
  {S.}~\bibnamefont {Longhi}}, \ and\ \bibinfo {author} {\bibfnamefont
  {L.}~\bibnamefont {Feng}},\ }\bibfield  {title} {\enquote {\bibinfo {title}
  {{Non-Hermitian topological light steering}}}, }\href {\doibase
  10.1126/science.aay1064} {\bibfield  {journal} {\bibinfo  {journal}
  {Science}\ }\textbf {\bibinfo {volume} {365}},\ \bibinfo {pages} {1163}
  (\bibinfo {year} {2019})}\BibitemShut {NoStop}%
\bibitem [{\citenamefont {Cerjan}\ \emph {et~al.}(2019)\citenamefont {Cerjan},
  \citenamefont {Huang}, \citenamefont {Wang}, \citenamefont {Chen},
  \citenamefont {Chong},\ and\ \citenamefont {Rechtsman}}]{Cerjan19Nphoto}%
  \BibitemOpen
  \bibfield  {author} {\bibinfo {author} {\bibfnamefont {A.}~\bibnamefont
  {Cerjan}}, \bibinfo {author} {\bibfnamefont {S.}~\bibnamefont {Huang}},
  \bibinfo {author} {\bibfnamefont {M.}~\bibnamefont {Wang}}, \bibinfo {author}
  {\bibfnamefont {K.~P.}\ \bibnamefont {Chen}}, \bibinfo {author}
  {\bibfnamefont {Y.}~\bibnamefont {Chong}}, \ and\ \bibinfo {author}
  {\bibfnamefont {M.~C.}\ \bibnamefont {Rechtsman}},\ }\bibfield  {title}
  {\enquote {\bibinfo {title} {{Experimental realization of a Weyl exceptional
  ring}}}, }\href {\doibase 10.1038/s41566-019-0453-z} {\bibfield  {journal}
  {\bibinfo  {journal} {Nature Photonics}\ }\textbf {\bibinfo {volume} {13}},\
  \bibinfo {pages} {623} (\bibinfo {year} {2019})}\BibitemShut {NoStop}%
\bibitem [{\citenamefont {Helbig}\ \emph {et~al.}(2020)\citenamefont {Helbig},
  \citenamefont {Hofmann}, \citenamefont {Imhof}, \citenamefont {Abdelghany},
  \citenamefont {Kiessling}, \citenamefont {Molenkamp}, \citenamefont {Lee},
  \citenamefont {Szameit}, \citenamefont {Greiter},\ and\ \citenamefont
  {Thomale}}]{Helbig20Nphys}%
  \BibitemOpen
  \bibfield  {author} {\bibinfo {author} {\bibfnamefont {T.}~\bibnamefont
  {Helbig}}, \bibinfo {author} {\bibfnamefont {T.}~\bibnamefont {Hofmann}},
  \bibinfo {author} {\bibfnamefont {S.}~\bibnamefont {Imhof}}, \bibinfo
  {author} {\bibfnamefont {M.}~\bibnamefont {Abdelghany}}, \bibinfo {author}
  {\bibfnamefont {T.}~\bibnamefont {Kiessling}}, \bibinfo {author}
  {\bibfnamefont {L.}~\bibnamefont {Molenkamp}}, \bibinfo {author}
  {\bibfnamefont {C.}~\bibnamefont {Lee}}, \bibinfo {author} {\bibfnamefont
  {A.}~\bibnamefont {Szameit}}, \bibinfo {author} {\bibfnamefont
  {M.}~\bibnamefont {Greiter}}, \ and\ \bibinfo {author} {\bibfnamefont
  {R.}~\bibnamefont {Thomale}},\ }\bibfield  {title} {\enquote {\bibinfo
  {title} {{Generalized bulk--boundary correspondence in non-Hermitian
  topolectrical circuits}}}, }\href {\doibase 10.1038/s41567-020-0922-9}
  {\bibfield  {journal} {\bibinfo  {journal} {Nature Phys.}\ }\textbf {\bibinfo
  {volume} {16}},\ \bibinfo {pages} {747} (\bibinfo {year} {2020})}\BibitemShut
  {NoStop}%
\bibitem [{\citenamefont {Weidemann}\ \emph {et~al.}(2020)\citenamefont
  {Weidemann}, \citenamefont {Kremer}, \citenamefont {Helbig}, \citenamefont
  {Hofmann}, \citenamefont {Stegmaier}, \citenamefont {Greiter}, \citenamefont
  {Thomale},\ and\ \citenamefont {Szameit}}]{Weidemann20science}%
  \BibitemOpen
  \bibfield  {author} {\bibinfo {author} {\bibfnamefont {S.}~\bibnamefont
  {Weidemann}}, \bibinfo {author} {\bibfnamefont {M.}~\bibnamefont {Kremer}},
  \bibinfo {author} {\bibfnamefont {T.}~\bibnamefont {Helbig}}, \bibinfo
  {author} {\bibfnamefont {T.}~\bibnamefont {Hofmann}}, \bibinfo {author}
  {\bibfnamefont {A.}~\bibnamefont {Stegmaier}}, \bibinfo {author}
  {\bibfnamefont {M.}~\bibnamefont {Greiter}}, \bibinfo {author} {\bibfnamefont
  {R.}~\bibnamefont {Thomale}}, \ and\ \bibinfo {author} {\bibfnamefont
  {A.}~\bibnamefont {Szameit}},\ }\bibfield  {title} {\enquote {\bibinfo
  {title} {Topological funneling of light}}, }\href {\doibase
  10.1126/science.aaz8727} {\bibfield  {journal} {\bibinfo  {journal}
  {Science}\ }\textbf {\bibinfo {volume} {368}},\ \bibinfo {pages} {311}
  (\bibinfo {year} {2020})}\BibitemShut {NoStop}%
\bibitem [{\citenamefont {Xiao}\ \emph {et~al.}(2020)\citenamefont {Xiao},
  \citenamefont {Deng}, \citenamefont {Wang}, \citenamefont {Zhu},
  \citenamefont {Wang}, \citenamefont {Yi},\ and\ \citenamefont
  {Xue}}]{Xiao20nphys}%
  \BibitemOpen
  \bibfield  {author} {\bibinfo {author} {\bibfnamefont {L.}~\bibnamefont
  {Xiao}}, \bibinfo {author} {\bibfnamefont {T.}~\bibnamefont {Deng}}, \bibinfo
  {author} {\bibfnamefont {K.}~\bibnamefont {Wang}}, \bibinfo {author}
  {\bibfnamefont {G.}~\bibnamefont {Zhu}}, \bibinfo {author} {\bibfnamefont
  {Z.}~\bibnamefont {Wang}}, \bibinfo {author} {\bibfnamefont {W.}~\bibnamefont
  {Yi}}, \ and\ \bibinfo {author} {\bibfnamefont {P.}~\bibnamefont {Xue}},\
  }\bibfield  {title} {\enquote {\bibinfo {title} {{Non-Hermitian
  bulk--boundary correspondence in quantum dynamics}}}, }\href {\doibase
  10.1038/s41567-020-0836-6} {\bibfield  {journal} {\bibinfo  {journal} {Nature
  Phys.}\ }\textbf {\bibinfo {volume} {16}},\ \bibinfo {pages} {761} (\bibinfo
  {year} {2020})}\BibitemShut {NoStop}%
\bibitem [{\citenamefont {Ghatak}\ \emph {et~al.}(2020)\citenamefont {Ghatak},
  \citenamefont {Brandenbourger}, \citenamefont {van Wezel},\ and\
  \citenamefont {Coulais}}]{Ghatak20PNAS}%
  \BibitemOpen
  \bibfield  {author} {\bibinfo {author} {\bibfnamefont {A.}~\bibnamefont
  {Ghatak}}, \bibinfo {author} {\bibfnamefont {M.}~\bibnamefont
  {Brandenbourger}}, \bibinfo {author} {\bibfnamefont {J.}~\bibnamefont {van
  Wezel}}, \ and\ \bibinfo {author} {\bibfnamefont {C.}~\bibnamefont
  {Coulais}},\ }\bibfield  {title} {\enquote {\bibinfo {title} {{Observation of
  non-Hermitian topology and its bulk{\textendash}edge correspondence in an
  active mechanical metamaterial}}}, }\href {\doibase 10.1073/pnas.2010580117}
  {\bibfield  {journal} {\bibinfo  {journal} {Proc. Nat. Acad. Sci.}\ }\textbf
  {\bibinfo {volume} {117}},\ \bibinfo {pages} {29561} (\bibinfo {year}
  {2020})}\BibitemShut {NoStop}%
\bibitem [{\citenamefont {Zhang}\ \emph {et~al.}(2021)\citenamefont {Zhang},
  \citenamefont {Ouyang}, \citenamefont {Huang}, \citenamefont {Wang},
  \citenamefont {Zhang}, \citenamefont {Yu}, \citenamefont {Chang},
  \citenamefont {Liu}, \citenamefont {Deng},\ and\ \citenamefont
  {Duan}}]{WGZhang21PRL}%
  \BibitemOpen
  \bibfield  {author} {\bibinfo {author} {\bibfnamefont {W.}~\bibnamefont
  {Zhang}}, \bibinfo {author} {\bibfnamefont {X.}~\bibnamefont {Ouyang}},
  \bibinfo {author} {\bibfnamefont {X.}~\bibnamefont {Huang}}, \bibinfo
  {author} {\bibfnamefont {X.}~\bibnamefont {Wang}}, \bibinfo {author}
  {\bibfnamefont {H.}~\bibnamefont {Zhang}}, \bibinfo {author} {\bibfnamefont
  {Y.}~\bibnamefont {Yu}}, \bibinfo {author} {\bibfnamefont {X.}~\bibnamefont
  {Chang}}, \bibinfo {author} {\bibfnamefont {Y.}~\bibnamefont {Liu}}, \bibinfo
  {author} {\bibfnamefont {D.-L.}\ \bibnamefont {Deng}}, \ and\ \bibinfo
  {author} {\bibfnamefont {L.-M.}\ \bibnamefont {Duan}},\ }\bibfield  {title}
  {\enquote {\bibinfo {title} {{Observation of Non-Hermitian Topology with
  Nonunitary Dynamics of Solid-State Spins}}}, }\href {\doibase
  10.1103/PhysRevLett.127.090501} {\bibfield  {journal} {\bibinfo  {journal}
  {Phys. Rev. Lett.}\ }\textbf {\bibinfo {volume} {127}},\ \bibinfo {pages}
  {090501} (\bibinfo {year} {2021})}\BibitemShut {NoStop}%
\bibitem [{\citenamefont {Xia}\ \emph {et~al.}(2021)\citenamefont {Xia},
  \citenamefont {Kaltsas}, \citenamefont {Song}, \citenamefont {Komis},
  \citenamefont {Xu}, \citenamefont {Szameit}, \citenamefont {Buljan},
  \citenamefont {Makris},\ and\ \citenamefont {Chen}}]{SQXia21Science}%
  \BibitemOpen
  \bibfield  {author} {\bibinfo {author} {\bibfnamefont {S.}~\bibnamefont
  {Xia}}, \bibinfo {author} {\bibfnamefont {D.}~\bibnamefont {Kaltsas}},
  \bibinfo {author} {\bibfnamefont {D.}~\bibnamefont {Song}}, \bibinfo {author}
  {\bibfnamefont {I.}~\bibnamefont {Komis}}, \bibinfo {author} {\bibfnamefont
  {J.}~\bibnamefont {Xu}}, \bibinfo {author} {\bibfnamefont {A.}~\bibnamefont
  {Szameit}}, \bibinfo {author} {\bibfnamefont {H.}~\bibnamefont {Buljan}},
  \bibinfo {author} {\bibfnamefont {K.~G.}\ \bibnamefont {Makris}}, \ and\
  \bibinfo {author} {\bibfnamefont {Z.}~\bibnamefont {Chen}},\ }\bibfield
  {title} {\enquote {\bibinfo {title} {{Nonlinear tuning of PT symmetry and
  non-Hermitian topological states}}}, }\href {\doibase
  10.1126/science.abf6873} {\bibfield  {journal} {\bibinfo  {journal}
  {Science}\ }\textbf {\bibinfo {volume} {372}},\ \bibinfo {pages} {72}
  (\bibinfo {year} {2021})}\BibitemShut {NoStop}%
\bibitem [{\citenamefont {Su}\ \emph {et~al.}(2021)\citenamefont {Su},
  \citenamefont {Estrecho}, \citenamefont {Biega\'nska}, \citenamefont {Huang},
  \citenamefont {Wurdack}, \citenamefont {Pieczarka}, \citenamefont {Truscott},
  \citenamefont {Liew}, \citenamefont {Ostrovskaya},\ and\ \citenamefont
  {Xiong}}]{RSu21SciAdv}%
  \BibitemOpen
  \bibfield  {author} {\bibinfo {author} {\bibfnamefont {R.}~\bibnamefont
  {Su}}, \bibinfo {author} {\bibfnamefont {E.}~\bibnamefont {Estrecho}},
  \bibinfo {author} {\bibfnamefont {D.}~\bibnamefont {Biega\'nska}}, \bibinfo
  {author} {\bibfnamefont {Y.}~\bibnamefont {Huang}}, \bibinfo {author}
  {\bibfnamefont {M.}~\bibnamefont {Wurdack}}, \bibinfo {author} {\bibfnamefont
  {M.}~\bibnamefont {Pieczarka}}, \bibinfo {author} {\bibfnamefont {A.~G.}\
  \bibnamefont {Truscott}}, \bibinfo {author} {\bibfnamefont {T.~C.~H.}\
  \bibnamefont {Liew}}, \bibinfo {author} {\bibfnamefont {E.~A.}\ \bibnamefont
  {Ostrovskaya}}, \ and\ \bibinfo {author} {\bibfnamefont {Q.}~\bibnamefont
  {Xiong}},\ }\bibfield  {title} {\enquote {\bibinfo {title} {{Direct
  measurement of a non-Hermitian topological invariant in a hybrid light-matter
  system}}}, }\href {\doibase 10.1126/sciadv.abj8905} {\bibfield  {journal}
  {\bibinfo  {journal} {Sci. Adv.}\ }\textbf {\bibinfo {volume} {7}},\ \bibinfo
  {pages} {eabj8905} (\bibinfo {year} {2021})}\BibitemShut {NoStop}%
\bibitem [{\citenamefont {Wang}\ \emph {et~al.}(2021)\citenamefont {Wang},
  \citenamefont {Dutt}, \citenamefont {Yang}, \citenamefont {Wojcik},
  \citenamefont {Vuckovi\'c},\ and\ \citenamefont {Fan}}]{KWang21Science}%
  \BibitemOpen
  \bibfield  {author} {\bibinfo {author} {\bibfnamefont {K.}~\bibnamefont
  {Wang}}, \bibinfo {author} {\bibfnamefont {A.}~\bibnamefont {Dutt}}, \bibinfo
  {author} {\bibfnamefont {K.~Y.}\ \bibnamefont {Yang}}, \bibinfo {author}
  {\bibfnamefont {C.~C.}\ \bibnamefont {Wojcik}}, \bibinfo {author}
  {\bibfnamefont {J.}~\bibnamefont {Vuckovi\'c}}, \ and\ \bibinfo {author}
  {\bibfnamefont {S.}~\bibnamefont {Fan}},\ }\bibfield  {title} {\enquote
  {\bibinfo {title} {{Generating arbitrary topological windings of a
  non-Hermitian band}}}, }\href {\doibase 10.1126/science.abf6568} {\bibfield
  {journal} {\bibinfo  {journal} {Science}\ }\textbf {\bibinfo {volume}
  {371}},\ \bibinfo {pages} {1240} (\bibinfo {year} {2021})}\BibitemShut
  {NoStop}%
\bibitem [{\citenamefont {Guo}\ \emph {et~al.}(2019)\citenamefont
  {Guo}, \citenamefont {Chen}, \citenamefont {Xie}, \citenamefont
  {Xiao}, \citenamefont {Deng},\ and\ \citenamefont
  {Yan}}]{Guo20PRL}%
  \BibitemOpen
  \bibfield  {author} {\bibinfo {author} {\bibfnamefont {W.}~\bibnamefont
  {Guo}}, \bibinfo {author} {\bibfnamefont {T.}~\bibnamefont {Chen}},
  \bibinfo {author} {\bibfnamefont {D.}~\bibnamefont {Xie}}, \bibinfo
  {author} {\bibfnamefont {T.~S.}~\bibnamefont {Deng}}, \bibinfo {author}
  {\bibfnamefont {B.}~\bibnamefont {Gadway}}, \bibinfo {author}
  {\bibfnamefont {W.}~\bibnamefont {Yi}}, \ and\ \bibinfo {author}
  {\bibfnamefont {B.}~\bibnamefont {Yan}},\ }\bibfield  {title} {\enquote
  {\bibinfo {title} {{Tunable Nonreciprocal Quantum Transport through a Dissipative Aharonov-Bohm Ring in Ultracold Atoms}}}, }\href {\doibase
  10.1103/PhysRevLett.124.070402} {\bibfield  {journal} {\bibinfo  {journal}
  {Phys. Rev. Lett.}\ }\textbf {\bibinfo {volume} {124}},\ \bibinfo {pages}
  {070402} (\bibinfo {year} {2020})}\BibitemShut {NoStop}%
\bibitem [{\citenamefont {{Liang}}\ \emph {et~al.}()\citenamefont {{Guo}},
  \citenamefont {{Liang}}, \citenamefont {{Xie}},\ and\ \citenamefont
  {{Yan}}}]{Liang22arXiv}%
  \BibitemOpen
  \bibfield  {author} {\bibinfo {author} {\bibfnamefont {Q.}\ \bibnamefont
  {{Liang}}}, \bibinfo {author} {\bibfnamefont {D.}~\bibnamefont {{Xie}}},
  \bibinfo {author} {\bibfnamefont {Z.}\ \bibnamefont {{Dong}}},
  \bibinfo {author} {\bibfnamefont {H.}\ \bibnamefont {{Li}}},
  \bibinfo {author} {\bibfnamefont {Z.}\ \bibnamefont {{Gadway}}},
  \bibinfo {author} {\bibfnamefont {W.}\ \bibnamefont {{Yi}}},\ and\
  \bibinfo {author} {\bibfnamefont {B.}\ \bibnamefont {{Yan}}},\
  }\bibfield  {title} {\enquote {\bibinfo {title} {{Observation of Non-Hermitian Skin Effect and Topology in Ultracold Atoms}}}, }\href@noop {}
  {\ }\Eprint {http://arxiv.org/abs/2201.09478} {arXiv:2201.09478} \BibitemShut
  {NoStop}%
\bibitem [{\citenamefont {Ren}\ \emph {et~al.}(2020)\citenamefont {Ren},
  \citenamefont {Liu}, \citenamefont {Zhao}, \citenamefont {He},
  \citenamefont {Pak}, \citenamefont {Li},\ and\ \citenamefont
  {Jo}}]{Ren2022chiral}%
  \BibitemOpen
  \bibfield  {author} {\bibinfo {author} {\bibfnamefont {Z.}~\bibnamefont
  {Ren}}, \bibinfo {author} {\bibfnamefont {D.}~\bibnamefont {Liu}}, \bibinfo
  {author} {\bibfnamefont {E.}~\bibnamefont {Zhao}}, \bibinfo {author}
  {\bibfnamefont {C.}~\bibnamefont {He}}, \bibinfo {author} {\bibfnamefont
  {K.~K.}~\bibnamefont {Pak}}, \bibinfo {author} {\bibfnamefont {J.}~\bibnamefont
  {Li}}, \ and\ \bibinfo {author} {\bibfnamefont {G.~B.}~\bibnamefont {Jo}},\
  }\bibfield  {title} {\enquote {\bibinfo {title} {{Chiral control of quantum states
   in non-Hermitian spin--orbit-coupled fermions}}}, }\href {\doibase
  10.1038/s41567-021-01491-x} {\bibfield  {journal} {\bibinfo  {journal} {Nature
  Phys.}\ }\textbf {\bibinfo {volume} {16}},\ \bibinfo {pages} {1-5} (\bibinfo
  {year} {2022})}\BibitemShut {NoStop}%
\bibitem [{\citenamefont {Haldane}(1981)}]{Haldane81JPC}%
  \BibitemOpen
  \bibfield  {author} {\bibinfo {author} {\bibfnamefont {F.}~\bibnamefont
  {Haldane}},\ }\bibfield  {title} {\enquote {\bibinfo {title} {{'Luttinger
  liquid theory'of one-dimensional quantum fluids. I. Properties of the
  Luttinger model and their extension to the general 1D interacting spinless
  Fermi gas}}}, }\href
  {https://iopscience.iop.org/article/10.1088/0022-3719/14/19/010/meta}
  {\bibfield  {journal} {\bibinfo  {journal} {J. Phys. C: Solid State Phys.}\
  }\textbf {\bibinfo {volume} {14}},\ \bibinfo {pages} {2585} (\bibinfo {year}
  {1981})}\BibitemShut {NoStop}%
\bibitem [{\citenamefont {Giamarchi}(2003)}]{Giamarchi03book}%
  \BibitemOpen
  \bibfield  {author} {\bibinfo {author} {\bibfnamefont {T.}~\bibnamefont
  {Giamarchi}},\ }\href@noop {} {\emph {\bibinfo {title} {Quantum physics in
  one dimension}}},\ Vol.\ \bibinfo {volume} {121}\ (\bibinfo  {publisher}
  {Clarendon press},\ \bibinfo {year} {2003})\BibitemShut {NoStop}%
\bibitem [{\citenamefont {Fradkin}(2013)}]{Fradkin13book}%
  \BibitemOpen
  \bibfield  {author} {\bibinfo {author} {\bibfnamefont {E.}~\bibnamefont
  {Fradkin}},\ }\href@noop {} {\emph {\bibinfo {title} {Field theories of
  condensed matter physics}}}\ (\bibinfo  {publisher} {Cambridge University
  Press},\ \bibinfo {year} {2013})\BibitemShut {NoStop}%
\bibitem [{\citenamefont {Nakagawa}\ \emph {et~al.}(2018)\citenamefont
  {Nakagawa}, \citenamefont {Kawakami},\ and\ \citenamefont
  {Ueda}}]{Nakagawa18prl}%
  \BibitemOpen
  \bibfield  {author} {\bibinfo {author} {\bibfnamefont {M.}~\bibnamefont
  {Nakagawa}}, \bibinfo {author} {\bibfnamefont {N.}~\bibnamefont {Kawakami}},
  \ and\ \bibinfo {author} {\bibfnamefont {M.}~\bibnamefont {Ueda}},\
  }\bibfield  {title} {\enquote {\bibinfo {title} {{Non-Hermitian Kondo Effect
  in Ultracold Alkaline-Earth Atoms}}}, }\href {\doibase
  10.1103/PhysRevLett.121.203001} {\bibfield  {journal} {\bibinfo  {journal}
  {Phys. Rev. Lett.}\ }\textbf {\bibinfo {volume} {121}},\ \bibinfo {pages}
  {203001} (\bibinfo {year} {2018})}\BibitemShut {NoStop}%
\bibitem [{\citenamefont {Yoshida}\ \emph
  {et~al.}(2019{\natexlab{b}})\citenamefont {Yoshida}, \citenamefont
  {Bessho},\ and\ \citenamefont {Sato}}]{Yoshida19SR}%
  \BibitemOpen
  \bibfield  {author} {\bibinfo {author} {\bibfnamefont {T.}~\bibnamefont
  {Yoshida}}, \bibinfo {author} {\bibfnamefont {K.}~\bibnamefont {Kudo}}, \
  and\ \bibinfo {author} {\bibfnamefont {Y.}~\bibnamefont {Hatsugai}},\ }\bibfield
  {title} {\enquote {\bibinfo {title} {{Non-Hermitian fractional quantum Hall states}}}, }\href {\doibase
  10.1038/s41598-019-53253-8} {\bibfield  {journal} {\bibinfo  {journal}
  {Sci. Rep.}\ }\textbf {\bibinfo {volume} {9}},\ \bibinfo {pages}
  {16895 } (\bibinfo {year} {2019}{\natexlab{b}})}\BibitemShut {NoStop}%
\bibitem [{\citenamefont {Hamazaki}\ \emph {et~al.}(2019)\citenamefont
  {Hamazaki}, \citenamefont {Kawabata},\ and\ \citenamefont
  {Ueda}}]{Hamazaki19PRL}%
  \BibitemOpen
  \bibfield  {author} {\bibinfo {author} {\bibfnamefont {R.}~\bibnamefont
  {Hamazaki}}, \bibinfo {author} {\bibfnamefont {K.}~\bibnamefont {Kawabata}},
  \ and\ \bibinfo {author} {\bibfnamefont {M.}~\bibnamefont {Ueda}},\
  }\bibfield  {title} {\enquote {\bibinfo {title} {{Non-Hermitian Many-Body
  Localization}}}, }\href {\doibase 10.1103/PhysRevLett.123.090603} {\bibfield
  {journal} {\bibinfo  {journal} {Phys. Rev. Lett.}\ }\textbf {\bibinfo
  {volume} {123}},\ \bibinfo {pages} {090603} (\bibinfo {year}
  {2019})}\BibitemShut {NoStop}%
\bibitem [{\citenamefont {Zhang}\ \emph
  {et~al.}(2020{\natexlab{b}})\citenamefont {Zhang}, \citenamefont {Zhang},
  \citenamefont {Li}, \citenamefont {Wang},\ and\ \citenamefont
  {Zhu}}]{GQZhang20PRB}%
  \BibitemOpen
  \bibfield  {author} {\bibinfo {author} {\bibfnamefont {G.-Q.}\ \bibnamefont
  {Zhang}}, \bibinfo {author} {\bibfnamefont {D.-W.}\ \bibnamefont {Zhang}},
  \bibinfo {author} {\bibfnamefont {Z.}~\bibnamefont {Li}}, \bibinfo {author}
  {\bibfnamefont {Z.~D.}\ \bibnamefont {Wang}}, \ and\ \bibinfo {author}
  {\bibfnamefont {S.-L.}\ \bibnamefont {Zhu}},\ }\bibfield  {title} {\enquote
  {\bibinfo {title} {{Statistically related many-body localization in the
  one-dimensional anyon Hubbard model}}}, }\href {\doibase
  10.1103/PhysRevB.102.054204} {\bibfield  {journal} {\bibinfo  {journal}
  {Phys. Rev. B}\ }\textbf {\bibinfo {volume} {102}},\ \bibinfo {pages}
  {054204} (\bibinfo {year} {2020}{\natexlab{b}})}\BibitemShut {NoStop}%
\bibitem [{\citenamefont {Zhai}\ \emph {et~al.}(2020)\citenamefont {Zhai},
  \citenamefont {Yin},\ and\ \citenamefont {Huang}}]{LJZhai20PRB}%
  \BibitemOpen
  \bibfield  {author} {\bibinfo {author} {\bibfnamefont {L.-J.}\ \bibnamefont
  {Zhai}}, \bibinfo {author} {\bibfnamefont {S.}~\bibnamefont {Yin}}, \ and\
  \bibinfo {author} {\bibfnamefont {G.-Y.}\ \bibnamefont {Huang}},\ }\bibfield
  {title} {\enquote {\bibinfo {title} {{Many-body localization in a
  non-Hermitian quasiperiodic system}}}, }\href {\doibase
  10.1103/PhysRevB.102.064206} {\bibfield  {journal} {\bibinfo  {journal}
  {Phys. Rev. B}\ }\textbf {\bibinfo {volume} {102}},\ \bibinfo {pages}
  {064206} (\bibinfo {year} {2020})}\BibitemShut {NoStop}%
\bibitem [{\citenamefont {Fayard}\ \emph {et~al.}(2021)\citenamefont {Fayard},
  \citenamefont {Henriet}, \citenamefont {Asenjo-Garcia},\ and\ \citenamefont
  {Chang}}]{Fayard21PRR}%
  \BibitemOpen
  \bibfield  {author} {\bibinfo {author} {\bibfnamefont {N.}~\bibnamefont
  {Fayard}}, \bibinfo {author} {\bibfnamefont {L.}~\bibnamefont {Henriet}},
  \bibinfo {author} {\bibfnamefont {A.}~\bibnamefont {Asenjo-Garcia}}, \ and\
  \bibinfo {author} {\bibfnamefont {D.~E.}\ \bibnamefont {Chang}},\ }\bibfield
  {title} {\enquote {\bibinfo {title} {Many-body localization in waveguide
  quantum electrodynamics}}, }\href {\doibase 10.1103/PhysRevResearch.3.033233}
  {\bibfield  {journal} {\bibinfo  {journal} {Phys. Rev. Research}\ }\textbf
  {\bibinfo {volume} {3}},\ \bibinfo {pages} {033233} (\bibinfo {year}
  {2021})}\BibitemShut {NoStop}%
\bibitem [{\citenamefont {Yamamoto}\ \emph {et~al.}(2019)\citenamefont
  {Yamamoto}, \citenamefont {Nakagawa}, \citenamefont {Adachi}, \citenamefont
  {Takasan}, \citenamefont {Ueda},\ and\ \citenamefont
  {Kawakami}}]{Yamamoto19PRL}%
  \BibitemOpen
  \bibfield  {author} {\bibinfo {author} {\bibfnamefont {K.}~\bibnamefont
  {Yamamoto}}, \bibinfo {author} {\bibfnamefont {M.}~\bibnamefont {Nakagawa}},
  \bibinfo {author} {\bibfnamefont {K.}~\bibnamefont {Adachi}}, \bibinfo
  {author} {\bibfnamefont {K.}~\bibnamefont {Takasan}}, \bibinfo {author}
  {\bibfnamefont {M.}~\bibnamefont {Ueda}}, \ and\ \bibinfo {author}
  {\bibfnamefont {N.}~\bibnamefont {Kawakami}},\ }\bibfield  {title} {\enquote
  {\bibinfo {title} {{Theory of Non-Hermitian Fermionic Superfluidity with a
  Complex-Valued Interaction}}}, }\href {\doibase
  10.1103/PhysRevLett.123.123601} {\bibfield  {journal} {\bibinfo  {journal}
  {Phys. Rev. Lett.}\ }\textbf {\bibinfo {volume} {123}},\ \bibinfo {pages}
  {123601} (\bibinfo {year} {2019})}\BibitemShut {NoStop}%
\bibitem [{\citenamefont {Lee}\ \emph {et~al.}(2020)\citenamefont {Lee},
  \citenamefont {Lee},\ and\ \citenamefont {Yang}}]{EWLee20PRB}%
  \BibitemOpen
  \bibfield  {author} {\bibinfo {author} {\bibfnamefont {E.}~\bibnamefont
  {Lee}}, \bibinfo {author} {\bibfnamefont {H.}~\bibnamefont {Lee}}, \ and\
  \bibinfo {author} {\bibfnamefont {B.-J.}\ \bibnamefont {Yang}},\ }\bibfield
  {title} {\enquote {\bibinfo {title} {{Many-body approach to non-Hermitian
  physics in fermionic systems}}}, }\href {\doibase
  10.1103/PhysRevB.101.121109} {\bibfield  {journal} {\bibinfo  {journal}
  {Phys. Rev. B}\ }\textbf {\bibinfo {volume} {101}},\ \bibinfo {pages}
  {121109} (\bibinfo {year} {2020})}\BibitemShut {NoStop}%
\bibitem [{\citenamefont {Mu}\ \emph {et~al.}(2020)\citenamefont {Mu},
  \citenamefont {Lee}, \citenamefont {Li},\ and\ \citenamefont
  {Gong}}]{SMu20PRB}%
  \BibitemOpen
  \bibfield  {author} {\bibinfo {author} {\bibfnamefont {S.}~\bibnamefont
  {Mu}}, \bibinfo {author} {\bibfnamefont {C.~H.}\ \bibnamefont {Lee}},
  \bibinfo {author} {\bibfnamefont {L.}~\bibnamefont {Li}}, \ and\ \bibinfo
  {author} {\bibfnamefont {J.}~\bibnamefont {Gong}},\ }\bibfield  {title}
  {\enquote {\bibinfo {title} {{Emergent Fermi surface in a many-body
  non-Hermitian fermionic chain}}}, }\href {\doibase
  10.1103/PhysRevB.102.081115} {\bibfield  {journal} {\bibinfo  {journal}
  {Phys. Rev. B}\ }\textbf {\bibinfo {volume} {102}},\ \bibinfo {pages}
  {081115} (\bibinfo {year} {2020})}\BibitemShut {NoStop}%
\bibitem [{\citenamefont {Zhang}\ \emph
  {et~al.}(2020{\natexlab{c}})\citenamefont {Zhang}, \citenamefont {Chen},
  \citenamefont {Zhang}, \citenamefont {Lang}, \citenamefont {Li},\ and\
  \citenamefont {Zhu}}]{DWZhang20PRB}%
  \BibitemOpen
  \bibfield  {author} {\bibinfo {author} {\bibfnamefont {D.-W.}\ \bibnamefont
  {Zhang}}, \bibinfo {author} {\bibfnamefont {Y.-L.}\ \bibnamefont {Chen}},
  \bibinfo {author} {\bibfnamefont {G.-Q.}\ \bibnamefont {Zhang}}, \bibinfo
  {author} {\bibfnamefont {L.-J.}\ \bibnamefont {Lang}}, \bibinfo {author}
  {\bibfnamefont {Z.}~\bibnamefont {Li}}, \ and\ \bibinfo {author}
  {\bibfnamefont {S.-L.}\ \bibnamefont {Zhu}},\ }\bibfield  {title} {\enquote
  {\bibinfo {title} {{Skin superfluid, topological Mott insulators, and
  asymmetric dynamics in an interacting non-Hermitian Aubry-Andr\'e-Harper
  model}}}, }\href {\doibase 10.1103/PhysRevB.101.235150} {\bibfield  {journal}
  {\bibinfo  {journal} {Phys. Rev. B}\ }\textbf {\bibinfo {volume} {101}},\
  \bibinfo {pages} {235150} (\bibinfo {year} {2020}{\natexlab{c}})}\BibitemShut
  {NoStop}%
\bibitem [{\citenamefont {Liu}\ \emph {et~al.}(2020)\citenamefont {Liu},
  \citenamefont {He}, \citenamefont {Yoshida}, \citenamefont {Xiang},\ and\
  \citenamefont {Nori}}]{TLiu20PRB}%
  \BibitemOpen
  \bibfield  {author} {\bibinfo {author} {\bibfnamefont {T.}~\bibnamefont
  {Liu}}, \bibinfo {author} {\bibfnamefont {J.~J.}\ \bibnamefont {He}},
  \bibinfo {author} {\bibfnamefont {T.}~\bibnamefont {Yoshida}}, \bibinfo
  {author} {\bibfnamefont {Z.-L.}\ \bibnamefont {Xiang}}, \ and\ \bibinfo
  {author} {\bibfnamefont {F.}~\bibnamefont {Nori}},\ }\bibfield  {title}
  {\enquote {\bibinfo {title} {{Non-Hermitian topological Mott insulators in
  one-dimensional fermionic superlattices}}}, }\href {\doibase
  10.1103/PhysRevB.102.235151} {\bibfield  {journal} {\bibinfo  {journal}
  {Phys. Rev. B}\ }\textbf {\bibinfo {volume} {102}},\ \bibinfo {pages}
  {235151} (\bibinfo {year} {2020})}\BibitemShut {NoStop}%
\bibitem [{\citenamefont {Panda}\ and\ \citenamefont
  {Banerjee}(2020)}]{Panda20PRB}%
  \BibitemOpen
  \bibfield  {author} {\bibinfo {author} {\bibfnamefont {A.}~\bibnamefont
  {Panda}}\ and\ \bibinfo {author} {\bibfnamefont {S.}~\bibnamefont
  {Banerjee}},\ }\bibfield  {title} {\enquote {\bibinfo {title} {{Entanglement
  in nonequilibrium steady states and many-body localization breakdown in a
  current-driven system}}}, }\href {\doibase 10.1103/PhysRevB.101.184201}
  {\bibfield  {journal} {\bibinfo  {journal} {Phys. Rev. B}\ }\textbf {\bibinfo
  {volume} {101}},\ \bibinfo {pages} {184201} (\bibinfo {year}
  {2020})}\BibitemShut {NoStop}%
\bibitem [{\citenamefont {Okuma}\ and\ \citenamefont
  {Sato}(2021)}]{OKuma21PRL}%
  \BibitemOpen
  \bibfield  {author} {\bibinfo {author} {\bibfnamefont {N.}~\bibnamefont
  {Okuma}}\ and\ \bibinfo {author} {\bibfnamefont {M.}~\bibnamefont {Sato}},\
  }\bibfield  {title} {\enquote {\bibinfo {title} {{Non-Hermitian Skin Effects
  in Hermitian Correlated or Disordered Systems: Quantities Sensitive or
  Insensitive to Boundary Effects and Pseudo-Quantum-Number}}}, }\href
  {\doibase 10.1103/PhysRevLett.126.176601} {\bibfield  {journal} {\bibinfo
  {journal} {Phys. Rev. Lett.}\ }\textbf {\bibinfo {volume} {126}},\ \bibinfo
  {pages} {176601} (\bibinfo {year} {2021})}\BibitemShut {NoStop}%
\bibitem [{\citenamefont {Yoshida}(2021)}]{Yoshida21PRB}%
  \BibitemOpen
  \bibfield  {author} {\bibinfo {author} {\bibfnamefont {T.}~\bibnamefont
  {Yoshida}},\ }\bibfield  {title} {\enquote {\bibinfo {title} {{Real-space
  dynamical mean field theory study of non-Hermitian skin effect for correlated
  systems: Analysis based on pseudospectrum}}}, }\href {\doibase
  10.1103/PhysRevB.103.125145} {\bibfield  {journal} {\bibinfo  {journal}
  {Phys. Rev. B}\ }\textbf {\bibinfo {volume} {103}},\ \bibinfo {pages}
  {125145} (\bibinfo {year} {2021})}\BibitemShut {NoStop}%
\bibitem [{\citenamefont {{Cao}}\ \emph {et~al.}()\citenamefont {{Cao}},
  \citenamefont {{Du}}, \citenamefont {{Wang}},\ and\ \citenamefont
  {{Kou}}}]{KCao21arXiv}%
  \BibitemOpen
  \bibfield  {author} {\bibinfo {author} {\bibfnamefont {K.}~\bibnamefont
  {{Cao}}}, \bibinfo {author} {\bibfnamefont {Q.}~\bibnamefont {{Du}}},
  \bibinfo {author} {\bibfnamefont {X.-R.}\ \bibnamefont {{Wang}}}, \ and\
  \bibinfo {author} {\bibfnamefont {S.-P.}\ \bibnamefont {{Kou}}},\ }\bibfield
  {title} {\enquote {\bibinfo {title} {{Physics of Many-body Nonreciprocal
  Model: Quantum system with Maxwell's Pressure Demon}}}, }\href@noop {} {\
  }\Eprint {http://arxiv.org/abs/2109.03690} {arXiv:2109.03690} \BibitemShut
  {NoStop}%
\bibitem [{\citenamefont {{Alsallom}}\ \emph {et~al.}()\citenamefont
  {{Alsallom}}, \citenamefont {{Herviou}}, \citenamefont {{Yazyev}},\ and\
  \citenamefont {{Brzezi{\'n}ska}}}]{Alsallom21arXiv}%
  \BibitemOpen
  \bibfield  {author} {\bibinfo {author} {\bibfnamefont {F.}~\bibnamefont
  {{Alsallom}}}, \bibinfo {author} {\bibfnamefont {L.}~\bibnamefont
  {{Herviou}}}, \bibinfo {author} {\bibfnamefont {O.~V.}\ \bibnamefont
  {{Yazyev}}}, \ and\ \bibinfo {author} {\bibfnamefont {M.}~\bibnamefont
  {{Brzezi{\'n}ska}}},\ }\bibfield  {title} {\enquote {\bibinfo {title} {{Fate
  of the non-Hermitian skin effect in many-body fermionic systems}}},
  }\href@noop {} {\ }\Eprint {http://arxiv.org/abs/2110.13164}
  {arXiv:2110.13164} \BibitemShut {NoStop}%
\bibitem [{\citenamefont {{Guo}}\ \emph {et~al.}()\citenamefont {{Guo}},
  \citenamefont {{Lieu}}, \citenamefont {{Tran}},\ and\ \citenamefont
  {{Gorshkov}}}]{YGuo21arXiv}%
  \BibitemOpen
  \bibfield  {author} {\bibinfo {author} {\bibfnamefont {A.~Y.}\ \bibnamefont
  {{Guo}}}, \bibinfo {author} {\bibfnamefont {S.}~\bibnamefont {{Lieu}}},
  \bibinfo {author} {\bibfnamefont {M.~C.}\ \bibnamefont {{Tran}}}, \ and\
  \bibinfo {author} {\bibfnamefont {A.~V.}\ \bibnamefont {{Gorshkov}}},\
  }\bibfield  {title} {\enquote {\bibinfo {title} {{Clustering of steady-state
  correlations in open systems with long-range interactions}}}, }\href@noop {}
  {\ }\Eprint {http://arxiv.org/abs/2110.15368} {arXiv:2110.15368} \BibitemShut
  {NoStop}%
\bibitem [{\citenamefont {Liu}\ \emph {et~al.}(2021)\citenamefont {Liu},
  \citenamefont {Xu},\ and\ \citenamefont {Li}}]{YGlIu21JPCM}%
  \BibitemOpen
  \bibfield  {author} {\bibinfo {author} {\bibfnamefont {Y.-G.}\ \bibnamefont
  {Liu}}, \bibinfo {author} {\bibfnamefont {L.}~\bibnamefont {Xu}}, \ and\
  \bibinfo {author} {\bibfnamefont {Z.}~\bibnamefont {Li}},\ }\bibfield
  {title} {\enquote {\bibinfo {title} {{Quantum phase transition in a
  non-Hermitian XY spin chain with global complex transverse field}}}, }\href
  {\doibase 10.1088/1361-648x/ac00dd} {\bibfield  {journal} {\bibinfo
  {journal} {J. Phys. Condens. Matter}\ }\textbf {\bibinfo {volume} {33}},\
  \bibinfo {pages} {295401} (\bibinfo {year} {2021})}\BibitemShut {NoStop}%
\bibitem [{\citenamefont {D\'ora}\ and\ \citenamefont
  {Moca}(2020)}]{Dora20PRL}%
  \BibitemOpen
  \bibfield  {author} {\bibinfo {author} {\bibfnamefont {B.}~\bibnamefont
  {D\'ora}}\ and\ \bibinfo {author} {\bibfnamefont {C.~P.}\ \bibnamefont
  {Moca}},\ }\bibfield  {title} {\enquote {\bibinfo {title} {{Quantum Quench in
  $\mathcal{PT}$-Symmetric Luttinger Liquid}}}, }\href {\doibase
  10.1103/PhysRevLett.124.136802} {\bibfield  {journal} {\bibinfo  {journal}
  {Phys. Rev. Lett.}\ }\textbf {\bibinfo {volume} {124}},\ \bibinfo {pages}
  {136802} (\bibinfo {year} {2020})}\BibitemShut {NoStop}%
\bibitem [{\citenamefont {Xi}\ \emph {et~al.}(2021)\citenamefont {Xi},
  \citenamefont {Zhang}, \citenamefont {Gu},\ and\ \citenamefont
  {Chen}}]{WJXi21SB}%
  \BibitemOpen
  \bibfield  {author} {\bibinfo {author} {\bibfnamefont {W.}~\bibnamefont
  {Xi}}, \bibinfo {author} {\bibfnamefont {Z.-H.}\ \bibnamefont {Zhang}},
  \bibinfo {author} {\bibfnamefont {Z.-C.}\ \bibnamefont {Gu}}, \ and\ \bibinfo
  {author} {\bibfnamefont {W.-Q.}\ \bibnamefont {Chen}},\ }\bibfield  {title}
  {\enquote {\bibinfo {title} {{Classification of topological phases in one
  dimensional interacting non-Hermitian systems and emergent unitarity}}},
  }\href {\doibase https://doi.org/10.1016/j.scib.2021.04.027} {\bibfield
  {journal} {\bibinfo  {journal} {Sci. Bull.}\ }\textbf {\bibinfo
  {volume} {66}},\ \bibinfo {pages} {1731} (\bibinfo {year}
  {2021})}\BibitemShut {NoStop}%
\bibitem [{\citenamefont {Pan}\ \emph {et~al.}(2020)\citenamefont {Pan},
  \citenamefont {Wang}, \citenamefont {Cui},\ and\ \citenamefont
  {Chen}}]{PLei20PRA}%
  \BibitemOpen
  \bibfield  {author} {\bibinfo {author} {\bibfnamefont {L.}~\bibnamefont
  {Pan}}, \bibinfo {author} {\bibfnamefont {X.}~\bibnamefont {Wang}}, \bibinfo
  {author} {\bibfnamefont {X.}~\bibnamefont {Cui}}, \ and\ \bibinfo {author}
  {\bibfnamefont {S.}~\bibnamefont {Chen}},\ }\bibfield  {title} {\enquote
  {\bibinfo {title} {{Interaction-induced dynamical $\mathcal{PT}$-symmetry
  breaking in dissipative Fermi-Hubbard models}}}, }\href {\doibase
  10.1103/PhysRevA.102.023306} {\bibfield  {journal} {\bibinfo  {journal}
  {Phys. Rev. A}\ }\textbf {\bibinfo {volume} {102}},\ \bibinfo {pages}
  {023306} (\bibinfo {year} {2020})}\BibitemShut {NoStop}%
\bibitem [{\citenamefont {Xu}\ and\ \citenamefont {Chen}(2020)}]{ZHXu21PRB}%
  \BibitemOpen
  \bibfield  {author} {\bibinfo {author} {\bibfnamefont {Z.}~\bibnamefont
  {Xu}}\ and\ \bibinfo {author} {\bibfnamefont {S.}~\bibnamefont {Chen}},\
  }\bibfield  {title} {\enquote {\bibinfo {title} {{Topological Bose-Mott
  insulators in one-dimensional non-Hermitian superlattices}}}, }\href
  {\doibase 10.1103/PhysRevB.102.035153} {\bibfield  {journal} {\bibinfo
  {journal} {Phys. Rev. B}\ }\textbf {\bibinfo {volume} {102}},\ \bibinfo
  {pages} {035153} (\bibinfo {year} {2020})}\BibitemShut {NoStop}%
\bibitem [{\citenamefont {Hyart}\ and\ \citenamefont {Hyart}(2020)}]{Hyart21arXiv}%
  \BibitemOpen
  \bibfield  {author} {\bibinfo {author} {\bibfnamefont {T.}~\bibnamefont
  {{Hyart}}}\ and\ \bibinfo {author} {\bibfnamefont {J.~L.}\ \bibnamefont
  {{Lado}}},\ }\bibfield   {title} {\enquote {\bibinfo {title} {{Non-Hermitian many-body topological excitations in interacting quantum dots
}}}, }\href
  {\doibase 10.1103/PhysRevResearch.4.L012006} {\bibfield  {journal} {\bibinfo
  {journal} {Phys. Rev. Research }\ }\textbf {\bibinfo {volume} {4}},\ \bibinfo
  {pages} {L012006 } (\bibinfo {year} {2022})}\BibitemShut {NoStop}%
\bibitem [{\citenamefont {Crippa}\ \emph {et~al.}(2021)\citenamefont {Crippa},
  \citenamefont {Budich},\ and\ \citenamefont {Sangiovanni}}]{Crippa21PRB}%
  \BibitemOpen
  \bibfield  {author} {\bibinfo {author} {\bibfnamefont {L.}~\bibnamefont
  {Crippa}}, \bibinfo {author} {\bibfnamefont {J.~C.}\ \bibnamefont {Budich}},
  \ and\ \bibinfo {author} {\bibfnamefont {G.}~\bibnamefont {Sangiovanni}},\
  }\bibfield  {title} {\enquote {\bibinfo {title} {Fourth-order exceptional
  points in correlated quantum many-body systems}}, }\href {\doibase
  10.1103/PhysRevB.104.L121109} {\bibfield  {journal} {\bibinfo  {journal}
  {Phys. Rev. B}\ }\textbf {\bibinfo {volume} {104}},\ \bibinfo {pages}
  {L121109} (\bibinfo {year} {2021})}\BibitemShut {NoStop}%
\bibitem [{\citenamefont {{Wang}}\ \emph {et~al.}()\citenamefont {{Wang}},
  \citenamefont {{Lang}},\ and\ \citenamefont {{He}}}]{WZuo21arxiv}%
  \BibitemOpen
  \bibfield  {author} {\bibinfo {author} {\bibfnamefont {Z.}~\bibnamefont
  {{Wang}}}, \bibinfo {author} {\bibfnamefont {L.-J.}\ \bibnamefont {{Lang}}},
  \ and\ \bibinfo {author} {\bibfnamefont {L.}~\bibnamefont {{He}}},\
  }\bibfield  {title} {\enquote {\bibinfo {title} {{Emergent Mott insulators at
  non-integer fillings and non-Hermitian conservation laws in an interacting
  bosonic chain with nonreciprocal hoppings}}}, }\href@noop {} {\ }\Eprint
  {http://arxiv.org/abs/2111.02911} {arXiv:2111.02911} \BibitemShut {NoStop}%
\bibitem [{\citenamefont {{Banerjee}}\ \emph {et~al.}()\citenamefont {{Banerjee}},
  \citenamefont {{Hegde}},\ and\ \citenamefont {{Narayan1}}}]{Banerjee21arxiv}%
  \BibitemOpen
  \bibfield  {author} {\bibinfo {author} {\bibfnamefont {A.}~\bibnamefont
  {{Banerjee}}}, \bibinfo {author} {\bibfnamefont {S.-S.}~\bibnamefont {{Hegde}}},
  \bibinfo {author} {\bibfnamefont {A.}~\bibnamefont {{Agarwala}}},
  \ and\ \bibinfo {author} {\bibfnamefont {A.}~\bibnamefont {{Narayan}}},\
  }\bibfield  {title} {\enquote {\bibinfo {title} {{Chiral metals and entrapped insulators in a
  one-dimensional topological non-Hermitian system}}}, }\href@noop {} {\ }\Eprint
  {https://ui.adsabs.harvard.edu/abs/2021arXiv211102223B} {arXiv:2111.02223} \BibitemShut {NoStop}%
\bibitem [{\citenamefont {Hatano}\ and\ \citenamefont
  {Nelson}(1996)}]{Hatano96PRL}%
  \BibitemOpen
  \bibfield  {author} {\bibinfo {author} {\bibfnamefont {N.}~\bibnamefont
  {Hatano}}\ and\ \bibinfo {author} {\bibfnamefont {D.~R.}\ \bibnamefont
  {Nelson}},\ }\bibfield  {title} {\enquote {\bibinfo {title} {{Localization
  Transitions in Non-Hermitian Quantum Mechanics}}}, }\href {\doibase
  10.1103/PhysRevLett.77.570} {\bibfield  {journal} {\bibinfo  {journal} {Phys.
  Rev. Lett.}\ }\textbf {\bibinfo {volume} {77}},\ \bibinfo {pages} {570}
  (\bibinfo {year} {1996})}\BibitemShut {NoStop}%
\bibitem [{Not({\natexlab{a}})}]{Note-defineGS}%
  \BibitemOpen
  \href@noop {} {\bibfield  {journal} {\bibinfo  {journal} {{Here, we are
  mainly interested in the low-$\text{Re}(E)$ regime of the spectrum. As in
  Hermitian systems, we refer the ground state to the state with the lowest
  $\text{Re}(E)$, and rank the excited states by their $\text{Re}(E)$ from low
  to high}}\ }}\BibitemShut {NoStop}%
\bibitem [{\citenamefont {{Zhang}}\ \emph {et~al.}()\citenamefont {{Zhang}},
  \citenamefont {{Yang}},\ and\ \citenamefont {{Fang}}}]{KZhang21arXiv}%
  \BibitemOpen
  \bibfield  {author} {\bibinfo {author} {\bibfnamefont {K.}~\bibnamefont
  {{Zhang}}}, \bibinfo {author} {\bibfnamefont {Z.}~\bibnamefont {{Yang}}}, \
  and\ \bibinfo {author} {\bibfnamefont {C.}~\bibnamefont {{Fang}}},\
  }\bibfield  {title} {\enquote {\bibinfo {title} {{Universal non-Hermitian
  skin effect in two and higher dimensions}}}, }\href@noop {} {\ }\Eprint
  {http://arxiv.org/abs/2102.05059} {arXiv:2102.05059} \BibitemShut {NoStop}%
\bibitem [{\citenamefont {Fukui}\ and\ \citenamefont
  {Kawakami}(1996)}]{Fukui98PRB}%
  \BibitemOpen
  \bibfield  {author} {\bibinfo {author} {\bibfnamefont {T.}~\bibnamefont
  {Fukui}}\ and\ \bibinfo {author} {\bibfnamefont {N.}\ \bibnamefont
  {Kawakami}},\ }\bibfield  {title} {\enquote {\bibinfo {title} {{Breakdown of the Mott insulator:
  Exact solution of an asymmetric Hubbard model}}}, }\href {\doibase
  10.1103/PhysRevB.58.16051} {\bibfield  {journal} {\bibinfo  {journal} {Phys.
  Rev. B}\ }\textbf {\bibinfo {volume} {58}},\ \bibinfo {pages} {16051}
  (\bibinfo {year} {1998})}\BibitemShut {NoStop}%
\bibitem [{Sup()}]{SuppInf}%
  \BibitemOpen
  \bibfield  {title} {{\bibinfo {title} {{See the Supplemental
  Material (SM) for details, which includes Refs.~\citep{Bergholtz21RMP,KZhang20PRL,Okuma20PRL,Byers61PRL,Kalthoff19PRB, Lindblad76CMP,Daley2014AP}}}}}\href@noop {} {\ }\BibitemShut {NoStop}%
\bibitem [{Not({\natexlab{a}})}]{Note-nonreciprocal}%
  \BibitemOpen
  \href@noop {} {\bibfield  {journal} {\bibinfo  {journal} {{Such a nonreciprocal hopping can be generated in cold-atom systems via synthetic magnetic flux and laser-induced loss, see e.g., Refs.~\cite{Guo20PRL,Liang22arXiv}}}\ }}\BibitemShut {NoStop}%
\bibitem [{Not({\natexlab{b}})}]{Note-failatsmallgamma}%
  \BibitemOpen
  \href@noop {} {\bibfield  {journal} {\bibinfo  {journal} {We note that for $\gamma <0.03$, we cannot extrapolate $U_{\protect \text {TD}}$ accurately with the exact-diagonalization results even up to the largest system size ($L=30$) that we can achieve in numerics. However, we observe that $U_{\protect \text {TD}}$ approaches $2t$ as $\gamma\rightarrow 0 $}}}\BibitemShut
  {NoStop}%
\bibitem [{\citenamefont {Affleck}\ and\ \citenamefont
  {Lieb}(1986)}]{Affleck86CMP}%
  \BibitemOpen
  \bibfield  {author} {\bibinfo {author} {\bibfnamefont {I.}~\bibnamefont
  {Affleck}}\ and\ \bibinfo {author} {\bibfnamefont {E.~H.}\ \bibnamefont
  {Lieb}},\ }\bibfield  {title} {\enquote {\bibinfo {title} {{A proof of part
  of Haldane's conjecture on spin chains}}}, }in\ \href@noop {} {\emph
  {\bibinfo {booktitle} {Condensed Matter Physics and Exactly Soluble
  Models}}}\ (\bibinfo  {publisher} {Springer},\ \bibinfo {year} {1986})\ pp.\
  \bibinfo {pages} {235--247}\BibitemShut {NoStop}%
\bibitem [{\citenamefont {Dalmonte}\ \emph {et~al.}(2015)\citenamefont
  {Dalmonte}, \citenamefont {Carrasquilla}, \citenamefont {Taddia},
  \citenamefont {Ercolessi},\ and\ \citenamefont {Rigol}}]{Dalmonte15PRB}%
  \BibitemOpen
  \bibfield  {author} {\bibinfo {author} {\bibfnamefont {M.}~\bibnamefont
  {Dalmonte}}, \bibinfo {author} {\bibfnamefont {J.}~\bibnamefont
  {Carrasquilla}}, \bibinfo {author} {\bibfnamefont {L.}~\bibnamefont
  {Taddia}}, \bibinfo {author} {\bibfnamefont {E.}~\bibnamefont {Ercolessi}}, \
  and\ \bibinfo {author} {\bibfnamefont {M.}~\bibnamefont {Rigol}},\ }\bibfield
   {title} {\enquote {\bibinfo {title} {{Gap scaling at
  Berezinskii-Kosterlitz-Thouless quantum critical points in one-dimensional
  Hubbard and Heisenberg models}}}, }\href {\doibase
  10.1103/PhysRevB.91.165136} {\bibfield  {journal} {\bibinfo  {journal} {Phys.
  Rev. B}\ }\textbf {\bibinfo {volume} {91}},\ \bibinfo {pages} {165136}
  (\bibinfo {year} {2015})}\BibitemShut {NoStop}%
\bibitem [{Not({\natexlab{c}})}]{Note-BKT-transition}%
  \BibitemOpen
  \href@noop {} {\bibfield  {journal} {\bibinfo  {journal} {{In the Hermitian
  limit, Eq.~(1) can be mapped to the XXZ model by Jordan-Wigner
  transformation}}\ }}\BibitemShut {NoStop}%
\bibitem [{\citenamefont {Byers}\ and\ \citenamefont
  {Yang}(1961)}]{Byers61PRL}%
  \BibitemOpen
  \bibfield  {author} {\bibinfo {author} {\bibfnamefont {N.}~\bibnamefont
  {Byers}}\ and\ \bibinfo {author} {\bibfnamefont {C.~N.}\ \bibnamefont
  {Yang}},\ }\bibfield  {title} {\enquote {\bibinfo {title} {Theoretical
  considerations concerning quantized magnetic flux in superconducting
  cylinders}}, }\href {\doibase 10.1103/PhysRevLett.7.46} {\bibfield  {journal}
  {\bibinfo  {journal} {Phys. Rev. Lett.}\ }\textbf {\bibinfo {volume} {7}},\
  \bibinfo {pages} {46} (\bibinfo {year} {1961})}\BibitemShut {NoStop}%
\bibitem [{Not({\natexlab{d}})}]{Pi-current}%
  \BibitemOpen
  \href@noop {} {\bibfield  {journal} {\bibinfo  {journal} {{{We note that a small magnetic flux $\phi$ will break the $\mathcal{PT}$ symmetry and generate a small imaginary free energy. Thus, the derivative of the free energy with respect to $\phi$ can be finite, leading to an imaginary persistent current. Imaginary persistent currents in non-Hermitian
  systems have been reported previously, see e.g.,~\protect \citep
  {Hatano96PRL,Hatano97prb,QMLi21PRB}.} However, they all focused on
  single-particle systems without interactions. A persistent current was also
  obtained within a field theoretical calculation in Ref.~\protect \citep
  {Kawabata21PRL}, which, in contrast to the current obtained in our
  tight-binding calculation, is real and quantized. Reconciling these results
  is an interesting question for future work}}\ }}\BibitemShut {NoStop}%
\bibitem [{\citenamefont {Cheung}\ \emph {et~al.}(1988)\citenamefont {Cheung},
  \citenamefont {Gefen}, \citenamefont {Riedel},\ and\ \citenamefont
  {Shih}}]{Cheung88PRB}%
  \BibitemOpen
  \bibfield  {author} {\bibinfo {author} {\bibfnamefont {H.-F.}\ \bibnamefont
  {Cheung}}, \bibinfo {author} {\bibfnamefont {Y.}~\bibnamefont {Gefen}},
  \bibinfo {author} {\bibfnamefont {E.~K.}\ \bibnamefont {Riedel}}, \ and\
  \bibinfo {author} {\bibfnamefont {W.-H.}\ \bibnamefont {Shih}},\ }\bibfield
  {title} {\enquote {\bibinfo {title} {Persistent currents in small
  one-dimensional metal rings}}, }\href {\doibase 10.1103/PhysRevB.37.6050}
  {\bibfield  {journal} {\bibinfo  {journal} {Phys. Rev. B}\ }\textbf {\bibinfo
  {volume} {37}},\ \bibinfo {pages} {6050} (\bibinfo {year}
  {1988})}\BibitemShut {NoStop}%
\bibitem [{Not({\natexlab{d}})}]{Note-configuration}%
  \BibitemOpen
  \href@noop {} {\bibfield  {journal} {\bibinfo  {journal} {{Namely, the number of bonds connecting two
  simultaneously occupied adjacent sites}}\ }}\BibitemShut {NoStop}%
 \bibitem [{Not({\natexlab{d}})}]{Note-winding}%
  \BibitemOpen
  \href@noop {} {\bibfield  {journal} {\bibinfo  {journal} {{For small $U<\Xi_R$, nontrivial winding numbers and the resulting many-body skin effect can also be found, as we have shown in Fig. S4 in the SM. However, these winding numbers may change as $U$ changes. The exact relation between these many-body winding numbers and a physical observable remains an open question}}\ }}\BibitemShut {NoStop}%
\bibitem [{Not({\natexlab{d}})}]{Note-Lee20PRB}%
  \BibitemOpen
  \href@noop {} {\bibfield  {journal} {\bibinfo  {journal} {{In many-body
  fermionic systems, the localization length of eigen wavefunctions in the
  presence of open boundaries depends on the particle number $N$. When the
  system is close to half-filling and for small $\gamma$, the
  localization length is comparable to $L$. In this
  case, it is hard to observe the localization behavior, as mentioned in
  Ref.~\citep{EWLee20PRB}. However, when the localization length is much
  smaller than $L$ (i.e., in the case with finite $N$ and large $L$ which we are
  considering), one can observe clearly the localization of all many-body
  wavefunctions to an open boundary. This localization behavior becomes constant
  for increasing $L$ (but fixed $N$), similar to the non-Hermitian skin effect
  in single-particle systems}}\ }}\BibitemShut {NoStop}%
\bibitem [{\citenamefont {Hatano}\ and\ \citenamefont
  {Nelson}(1997)}]{Hatano97prb}%
  \BibitemOpen
  \bibfield  {author} {\bibinfo {author} {\bibfnamefont {N.}~\bibnamefont
  {Hatano}}\ and\ \bibinfo {author} {\bibfnamefont {D.~R.}\ \bibnamefont
  {Nelson}},\ }\bibfield  {title} {\enquote {\bibinfo {title} {Vortex pinning
  and non-Hermitian quantum mechanics}}, }\href {\doibase
  10.1103/PhysRevB.56.8651} {\bibfield  {journal} {\bibinfo  {journal} {Phys.
  Rev. B}\ }\textbf {\bibinfo {volume} {56}},\ \bibinfo {pages} {8651}
  (\bibinfo {year} {1997})}\BibitemShut {NoStop}%
\bibitem [{\citenamefont {Li}\ \emph {et~al.}(2021)\citenamefont {Li},
  \citenamefont {Liu},\ and\ \citenamefont {Zhang}}]{QMLi21PRB}%
  \BibitemOpen
  \bibfield  {author} {\bibinfo {author} {\bibfnamefont {Q.}~\bibnamefont
  {Li}}, \bibinfo {author} {\bibfnamefont {J.-J.}\ \bibnamefont {Liu}}, \ and\
  \bibinfo {author} {\bibfnamefont {Y.-T.}\ \bibnamefont {Zhang}},\ }\bibfield
  {title} {\enquote {\bibinfo {title} {{Non-Hermitian Aharonov-Bohm effect in
  the quantum ring}}}, }\href {\doibase 10.1103/PhysRevB.103.035415} {\bibfield
   {journal} {\bibinfo  {journal} {Phys. Rev. B}\ }\textbf {\bibinfo {volume}
  {103}},\ \bibinfo {pages} {035415} (\bibinfo {year} {2021})}\BibitemShut
  {NoStop}%
\bibitem [{\citenamefont {Kawabata}\ \emph {et~al.}(2021)\citenamefont
  {Kawabata}, \citenamefont {Shiozaki},\ and\ \citenamefont
  {Ryu}}]{Kawabata21PRL}%
  \BibitemOpen
  \bibfield  {author} {\bibinfo {author} {\bibfnamefont {K.}~\bibnamefont
  {Kawabata}}, \bibinfo {author} {\bibfnamefont {K.}~\bibnamefont {Shiozaki}},
  \ and\ \bibinfo {author} {\bibfnamefont {S.}~\bibnamefont {Ryu}},\ }\bibfield
   {title} {\enquote {\bibinfo {title} {{Topological Field Theory of
  Non-Hermitian Systems}}}, }\href {\doibase 10.1103/PhysRevLett.126.216405}
  {\bibfield  {journal} {\bibinfo  {journal} {Phys. Rev. Lett.}\ }\textbf
  {\bibinfo {volume} {126}},\ \bibinfo {pages} {216405} (\bibinfo {year}
  {2021})}\BibitemShut {NoStop}%
\bibitem [{\citenamefont {Kalthoff}\ \emph {et~al.}(2019)\citenamefont
  {Kalthoff}, \citenamefont {Kennes},\ and\ \citenamefont
  {Sentef}}]{Kalthoff19PRB}%
  \BibitemOpen
  \bibfield  {author} {\bibinfo {author} {\bibfnamefont {M.~H.}\ \bibnamefont
  {Kalthoff}}, \bibinfo {author} {\bibfnamefont {D.~M.}\ \bibnamefont
  {Kennes}}, \ and\ \bibinfo {author} {\bibfnamefont {M.~A.}\ \bibnamefont
  {Sentef}},\ }\bibfield  {title} {\enquote {\bibinfo {title}
  {Floquet-engineered light-cone spreading of correlations in a driven quantum
  chain}}, }\href {\doibase 10.1103/PhysRevB.100.165125} {\bibfield  {journal}
  {\bibinfo  {journal} {Phys. Rev. B}\ }\textbf {\bibinfo {volume} {100}},\
  \bibinfo {pages} {165125} (\bibinfo {year} {2019})}\BibitemShut {NoStop}%
\bibitem [{\citenamefont {Lindblad}\ \emph {et~al.}(2019)\citenamefont
  {Lindblad}}]{Lindblad76CMP}%
  \BibitemOpen
  \bibfield  {author} {\bibinfo {author} {\bibfnamefont {G.}\ \bibnamefont
  {Lindblad}},\ }\bibfield  {title} {\enquote {\bibinfo {title}
  {On the generators of quantum dynamical semigroups}}, }\href {\doibase 10.1007/BF01608499} {\bibfield  {journal}
  {\bibinfo  {journal} {Commun. Math. Phys.}\ }\textbf {\bibinfo {volume} {48}},\
  \bibinfo {pages} {119} (\bibinfo {year} {1976})}\BibitemShut {NoStop}%
\bibitem [{\citenamefont {Daley}\ \emph {et~al.}(2019)\citenamefont
  {Daley}}]{Daley2014AP}%
  \BibitemOpen
  \bibfield  {author} {\bibinfo {author} {\bibfnamefont {A.~J}\ \bibnamefont
  {Daley}},\ }\bibfield  {title} {\enquote {\bibinfo {title}
  {Quantum trajectories and open many-body quantum systems}}, }\href {\doibase 10.1080/00018732.2014.933502} {\bibfield  {journal}
  {\bibinfo  {journal} {Adv. Phys.}\ }\textbf {\bibinfo {volume} {63}},\
  \bibinfo {pages} {77} (\bibinfo {year} {2014})}\BibitemShut {NoStop}%
\bibitem [{\citenamefont {Ferry}\ \emph {et~al.}(2019)\citenamefont
  {Ferry}, \citenamefont {Akis},\ and\ \citenamefont
  {Sentef}}]{Ferry12FP}%
  \BibitemOpen
  \bibfield  {author} {\bibinfo {author} {\bibfnamefont {D.~K.}\ \bibnamefont
  {Ferry}}, \bibinfo {author} {\bibfnamefont {R.}\ \bibnamefont  {Akis}},
  \bibinfo {author} {\bibfnamefont {A.~M.}\ \bibnamefont  {Burke}},
  \bibinfo {author} {\bibfnamefont {I.}\ \bibnamefont  {Knezevis}},
  \bibinfo {author} {\bibfnamefont {R.}\ \bibnamefont  {Brunner}},
  \bibinfo {author} {\bibfnamefont {R.}\ \bibnamefont  {Meisels}},
  \bibinfo {author} {\bibfnamefont {F.}\ \bibnamefont  {Kuchar}},
  \ and\ \bibinfo {author} {\bibfnamefont {J.~P.}\ \bibnamefont
  {Bird}},\ }\bibfield  {title} {\enquote {\bibinfo {title}
  {Open quantum dots: Physics of the non-Hermitian Hamiltonian}}, }\href {\doibase 10.1002/prop.201200065} {\bibfield  {journal}
  {\bibinfo  {journal} {Fortschr. Phys.}\ }\textbf {\bibinfo {volume} {61}},\
  \bibinfo {pages} {291} (\bibinfo {year} {2013})}\BibitemShut {NoStop}%
\bibitem [{\citenamefont {Zhang}\ \emph {et~al.}(2019)\citenamefont
  {Zhang}, \citenamefont {Zhan},\ and\ \citenamefont
  {Wu}}]{LLZhang18SM}%
  \BibitemOpen
  \bibfield  {author} {\bibinfo {author} {\bibfnamefont {L.~L.}\ \bibnamefont
  {Zhang}}, \bibinfo {author} {\bibfnamefont {G.~H.}\ \bibnamefont  {Zhan}},
  \bibinfo {author} {\bibfnamefont {D.~Q.}\ \bibnamefont  {Yu}},
  \ and\ \bibinfo {author} {\bibfnamefont {W.~J.}\ \bibnamefont
  {Gong}},\ }\bibfield  {title} {\enquote {\bibinfo {title}
  {Transport through a non-Hermitian parallel double-quantum-dot structure in the presence of interdot Coulomb interaction}}, }\href {\doibase 10.1016/j.spmi.2017.11.040} {\bibfield  {journal}
  {\bibinfo  {journal} {Fortschr. Phys.}\ }\textbf {\bibinfo {volume} {113}},\
  \bibinfo {pages} {558} (\bibinfo {year} {2018})}\BibitemShut {NoStop}%
\bibitem [{\citenamefont {Zajac}\ \emph {et~al.}(2019)\citenamefont
  {Zajac}, \citenamefont {Zajac},\ and\ \citenamefont
  {Petta}}]{Zajac16PRApplied}%
  \BibitemOpen
  \bibfield  {author} {\bibinfo {author} {\bibfnamefont {D.~M.}\ \bibnamefont
  {Zajac}}, \bibinfo {author} {\bibfnamefont {T.~M.}\ \bibnamefont  {Hazard}},
  \bibinfo {author} {\bibfnamefont {X.}\ \bibnamefont  {Mi}},
  \bibinfo {author} {\bibfnamefont {E.}\ \bibnamefont  {Nielsen}},
  \ and\ \bibinfo {author} {\bibfnamefont {J. R.}\ \bibnamefont
  {Petta}},\ }\bibfield  {title} {\enquote {\bibinfo {title}
  {Scalable Gate Architecture for a One-Dimensional Array of Semiconductor Spin Qubits}}, }\href {\doibase 10.1103/PhysRevApplied.6.054013} {\bibfield  {journal}
  {\bibinfo  {journal} {Phys. Rev. Applied}\ }\textbf {\bibinfo {volume} {6}},\
  \bibinfo {pages} {054013} (\bibinfo {year} {2016})}\BibitemShut {NoStop}%
\bibitem [{\citenamefont {Hensgens}\ \emph {et~al.}(2019)\citenamefont
  {Hensgens}, \citenamefont {Hensgens},\ and\ \citenamefont
  {Hensgens}}]{Hensgens17Nature}%
  \BibitemOpen
  \bibfield  {author} {\bibinfo {author} {\bibfnamefont {T.}\ \bibnamefont
  {Hensgens}},  \bibinfo {author} {\bibfnamefont {T.}\ \bibnamefont  {Fujita}},
  \bibinfo {author} {\bibfnamefont {L.}\ \bibnamefont  {Janssen}},
  \bibinfo {author} {\bibfnamefont {X.}\ \bibnamefont  {Li}},
  \bibinfo {author} {\bibfnamefont {C.~J.}\ \bibnamefont  {Van Diepen}},
  \bibinfo {author} {\bibfnamefont {C.}\ \bibnamefont  {Reichl}},
  \bibinfo {author} {\bibfnamefont {W.}\ \bibnamefont  {Wegscheider}},
  \bibinfo {author} {\bibfnamefont {S.}\ \bibnamefont  {Das Sarma}},
  \ and\ \bibinfo {author} {\bibfnamefont {J. R}\ \bibnamefont
  {Petta}},\ }\bibfield  {title} {\enquote {\bibinfo {title}
  {Quantum simulation of a Fermi-Hubbard model using a semiconductor quantum dot array}}, }\href {\doibase 10.1038/nature23022} {\bibfield  {journal}
  {\bibinfo  {journal} {Nature}\ }\textbf {\bibinfo {volume} {548}},\
  \bibinfo {pages} {70-73} (\bibinfo {year} {2017})}\BibitemShut {NoStop}%
\bibitem [{\citenamefont {Chin}\ \emph {et~al.}(2019)\citenamefont
  {Chin}, \citenamefont {Chin},\ and\ \citenamefont
  {Chin}}]{Chin10RMP}%
  \BibitemOpen
  \bibfield  {author} {\bibinfo {author} {\bibfnamefont {C.}\ \bibnamefont
  {Chin}}, \bibinfo {author} {\bibfnamefont {R.}\ \bibnamefont  {Grimm}},
  \bibinfo {author} {\bibfnamefont {P.}\ \bibnamefont  {Julienne}},
  \ and\ \bibinfo {author} {\bibfnamefont {E.}\ \bibnamefont
  {Tiesinga}},\ }\bibfield  {title} {\enquote {\bibinfo {title}
  {Feshbach resonances in ultracold gases}}, }\href {\doibase 10.1103/RevModPhys.82.1225} {\bibfield  {journal}
  {\bibinfo  {journal} {Rev. Mod. Phys.}\ }\textbf {\bibinfo {volume} {82}},\
  \bibinfo {pages} {1225} (\bibinfo {year} {2010})}\BibitemShut {NoStop}%
\bibitem [{\citenamefont {{Zhou}}\ \emph {et~al.}()\citenamefont {{Guo}},
  \citenamefont {{Zhou}}, \citenamefont {{Li}},\ and\ \citenamefont
  {{Yi}}}]{Zhou21arXiv}%
  \BibitemOpen
  \bibfield  {author} {\bibinfo {author} {\bibfnamefont {L.}\ \bibnamefont
  {{Zhou}}}, \bibinfo {author} {\bibfnamefont {D.}~\bibnamefont {{Liu}}},
  \bibinfo {author} {\bibfnamefont {H.}\ \bibnamefont {{Li}}},
  \bibinfo {author} {\bibfnamefont {W.}\ \bibnamefont {{Yi}}},\ and\
  \bibinfo {author} {\bibfnamefont {X.~L.}\ \bibnamefont {{Cui}}},\
  }\bibfield  {title} {\enquote {\bibinfo {title} {{Engineering Non-Hermitian Skin Effect with Band Topology in Ultracold Gases
}}}, }\href@noop {}
  {\ }\Eprint {https://arxiv.org/abs/2111.04196} {arXiv:2111.04196} \BibitemShut
  {NoStop}%
\bibitem [{\citenamefont {{Guo}}\ \emph {et~al.}()\citenamefont {{Guo}},
  \citenamefont {{Guo}}, \citenamefont {{Li}},\ and\ \citenamefont
  {{Yang}}}]{Guo21arXiv}%
  \BibitemOpen
  \bibfield  {author} {\bibinfo {author} {\bibfnamefont {S.}\ \bibnamefont
  {{Guo}}}, \bibinfo {author} {\bibfnamefont {C.}~\bibnamefont {{Dong}}},
  \bibinfo {author} {\bibfnamefont {F.}\ \bibnamefont {{Zhang}}},
  \bibinfo {author} {\bibfnamefont {J.}\ \bibnamefont {{Hu}}},\ and\
  \bibinfo {author} {\bibfnamefont {Z.}\ \bibnamefont {{Yang}}},\
  }\bibfield  {title} {\enquote {\bibinfo {title} {{Theoretical Prediction of Non-Hermitian Skin Effect in Ultracold Atom Systems}}}, }\href@noop {}
  {\ }\Eprint {https://arxiv.org/abs/2111.04220} {arXiv:2111.04220} \BibitemShut
  {NoStop}%
\bibitem [{\citenamefont {{Kells}}\ \emph {et~al.}()\citenamefont {{Guo}},
  \citenamefont {{Kells}}, \citenamefont {{Meidan}},\ and\ \citenamefont
  {{Romito}}}]{Kells21arXiv}%
  \BibitemOpen
  \bibfield  {author} {\bibinfo {author} {\bibfnamefont {G.}\ \bibnamefont
  {{Kells}}}, \bibinfo {author} {\bibfnamefont {D.}~\bibnamefont {{Meidan}}},\ and\
  \bibinfo {author} {\bibfnamefont {A.}\ \bibnamefont {{Romito}}},\
  }\bibfield  {title} {\enquote {\bibinfo {title} {{Topological transitions with continuously monitored free fermions}}}, }\href@noop {}
  {\ }\Eprint {https://arxiv.org/abs/2112.09787} {arXiv:2112.09787} \BibitemShut
  {NoStop}%
\bibitem [{\citenamefont {{Fleckenstein}}\ \emph {et~al.}()\citenamefont {{Guo}},
  \citenamefont {{Fleckenstein}}, \citenamefont {{Fleckenstein}},\ and\ \citenamefont
  {{Fleckenstein}}}]{Fleckenstein21arXiv}%
  \BibitemOpen
  \bibfield  {author} {\bibinfo {author} {\bibfnamefont {C.}\ \bibnamefont
  {{Fleckenstein}}},
  \bibinfo {author} {\bibfnamefont {A.}~\bibnamefont {{Zorzato}}},
  \bibinfo {author} {\bibfnamefont {D.}~\bibnamefont {{Varjas}}},
  \bibinfo {author} {\bibfnamefont {E.~J.}~\bibnamefont {{Bergholtz}}},
  \bibinfo {author} {\bibfnamefont {J.~H.}~\bibnamefont {{Bardarson}}}, \ and\
  \bibinfo {author} {\bibfnamefont {A.}\ \bibnamefont {{Tiwari}}},\
  }\bibfield  {title} {\enquote {\bibinfo {title} {{Non-Hermitian topology in monitored quantum circuits}}}, }\href@noop {}
  {\ }\Eprint {https://arxiv.org/abs/2201.05341} {arXiv:2201.05341} \BibitemShut
  {NoStop}%
\bibitem [{\citenamefont {Kawabata}\ \emph {et~al.}(2019)\citenamefont
  {Kawabata}, \citenamefont {Kawabata},\ and\ \citenamefont
  {Kawabata}}]{Kawabata22PRR}%
  \BibitemOpen
  \bibfield  {author} {\bibinfo {author} {\bibfnamefont {K.}\ \bibnamefont
  {Kawabata}},
  \bibinfo {author} {\bibfnamefont {K.}\ \bibnamefont  {Shiozaki}},
  \ and\ \bibinfo {author} {\bibfnamefont {S.}\ \bibnamefont
  {Ryu}},\ }\bibfield  {title} {\enquote {\bibinfo {title}
  {Many-body topology of non-Hermitian systems}}, }\href {\doibase 10.1103/PhysRevB.105.165137} {\bibfield  {journal}
  {\bibinfo  {journal} {Phys. Rev. B}\ }\textbf {\bibinfo {volume} {105}},\
  \bibinfo {pages} {165137} (\bibinfo {year} {2022})}\BibitemShut {NoStop}%
\end{thebibliography}
\end{document}